\documentclass[epj,nopacs]{svjour}
\usepackage{latexsym}
\usepackage{graphics}
\usepackage{amssymb}
\usepackage{amsmath}
\usepackage[english]{babel}
\usepackage{hyperref}
\usepackage{color}
\usepackage{graphicx}
\usepackage{subfigure}
\usepackage{longtable}
\usepackage{bm}
\usepackage{mwe}
\usepackage{float}
\usepackage{textcomp}
\usepackage{mathtools}
\usepackage{dcolumn}
\usepackage{epsf}
\usepackage[T1]{fontenc}
\usepackage{cuted}
\usepackage{microtype}
\usepackage{cite}
\usepackage{balance}

\begin{document}

\title{Tilt in quadratic gravity}
\author{Waleska P. F. de Medeiros\inst{1}\thanks{\emph{Present address:} waleskademedeiros@gmail.com}%
 \and Matheus J. Lazo\inst{1}\thanks{\emph{Present address:} matheusjlazo@gmail.com}%
\and Daniel M{\"u}ller\inst{2}\thanks{\emph{Present address:} dmuller@unb.br}%
\and Dinalva A. Sales\inst{1}\thanks{\emph{Present address:} dinalvaires@gmail.com}%
}                     
\offprints{Waleska P. F. de Medeiros}\mail{waleskademedeiros@gmail.com}          
\institute{Instituto de Matem{\'a}tica, Estat{\'i}stica e F{\'i}sica, Universidade Federal do Rio Grande, Caixa Postal 474, Rio Grande,\\ Rio Grande do Sul 96201-900, Brazil. \and Instituto de F{\'i}sica, Universidade de Bras{\'i}lia, Caixa Postal 04455, 70919-970, Bras{\'i}lia, Brazil.}
\date{Received: 19 March 2024 / Revised version: 30 July 2024}
%
\abstract{In this work, tilted source solutions in both Einstein-Hilbert General Relativity (GR) and Quadratic Gravity (QG) for the anisotropic Bianchi V model are addressed. Since the excellent CMBR match of Starobinsky's inflation with Planck's team measurements data, QG has acquired a prominent status in the effective sense, for sufficiently strong gravity fields. The main interest is in the numeric time evolution to the {\it past} towards the singularity and the behavior of the kinematic variables, vorticity, acceleration, and the expansion of this source substance. In QG we found that for universes with higher and smaller matter densities fall into the Kasner or isotropic singularity attractors to the past, respectively. We also found that the Kasner singularity attractor to the past has always zero vorticity, for both GR and QG theories. While for QG the isotropic singularity attractor may have divergent vorticity. For the set of assumptions and conditions supposed in this work, the isotropic singularity attractor, favors QG as compared to GR. Only in QG we were able to find a geometric singularity with divergences in all of the kinematic variables of the substance, decreasing to finite values to the {\it future, upon time reversing}. That is, we obtained an initial kinematic singularity substance, that approaches a perfect fluid source.
\PACS{
      {PACS-key}{}\and
      {PACS-key}{}\and
      {PACS-key}{}
     } 
} 
\maketitle
\global\long\def\imsize{0.9\columnwidth}%
\global\long\def\halfsize{0.50\columnwidth}%
\global\long\def\big{1.2\columnwidth}%
\global\long\def\peq{0.65\columnwidth}%

\section{Introduction}
\label{intro}
In this work, tilted source solutions in both Einstein-Hilbert (GR) and Quadratic Gravity (QG) are addressed. The approach is numerical, and only evolution to the past is considered. Albeit numeric, the procedure is non perturbative in the sense that the results depend only on the precision of the machine. The intention is to investigate the behavior of the kinematic variables of this source substance, for instance, vorticity, acceleration, and expansion toward the singularity, and expansion normalized variables (ENV) are the appropriate ones, since they remain finite at the singularity.

It must be emphasized that non tilted source is a particular case of the generalized tilted one and in this context it is a better physical candidate for an initial state of the Universe. In the cosmological scenario, it is expected that the most general initial choices to converge to physical acceptable space times, homogeneous isotropic with zero tilt. In this article, we show that this occurs for QG.

Tilt is not a result of a coordinate transformation or frame choice, it occurs when the four-velocity of the source substance is not orthonormal to the homogeneous surface, see for instance, \cite{King:1972td,ellisking}. Restricted to the spatially homogeneous space-times in the non tilted case, the constituent particles of the cosmological source follow geodesic motion. On the other hand, in the presence of tilt, the particles may not follow geodesics. In tilted sources the constituent particles can have non zero acceleration, vorticity, and expansion \cite{King:1972td,ellisking}. 

Past evolution towards singularity has been investigated in non tilted source solutions for quadratic gravity in Refs. \cite{Barrow:2006xb,Toporensky:2016kss}. It is already known that the isotropic singularity or false radiation solution, is a past attractor for the anisotropic solutions of quadratic gravity. In addition, it is also known that Kasner solution is a past attractor. To our knowledge, the behavior of a tilted source for quadratic gravity has not yet been investigated. 

As is well known, curvature scalars for instance, $R$, $R_{ab}R^{ab}$, $R_{abcd}R^{abcd}$, and analogous scalars obtained by polynomial expressions in derivatives of the curvature tensor diverge at the geometric singularity \cite{Hawking:1973uf}. However, a different type of singularity may occur, in which the kinematic variables diverge for tilted Bianchi models \cite{Coley_2006,Lim_2006,Coley:2008zz,Coley_2009}. Considering GR, Coley, Hervik, and Lim showed that for the Equation of State EoS parameter $1/3<w\leq 1$, for $p=w\rho$, the Bianchi V model solutions are future attracted to the Milne's solution, and tilt increases for dynamic time evolution to the future, see Refs. \cite{Coley_2006,Coley:2008zz}. In their analysis, the tilt increases, resulting in a divergence in the acceleration, in the shear, and in the expansion of the source. Also, the authors mention that according to GR, tilt can suppress the need for dark energy \cite{Coley_2006,Coley:2008zz}. Also, considering GR, the authors Allahyari, Ebrahimian, Krishnan, Mondol, and Sheikh-Jabbari are in good agreement with the tilt increase aforementioned for the EoS parameter $1/3<w\leq1$ \cite{Krishnan:2022qbv,Krishnan:2022uar,Ebrahimian:2023svi,Allahyari:2023kfm}. In addition, it is shown in \cite{Krishnan:2022qbv,Krishnan:2022uar} that the tilt can increase even when the anisotropic Bianchi V model becomes isotropic for time evolution to the future. Numerical time past evolution towards the singularity was addressed in \cite{Allahyari:2023kfm}.

The model of effective quadratic gravity occurs in a scenario in which quantized fields are taken into account on a classical gravitational background. Schwinger-DeWitt developed the method where the divergences present in the effective action can be eliminated by redefining some renormalized parameters \cite{Grib:330376}; see, for instance, \cite{dewitt1965dynamical,article,Christensen:1976vb,Christensen:1978yd,Birrell:1982ix}. 
The counter-terms required in order to have a consistent theory are 
\begin{align}
    S=&\int \mathrm{d}^4x\sqrt{-g}\frac{m^2_{p}}{2}\left [ \left(R-2\,\Lambda\right)+\beta R^2+\alpha  \left ( R_{ab}R^{ab} \right . \right.\nonumber\\&\left.\left.-{1}/{3}\,R^2\right ) \right ],\label{S}
\end{align} 
where $g$ is the determinant of the metric, $g_{ab}$, $\Lambda$ is the cosmological constant, $\alpha $ and $\beta$ are the renormalization parameters, and $m_p=(\hbar c^5/G)^{1/2}$ is the Planck mass, $c$ is the speed of light, $\hbar$ is the Planck constant, and $G$ is the gravitational constant. As usual, $R_{ab}$ is the Ricci tensor, and $R$ is the Ricci scalar. Metric variation on the action \eqref{S} results in field equations of fourth-order derivatives. GR is recovered when both $\beta=0$ and $\alpha =0$.

It is already well known the model of effective quadratic gravity for a homogeneous and isotropic Friedmann metric. Since the metric is isotropic, the field equations do not depend on $\alpha$, it occurs for all conformally flat spacetime because the term $ R_{ab}R^{ab}-{1}/{3}R^2$ in \eqref{S} is conformal invariant \cite{Ginzburg1971} and \cite{Parker:1993dk}. 

Starobinsky inflation \cite{Starobinsky:1980te} is now the best inflation model according to CMBR observations \cite{Ade:2015lrj}. It is an asymptotic isotropic solution of quadratic gravity. In $4$ dimensions and isotropic space-times, QG \eqref{S} reduces to a particular type of $f(R) \sim R+R^2$. $f(R)$ are gravity theories in which the action depends only on the Ricci scalar; see Refs. \cite{Sotiriou:2008rp,DeFelice2010,Nojiri:2010wj,Nojiri:2017ncd}. As it is equivalent to $f(R)$, Starobinsky inflation can be converted by the appropriate conformal metric transformation to Einstein-Hilbert gravity with an additional scalar field coupled to matter fields \cite{Maeda:1988ab,Cotsakis_2008,Sotiriou:2008rp,DeFelice2010,Nojiri:2010wj,Nojiri:2017ncd}.

Higher-order derivative theories have been known to be unstable since Ostrogradsky's time \cite{Ostrogradsky}. This type of instability occurs when there are kinetic energy terms for distinct degrees of freedom with opposite signs. These can lead to unlimited energy transfers between these degrees of freedom through self couplings, while maintaining the overall energy conservation of the system. Nowadays, ghost is the term used to describe this kind of instability \cite{PhysRev.79.145,Stelle:1977ry,Woodard:2007,trodden2016theoretical}. As it is known, linearization of the theory near Minkowski space shows a massless spin 2 field, a massive spin 0 field, and a massive spin 2 field that has energy with the wrong sign \cite{Stelle:1977ry}. It should be noted that ghosts in quadratic gravity cannot be eliminated by a suitable choice of theory parameters. However, despite the presence of the ghost in QG, it can be treated as a healthy classical theory of gravity, as the QG is free from causality issues \cite{Edelstein:2021jyu}. On the other hand, also related to linearization of the theory, the tachyon in the action \eqref{S} is eliminated for the following choice of the regularized parameters $\alpha<0$ and $\beta>0$, as pointed out by \cite{PhysRevD.32.379,MULLER:2014jaa}.

Nevertheless, we must mention that the gravitational $f(R)$ theory \cite{Woodard:2007,Schmidt:1994iz} is free from ghosts. This particular case violates the nondegeneracy assumption for Ostrogradsky's instability \cite{Ostrogradsky}. As pointed out in \cite{Woodard:2007}, the field equations from the gravitational $f(R)$ theory result in only a single higher-order derivative equation that carries the dynamics. Also, according to \cite{Woodard:2007}, the lower-order derivative equation results in a constraint that limits the lower derivative degrees of freedom.

In this work, anisotropic metrics are addressed; the effective quadratic action will also depend on $\alpha$, and it is not possible to disregard the Ricci tensor term $R_{ab}R^{ab}$ in \eqref{S} as occurs in the isotropic case. The past evolution toward the singularity is analyzed for the anisotropic Bianchi V model for both GR and QG. Bianchi V model is chosen because it is among the simplest ones that have a non-trivial time evolution of tilted matter. We discovered that in QG the solutions fall into the false radiation attractor, or Kasner attractor, respectively, for smaller or higher densities, $\Omega_m$. In QG the isotropic singularity solution allows for a complete singularity in all of the kinematic variables as in Coley, Hervik, Lim, et al. sense \cite{Coley_2006,Lim_2006,Coley:2008zz,Coley_2009}. While for GR, all singularities have zero vorticity. Since only evolution to the past is considered, inverting time otherwise to the future, this means that in GR any infinitesimally small amount of vorticity at the singularity will grow, while in QG it will decrease instead. The decrease of the vorticity in QG occurs for the false radiation singularity, a solution which does not exist in GR. 

This paper is organized as follows: In Sect. \ref{sec2}, the field equations for a tilted source are shown. In Sect. \ref{sec3}, the linear stability of some solutions in GR and QG is analyzed. In Sect. \ref{sec4}, the numeric results are presented. The behavior of kinematic variables is analyzed for both theories of gravity. Conclusions are shown in Sect. \ref{summary}.

For the numerical codes, it is used the GNU/GSL ode package with the Implicit 4th order Runge-Kutta at Gaussian points method on Linux. The computational codes were obtained using the algebraic manipulator Maple 17. The following conventions and units are taken: the metric signature $-+++$, the Latin indices $a$, $b$, ... run from $0-3$, and $G=\hbar=c=1$. The constraints fluctuate numerically, always smaller than $10^{-9}$.

\section{Field equations}\label{sec2}

The field equations following the metric variation of the effective action \eqref{S} are
\begin{align}
    & E_{ab} \equiv \nonumber\\&\left( G_{ab}+g_{ab}\Lambda \right)+\left ( \beta -\frac{1}{3}\,\alpha  \right )H^{(1)}_{ab}+\alpha H^{(2)}_{ab}-\kappa T_{ab}=0,\label{eqdemov}
\end{align}
where
\begin{align}
    G_{ab}=&\,R_{ab}-\frac{1}{2}\,g_{ab}R,\nonumber\\
    H^{(1)}_{ab}=&- \frac{1}{2}\,g_{ab}R^2+2\,RR_{ab}+2\,g_{ab}\square R-2\,{R}_{;ab},\nonumber\\
    H^{(2)}_{ab}=&-\frac{1}{2}\,g_{ab}R_{cd}R^{cd}-R_{;ab}+2\,R^{cd}{{{{R_c}_{bda}}}}+\square R_{ab}\nonumber\\&+\frac{1}{2}\,\square Rg_{ab}.
\end{align}
Here, $G_{ab}$ is the Einstein-Hilbert tensor, $H^{(1)}_{ab}$ results from metric variation on the term $R^2$, $H^{(2)}_{ab}$ results from metric variation on the term $R_{ab}R^{ab}$, $T_{ab }$ is the energy-momentum tensor, and $\kappa=1/m_p^2$. The field equations \eqref{eqdemov} have partial differential equations in order 4 on the metric. As it is well known, all vacuum solutions of GR field equation are also exact solutions of the effective quadratic action \eqref{S}. 

The covariant divergence of the equation of motion \eqref{eqdemov} vanishes, which implies that $E_{00}$ and $E_{0i}=0$ are constraints, while the dynamic equation is contained in the spatial part $E_{ij}$; see, for example, \cite{Weinberg:1972kfs,Stephanibook,Stelle:1977ry}. Thus, the temporal field equations, as the lowest-order equation, are a constraint that is dynamically preserved and is used as a numerical check.

In this work, dynamical time $t$, which is dimensionless, is chosen instead of the proper time $\tau$, $d\tau=dt/H(t)$ \cite{Coley:2008zz,Barrow:2006xb,Toporensky:2016kss}. 

It is used an orthogonal, non-rigid base $e_a^c$ where the index $a$ numerates the tetrad and a corresponding lapse function $1/H^2$ which results in the projected metric tensor $g_{ab}=-\delta_{a0}\delta_{b0}1/H^2 +\delta_{ij}\omega^i\otimes \omega^j$ where the $\omega^i$ are the dual of the spatial part of the basis $e_j$ such that as usual $\omega^i e_j=\delta^i_j$. In other words,
\begin{equation}
    {g_{ab}=\begin{pmatrix}
-H(t)^{-2} &0  &0  &0 \\ 
 0&1  & 0 &0 \\ 
 0& 0 &1  &0 \\ 
 0& 0 & 0 &1 \label{mett}
\end{pmatrix}},
\end{equation}
tetrads are chosen only for the spatial part of the metric.

First, the temporal part of the connection is determined. The time-like vector is chosen as $u^a=(H,0,0,0)$, with normalization, $g_{ab}u^a u^b=-1$, and the projection of the metric on the orthogonal three-dimensional space to $u^a$ as $h_{ab}=u_au_b+g_{ab}$, where ${h^c}_bg_{ca}=h_{ab}$, and ${h^c}_au_c=0$. 
 
Its covariant derivative $\nabla _au_b$ can be decomposed into its irreducible parts \cite{Stephanibook}: an anti symmetric part, which is called the rotation tensor $\omega_{ab}$, a symmetric traceless part, $\sigma_{ab}$, which is called the shear tensor, and another part containing only its trace $\Theta$
\begin{align}
    &\nabla _au_b=\sigma _{ab}+\omega _{ab}+\frac{1}{3}\,\Theta \,\delta _{ab}-\dot{u}_au_b,\nonumber\\
    &\sigma _{ab}=u_{(a;b)}-\frac{1}{3}\,\Theta \,\delta _{ab}+\dot{u}(_{a}u_b),\nonumber\\
    &\omega  _{ab}=u_{[a;b]}+\dot{u}[_{a}u_b],\nonumber\\
    &\dot{u}_a=u^b\nabla _bu_a,\nonumber\\
    &\Theta =\nabla _au^a \label{cinva},
\end{align}
where $\dot{u}_a$ is the acceleration vector.

The vorticity vector is related to the rotation tensor as follows
\begin{equation}
    \omega_a=\frac{1}{2}\,\epsilon_{abcd}\,u^b\,\omega^{cd},\label{vortve}
\end{equation}
where $\epsilon^{abcd}$ is the levi-civita tensor. It can be seen that this vector $u^a$ is geodesic with zero vorticity for spatially homogeneous geometries.

On the other hand, as any covariant derivative in the base $e_a^c$ can be written with the appropriate connection \footnote{see below.}
\begin{equation}
    \nabla _au_b=e^c_a\partial_c u_b-\Gamma^c_{ba}u_c\label{nabla}
\end{equation}
for this specific vector $u_a=(-1/H,0,0,0)$ becomes 
\begin{equation}
    {\nabla _au_b=\delta _{a0}\delta _{b0}\frac{\dot{H}}{H^2}-\Gamma ^c_{ab}u_c=\delta _{a0}\delta _{b0}\frac{\dot{H}}{H^2}+\frac{\Gamma ^0_{ab}}{H}},\label{tri}
\end{equation}
which in accordance to \eqref{cinva}
\begin{equation}
    {\Gamma ^0_{ab}=\left\{\begin{array}{ll}
&0,  \,  \mbox{if} \,  a=0 \,\mbox{ or} \, b=0,  \\
& H\sigma _{ij}+H^2\delta _{ij}, \,\mbox{if} \, a \neq 0 \,  \mbox{and} \,  b\neq 0.  
\end{array}\right.} \label{Gamma_0ab}
\end{equation}
Also, taking into account \eqref{cinva}, \eqref{tri}, \eqref{Gamma_0ab} and remembering that the shear $\sigma_{ij}$ has zero trace, it is immediately seen that using this connection, $\Theta=\nabla_a u^a=3H$. The shear in \eqref{Gamma_0ab} is chosen as
\begin{equation}
      {{\sigma _{ij}}= \left( \begin {array}{ccc} \,- {2\,\sigma_{{+}}
}&0& {\phi_{{3}} }\\ \noalign{\medskip}0&{\sigma_{{+}}  +\sqrt {3}\,
\sigma_{{-}}  }&{0}\\ \noalign{\medskip} {\phi_{{3}} }&{0}& {\sigma_{{+
}} -\sqrt {3}\,\sigma_{{-}} }\end {array} \right) }.\label{sig1}
  \end{equation}
Here, zero shear corresponds to the isotropic case, see Section \ref{subsec21} and Appendix \ref{appa}. 

Now we turn to the spatial part of the connection. 
Since the basis $e_a^c$ are non-coordinate, the connection is uniquely determined by metricity and zero torsion respectively
\begin{align}
   &\nabla _cg_{ab}=0 \implies e_c^d\partial_d g_{ab}-\Gamma^d_{ac}g_{db}-\Gamma^d_{bc}g_{ad}=0,\nonumber\\
   &\nabla _ae_{b}-\nabla _be_{a}=[e _{a},e_{b}]\implies (\Gamma^d_{ba}-\Gamma^d_{ab})e_d=[e _{a},e_{b}], \label{mtr_torc}
\end{align}
where the metric is given by \eqref{mett}.

For spatially homogeneous models, the commutator of the spatial part of the basis is,
\begin{equation}
    [e_i,e_j]=-C^k_{ij}e_k,\label{sc1}
\end{equation}
where $C^k_{ij}=-C^k_{ji}$.

The spatial derivatives of the metric \eqref{mett} are zero and do not contribute to the connection, the metricity condition implies for the spatial part that $\Gamma_{ijk}=-\Gamma_{jik}$ \eqref{mtr_torc} while zero torsion implies that $\Gamma_{c\,ba}-\Gamma_{c\,ab}=C_{c\,ab}$ \eqref{mtr_torc} resulting for the pure spatial part 
\begin{equation}
{\Gamma _{i\,jk}=\frac{1}{2}(C_{jki}+C_{kji}-C_{ikj}).}\label{cestru}
        \end{equation}
The Bianchi V model is chosen because it is among the simplest ones that contains a non-trivial behavior for tilted sources. The only non-null structure constants appropriate for Bianchi V geometries are then \cite{Stephani:2003tm}
\begin{align}
    &C^2_{12}=C^3_{13}=b(t),\,C^2_{21}=C^3_{31}=-b(t).\label{ce}
\end{align}
Here, time dependence means that the structure constants are preserved at the slices of constant time homogeneous hypersurfaces. These structure constants \eqref{ce} for Bianchi V result in the only non-null components for the connection 
\begin{align}
    &\Gamma _{212}=\Gamma _{313}=b(t), &\Gamma _{122}=\Gamma _{133}=-b(t). \label{Gamma_ijk_q}
\end{align}
The remaining component of the connection is determined by metricity $\partial_0g_{00}-2\Gamma_{0\,00}=0$,
 \begin{align}
    \Gamma ^0_{00}=-{\dot{H}}/{H}. \label{Gamma_000}
\end{align}

The connection is not completely determined by \eqref{Gamma_000}, \eqref{Gamma_0ab} with shear \eqref{sig1} and \eqref{Gamma_ijk_q}. This happens because instead of supposing the metric as the dynamical variable, we are supposing the connection and there is an additional condition which must be fulfilled in order for consistency shown in the following.

As usual, the Riemann tensor results from the commutator of the covariant derivatives
\begin{equation}
    {\left [ \nabla _c,\nabla _d \right ]V^a={R^a}_{bcd}V^b},
\end{equation}
        \begin{align}
            &{{R^{a}}_{bcd}=\Gamma ^{a}_{bd|c}-\Gamma ^{a}_{bc|d}+\Gamma ^{a}_{fc}\Gamma ^{f}_{bd}-\Gamma ^{a}_{fd}\Gamma ^{f}_{bc}+C^f_{cd}\Gamma ^{a}_{bf}}.
        \end{align}
The Riemann tensor must satisfy the Jacobi identity $R_{abcd}+R_{acdb}+R_{adbc}=0$, the above-mentioned condition, which results in a first-order differential equation
\begin{equation}
    ( \dot{b} +b)H -2b \sigma_{{+}}  =0.\label{jacobi}
\end{equation}

\subsection{Tilted source and Einstein-Hilbert gravity}\label{subsec21}

We begin with the isotropic Bianchi V GR non-tilted case, and as is well known, this reproduces Friedmann's open model.
The spatial isotropic line element $d\sigma^2=a(t)^2\delta_{ab}\omega^a\otimes \omega^b$ which for the appropriate Bianchi V 1-form basis $\omega^a$ \cite{Stephani:2003tm} results 
\[
 d\sigma ^2=a^2(t)\left(dx^2+e^{2x}dy^2+e^{2x}dz^2\right).
\]
The following coordinate transformation 
\begin{align}
&x=\ln\left ( \cosh\chi - \sinh\chi \cos\theta  \right ),\nonumber\\
&y=\frac{\sin\theta \cos\phi }{\coth \chi -\cos\theta },\nonumber\\
&z=\frac{\sin\theta \sin\phi }{\coth \chi -\cos\theta },
\end{align}
leads to the \textit{Friedmann-Lemaître-Robertson-Walker} (FLRW) line element 
\begin{equation}
    d\sigma^2=a^2(t)\left[d\chi^2+\sinh^2\chi\left ( d \theta^2+\sin^2\phi \, d \phi ^2\right )\right]
\end{equation}
for negative curvature, $K=-1$.

The field equations for GR are obtained by setting $\alpha=0$ and $\beta=0$ in \eqref{eqdemov}
\begin{equation}
    E_{ab }\equiv G_{ab }-\kappa T_{ab }=0,
\end{equation}
with source the perfect fluid $T_{ab}=(\rho+p)u_au_b+pg_{ab}$, where $p=w\rho$ with equation of state parameter EoS $w$, metric \eqref{mett} and fluid velocity $u_a=(-1/H,0,0,0)$. This altogether results in Friedmann's equations 
\begin{align*}
   & -\frac{\kappa \rho}{H^2}+\frac{-3b^2+3H^2}{H^2}=0,\nonumber\\
   &-2H\dot{H}-3H^2+b^2-w\kappa\rho=0,
\end{align*}
such that defining the usual density and curvature parameters
\begin{align*}
&\Omega_{m }=\frac{\kappa\rho}{3H^2}, & \Omega_K=b^2/H^2, 
 \end{align*}
we get the following Friedmann equations with respect to dynamical time $t$, $dt/H=d\tau$ instead of proper time $\tau$
\begin{align}
   & \Omega_m+\Omega_K-1=0,\nonumber\\
   &2\frac{\dot{H}}{H}-\Omega_K+3(1+w\Omega_m)=0,\label{fried}
\end{align}
together with the additional equation to be satisfied by $\Omega_K$
\begin{equation}
     \dot{\Omega}_{K} =- 2\,\Omega_{K} +4\,\Omega_{{K}} \Sigma_{+} -2\,\frac{\Omega_{{K}} \dot{H}  }{{H  }},\label{deromegak}
\end{equation}
which is Jacobi identity \eqref{jacobi} written for the curvature density $\Omega_K$. This above equation is one of the dynamical equations that must be satisfied throughout this article. 

Now the connection is completely defined by  \eqref{Gamma_000}, \eqref{Gamma_0ab} with shear \eqref{sig1} and 
\begin{align}
    &\Gamma _{212}=\Gamma _{313}=H\sqrt{\Omega_K}, &\Gamma _{122}=\Gamma _{133}=-H\sqrt{\Omega_K}. \label{Gamma_ijk}
\end{align}
As already mentioned, expansion-normalized variables ENV are good variables to describe time evolution of the Universe to the past since they remain finite at the cosmological singularity. They are dimensionless and are obtained by dividing the shear components by $H$. For GR, the necessary ENV 
\begin{align}
&\Omega_{m }=\frac{\kappa\rho}{3H^2},  &\Phi_3=\frac{\phi _{3 }}{H},     \nonumber\\ 
      &\Sigma_{\pm}=\frac{\sigma _{\pm }}{H},
      & \Omega_K.\label{ENV1}
 \end{align}

Now we turn to non-isotropic tilted Bianchi V solutions in GR. First, in Appendix \ref{appa} it is provided a detailed overview of the anisotropic Bianchi V model. Remind the well known relation for the Hubble parameter $H$, $H=a^\prime/a$, where $\prime$ is the derivative with respect to proper time $\tau$ and $a$ is the scale factor. For non-isotropic, or the non zero shear case, there is no scale factor, however it can be defined as
\[ 
H=a^\prime/a \implies a=a_0e^{\int H d\tau}=a_0e^t,
\]
where $a_0$ is a constant of integration and remembering that the dynamical time relates to proper time as $d\tau=dt/H$. Then, the deceleration parameter can be rewritten with respect to the dynamical time $t$
\begin{equation}
    q=-\frac{a^{\prime\prime}a}{(a^\prime)^2}=-(\dot{H}/H+1)=-Q_1-1,\label{decpar}
\end{equation} 
where $Q_1$ is defined in Section \ref{subsec22}. 

Tilt is described by the energy-momentum tensor 
\begin{equation}
    T_{ab}=\frac{1}{\kappa}\left [ (1+w)\,3\,H^2\Omega_m  \hat{u}_a \hat{u}_b +3\,wH^2\Omega_m\,g_{ab}\right],\label{tem}
\end{equation}
where the metric is \eqref{mett}, $w$ the Equation of State parameter EoS, $p=w\rho$ and the time-like vector $\hat{u}^a$ 
\begin{equation}    
\hat{u}^a=\left[H\cosh(r),\sinh(r)\cos(\eta),0,\sinh(r)\sin(\eta)\right],\label{timeve}
\end{equation}
with normalization $\hat{u}^a\hat{u}^bg_{ab}=-1$. Tilted sources are non-perfect fluids types that have energy and momentum fluxes \cite{King:1972td}. The tilt variable $r$ is dimensionless, and the direction of the tilt is given by $\eta$ in radians, so that the time evolution of $r$ describes tilt changes in time while $\eta$ defines the direction of this tilt. The first thing to note is that when tilt is zero, $r=0$, it is recovered the usual perfect fluid source, and $\hat{u}^a$ coincides with the geodesic vector $u^a=(H,0,0,0)$. Then the tilt $r$ is with respect to the geodesic and vorticity free vector $u^a$.

Since the tilted substance moves according to $\hat{u}^a$ as chosen in \eqref{timeve}, there is the appropriate set of kinematic variables instead of \eqref{cinva}-\eqref{vortve}. To obtain these kinematic variables for tilted matter, $\nabla_a\hat{u}_b$, it is used metric \eqref{mett} and connection defined by \eqref{Gamma_000}, \eqref{Gamma_0ab} with \eqref{sig1}, \eqref{Gamma_ijk} and \eqref{deromegak} which gives
\begin{align}
\hat{\dot{u}}_0=&- {H}^{2}w\sinh \left( r  \right)  D,\nonumber\\
\hat{\dot{u}}_1=&- 3\,H w \cos \left( \eta \right) \cosh \left( r  \right)  D,\nonumber\\
\hat{\dot{u}}_3=&-3\,H w\cosh \left( r  \right) \sin \left( \eta \right) D,\nonumber\\
\hat{\Theta}=&-\left\{ \cosh \left( r  \right) H \left[ 2\,\sinh \left( r  \right) \cos \left( \eta \right) \cosh
 \left( r  \right) \sqrt {\Omega_{{K}}} \right.\right.\nonumber\\&\left.\left. +A
 \right]
\right\}/\left\{ B \right\},\nonumber\\
   \hat{\omega}^2 =&-
{\frac { \left| H \right| \sinh \left( r   \right) 
\sin \left( \eta  \right) \sqrt {\Omega_{{K}}}}{\cosh
 \left( r  \right) }},\label{cininc}
\end{align}
where $\hat{\dot{u}}_2$ and the other components of the vorticity are identically zero. The terms $A$, $B$, $C$, and $D$ are defined to simplify the notation as
\begin{align}
   A  =  &-\Sigma_- \left(\left( \cos
 \left( \eta \right)\right)^2  -1 \right) \left( \left( \cosh \left( r  \right)\right)^2 -1 \right) \sqrt {3}\nonumber\\&+ \left( 3\, \left( \cos \left( \eta \right)  \right) ^{2}\Sigma_{+}-2\,\cos \left( \eta \right) \sin \left( \eta \right) \Phi_3+2\right. \nonumber\\&\left.-\Sigma_{+}\right) \left( \cosh \left( r  \right)  \right) ^{2}+2\,\cos \left( \eta \right) \sin \left( \eta \right)\Phi_3+1\nonumber\\&+\Sigma_{+}\left(1-3\, \left( \cos \left( \eta \right)  \right) ^{2}\right),\nonumber\\
B = &\left( w-1 \right)  \left( \cosh \left( r  \right)  \right) ^{2}-w,\nonumber\\
C = & \,2\,\cosh \left( r  \right) \cos \left( \eta \right)\sqrt {\Omega_{{K}}}\left( \left( \cosh \left(r  \right)  \right) ^{2} -1\right),\nonumber\\
D =&  \left[  \,C +A \sinh \left( r  \right)  \right] /\left(B\right).
\end{align}
It is straightforward to see that both acceleration $\hat{\dot{u}}^a=0$ and vorticity $\hat{\omega}^a=0$ in \eqref{cininc} are zero when $r=0$ while $\hat{\Theta}=3H$, such that geodesic motion of matter with zero vorticity is recovered when tilt is absent. For dust-dominated Universe with EoS parameter $w=0$ in \eqref{cininc}, $\hat{\dot{u}}^a=0$, that is, the motion of the substance is geodesic with non-zero vorticity, as long as $\Omega_K \neq 0$. Also \eqref{cininc}, the vorticity vector $\hat{\omega}^a$ is also zero when $\eta=0$, $\eta=\pi$, or whenever $\Omega_K$ is zero. 

Consider this particular $\hat{u}^a$ given in \eqref{timeve} and energy-momentum source \eqref{tem}, metric \eqref{mett} and connection defined by \eqref{Gamma_000}, \eqref{Gamma_0ab} with \eqref{sig1}, \eqref{Gamma_ijk} and \eqref{deromegak} the field equations $E_{02}\equiv 0$, $E_{21}\equiv 0$ and $E_{23}\equiv 0$ in \eqref{eqdemov} are identically satisfied for whichever values of $\alpha$ and $\beta$, i.e. $E_{02}\equiv 0$, $E_{21}\equiv 0$ and $E_{23}\equiv 0$ both in GR and QG.

On the other hand, energy-momentum covariant conservation
\[\nabla_bT^{ab}=0,\]
must be obeyed independent of the gravitational theory. Again, considering the same setting, namely $\hat{u}^a$ in \eqref{timeve} and energy-momentum source \eqref{tem}, metric \eqref{mett} and connection defined by \eqref{Gamma_000}, \eqref{Gamma_0ab} with \eqref{sig1}, \eqref{Gamma_ijk} and \eqref{deromegak} results in the following non-trivial differential equations for the variables $H$, $\Omega_m$, $\eta$ and $r$:
\begin{align}
\nabla_bT^{0b}=&\, 3\,H  \left\{ -2\,\cos \left( \eta  \right) \Omega_{{m}}H  \cosh\left( r  \right) \sinh \left( r  \right)  \left( 1+w \right) \right.\nonumber\\&\left.\sqrt {\Omega_{{K}}  }-H  \left(  \left( 1+w \right)  \left( \cosh \left( r  \right)  \right) ^{2}-w \right) \dot{\Omega}_{{m}} + \left[2\,\right.\right.\nonumber\\&\left.\left.\left( w -\left( 1+w \right)  \left( \cosh \left( r  \right)  \right) ^{2}\right) \dot{H} + \left( 1+w \right) H  \left( -2\, \right.\right.\right.\nonumber\\&\left.\left.\left. \dot{r}\cosh \left( r  \right) \sinh \left( r  \right) + \left(  \left( - \Sigma_-  \sqrt {3}+ 3\,\Sigma_{+}   \right) \right.\right.\right.\right.\nonumber\\&\left.\left.\left.\left. \left( \cos \left( \eta  \right)  \right) ^{2}-2\,\cos \left( \eta   \right) \sin \left( \eta  \right)  \Phi_3   -  \Sigma_{+}   -4\,\right. \right.\right.\right.\nonumber\\&\left.\left.\left.\left. +\Sigma_-   \sqrt {3} \right)\left( \cosh \left( r \right)  \right) ^{2}+ \left( -  3\,\Sigma_{+}   +  \Sigma_-   \sqrt {3} \right)  \right.\right.\right.\nonumber\\&\left.\left.\left.\left( \cos \left( \eta  \right)  \right) ^{2}+2\,\cos \left( \eta \right) \sin \left( \eta   \right)  \Phi_3    +\left( \Sigma_{+}    +1\right.\right.\right.\right.\nonumber\\&\left.\left.\left.\left.- \Sigma_-   \sqrt {3} \right)\right)  \right]\Omega_{{m}}  \right\},\label{GRv1} \\
   \nabla_bT^{1b}=&-3\, H ^{2} \left( 1+w \right)  \left\{ -\Omega_{{m}}   \left( 3\, \left( \cos \left( \eta  \right)  \right) ^{2}-1 \right)  \sqrt {\Omega_{{K}}  }\right.\nonumber\\&\left. H\left(\left( \cosh \left( r   \right)\right)^2 -1 \right)   -\Omega_{{m}}  \left( 2\, \left( \cosh \left( r   \right)  \right) ^{2}-1 \right) \right.\nonumber\\&\left.\cos \left( \eta  \right) H \dot{r}  +\cosh \left( r  \right)  \left[ -H  \cos \left( \eta \right) \dot{\Omega}_{{m}}   +\Omega_{{m}}   \right.\right.\nonumber\\&\left.\left. \left( \sin \left( \eta   \right) \dot{\eta} H-2\,\cos \left( \eta   \right) \dot{H}  -2\,H   \left(  \left( -  \Sigma_{+}    +2 \right)\right.  \right. \right.\right.\nonumber\\&\left.\left.\left.\left.\cos \left( \eta  \right)+\sin \left( \eta \right)  \Phi_{3}  \right) \right)  \right] \sinh \left( r  \right)  \right\}\label{GRv2},\\
    \nabla_bT^{3b}=&\,3\,H ^{2} \left( 1+w \right)  \left\{ 3\,\sin \left( \eta  \right) \Omega_{{m}} H  \left( \cosh \left( r  \right) -1 \right) \right.\nonumber\\&\left. \left( \cosh \left( r  \right) +1 \right) \cos \left( \eta \right)\sqrt {\Omega_{{K}}  }+ \Omega_{{m}}  \left( 2\, \left( \cosh \left( r   \right)  \right) ^{2}\right.  \right.\nonumber\\&\left.\left.-1 \right) \sin \left( \eta  \right) H \dot{r}+\cosh \left( r  \right)  \left[ \sin \left( \eta  \right) H \dot{\Omega}_{{m}}  +\Omega_{{m}}     \right.\right.\nonumber\\&\left.\left.\left( H \cos \left( \eta  \right)  \dot{\eta}+\sin \left( \eta   \right)\left( 2\,\dot{H} +H  \left( - \Sigma_- \sqrt {3}  \right.\right.\right.\right.\right.\nonumber\\&\left.\left.\left.\left.\left.  +  \Sigma_{+} +4 \right)  \right)  \right)  \right] \sinh \left( r\right)  \right\},\label{GRv3}
\end{align}
while $\nabla_bT^{2b}\equiv 0$ is identically null.

The de Sitter Universe is obtained as the solution to the Friedmann equations with the cosmological constant as source \cite{Mukhanov:2005sc}. The cosmological constant is a perfect fluid with the EoS parameter $w=-1$. In de Sitter Universe, the energy-momentum tensor, \eqref{tem}, is independent of the time-like vector, \eqref{timeve}, and the tilted behavior of the theory could not be investigated. For this reason, de Sitter Universe will not be considered in this work.

Of course, also considering the same settings, that is, $\hat{u}^a$ in \eqref{timeve} and energy-momentum source \eqref{tem}, metric \eqref{mett} and connection defined by \eqref{Gamma_000}, \eqref{Gamma_0ab} with \eqref{sig1}, \eqref{Gamma_ijk} and \eqref{deromegak} results in the non-trivially satisfied Einstein-Hilbert field equations with shear and tilt
\begin{align}
E_{11}=&- H \left\{  \left( 2\, \Sigma_{+}+2 \right) \dot{H} + \left[ 2\, \dot{\Sigma}_{+}+3\,\Omega_{{m}}\left( 1+w \right)   \right.\right.\nonumber\\&\left.\left. \left(\left( \cosh \left( r   \right)\right)^2 -1 \right)    \left( \cos \left( \eta   \right)  \right) ^{2}+3\,w\,\Omega_{{m}}  +3\,  \left( \Sigma_{+}    \right) ^{2}\right.\right.\nonumber\\&\left.\left.+3\, \left(  \Sigma_-  \right) ^{2}- \left(  \Phi_3   \right) ^{2}+6\,  \Sigma_{+}   -\Omega_{{K}}+3 \right] H  \right\} ,\label{GReq1}\\
E_{22}=&- H\left\{  \left( -  \Sigma_- \sqrt {3}-  \Sigma_{+} +2 \right) \dot{H} + \left[ 3\, \left( \Sigma_{-} \right)   ^{2}\right.\right.\nonumber\\&\left.\left.-3\,  \Sigma_{-}  \sqrt {3}- \dot{\Sigma}_{-}  \sqrt {3}+3\,w\,\Omega_{{m}}  + \left( \Phi_3   \right) ^{2}\right.\right.\nonumber\\&\left.\left.+3\, \left(  \Sigma_{+}   \right) ^{2}-\dot{\Sigma}_{+}  -\Omega_{{K}} -3\,  \Sigma_{+}    +3 \right] H   \right\},\label{GReq2}\\
  E_{33}=&\left\{  \left( -  \Sigma_- \sqrt {3}+ H \Sigma_{+}  -2 \right) \dot{H} + \left[ \dot{ \Sigma}_{+}   -  \dot{ \Sigma}_{-}  \sqrt {3}\right.\right.\nonumber\\&\left.\left.+3\,\Omega_{{m}}   \left( \cosh \left( r  \right) -1 \right)  \left( \cosh \left( r   \right) +1 \right)  \left( 1+w \right)  \right.\right.\nonumber\\&\left.\left.\left( \cos \left( \eta   \right)  \right) ^{2}-3\,\Omega_{{m}}  \left( 1+w \right)  \left( \cosh \left( r  \right)  \right) ^{2}-3\, \left(   \Sigma_{+} \right)  ^{2}\right.\right.\nonumber\\&\left.\left.-3\, \left(   \Sigma_{-} \right)  ^{2}-3\,  \Sigma_-  \sqrt {3}-3\, \left(   \Phi_3 \right)   ^{2}+\Omega_{{K}}    \right.\right.\nonumber\\&\left.\left.   +3\,\Omega_{{m}}+3\,  \Sigma_{+}-3 \right] H   \right\}  ,\label{GReq3}\\
  E_{31}=&\,H \left\{ H  \dot{\Phi}_3   + \dot{H}    \Phi_3   -3\, \left[ \sin \left( \eta   \right) \Omega_{{m}}  \left( \cosh \left( r   \right) -1 \right) \right.\right.\nonumber\\&\left.\left. \left( \cosh \left( r   \right) +1 \right)  \left( 1+w \right) \cos \left( \eta   \right) + \left( - \Sigma_-   \sqrt {3}  \right.\right.\right.\nonumber\\&\left.\left.\left. +3\, \Sigma_{+}  +3 \right)  \Phi_3   \right] H   \right\},\label{merg}
\end{align}
while
\begin{align}
E_{00}=&\,3\,\left[-\Omega_{{m}}   \left( 1+w \right)  \left( \cosh \left( r   \right)  \right) ^{2}+w\Omega_{{m}}  -   \left( \Sigma_{+} \right)  ^{2}\right.\nonumber\\&\left.- \left( \Sigma_- \right)   ^{2}-\Omega_{{K}}  +1\right]- \left(   \Phi_3  \right) ^{2},\nonumber\\
 E_{01}=&\,3\,H   \left[ -2\,\sqrt {\Omega_{{K}}  }  \Sigma_{+}    +\Omega_{{m}}  \cosh \left( r   \right) \sinh \left( r   \right) \cos \left( \eta   \right)  \right.\nonumber\\&\left.\left( 1+w \right)  \right],\nonumber
\\
E_{03}=&\,3\,H  \left[ \sqrt {\Omega_{{K}}  }  \Phi_3  +\Omega_{{m}} \cosh \left( r  \right) \sinh \left( r   \right) \sin \left( \eta  \right) \right.\nonumber\\&\left. \left( 1+w \right)  \right]
\label{rgvinc2}
\end{align}
are the constraints which must be satisfied by the initial conditions. Once these constraints are initially satisfied, they must be maintained during time evolution and are used as numerical check throughout the article. 

The full dynamical system for GR is defined by \eqref{deromegak}, \eqref{GRv1}-\eqref{merg}.

\subsection{Quadratic gravity}\label{subsec22}

Quadratic gravity QG mentioned in \eqref{eqdemov} has higher-order derivatives terms. Therefore, in addition to the variables defined in \eqref{ENV1}, new ones will be needed. Following \cite{Barrow:2006xb}, they are defined as
 \begin{align}
 &B=\frac{\chi}{3\,\beta\, H^2}, &Q _{ 1}=\frac{\dot{H} }{H},\nonumber\\
&Q _{ 2}=\frac{\ddot{H} }{H}, &\Phi_{3 ,0}=\frac{\dot{\phi} _{3 }}{H},\nonumber \\
 &\Phi_{3,1}=\frac{\ddot{\phi} _{3 }}{H}, &\Sigma_{\pm1}=\frac{\dot{\sigma} _{\pm }}{H},\nonumber \\
 &\Sigma_{\pm2}=\frac{\ddot{\sigma} _{\pm}}{H}\label{vnq1}.
 \end{align}
where $\chi=\beta/(3\alpha+\beta)$. The coupling constants $\beta>0$ and $\chi<0$ or $\chi>1$ to avoid tachyon behavior \cite{MULLER:2014jaa}. Again, the dynamical time is used instead of the proper time, as in\cite{Barrow:2006xb}. 

According to their definition, the dimensionless variables \eqref{vnq1} must satisfy the first-order differential equations:
\begin{align}
    &\dot{\Sigma }_{\pm}=\Sigma_{\pm1}-\Sigma_{\pm}\,Q_1,
&\dot{\Sigma }_{\pm1}=\Sigma_{\pm2}-\Sigma_{\pm1}\,Q_1,\nonumber\\
 &\dot{\Phi }_{3}=\Phi_{3 ,0}-\Phi_{3}\,Q_1,
        &\dot{\Phi }_{3, 0}=\Phi_{3,1}-\Phi_{3 ,0}\,Q_1,\nonumber\\
    &\dot{B}=-2\,B\,Q_1,
   &\dot{Q}_1=Q_2-(Q_1)^2.\label{senv}
\end{align}

The time evolution of the higher derivatives ENV $Q_2$, $\Phi_{3,1}$, and $\Sigma_{\pm2 }$ for quadratic gravity are obtained through the field equations \eqref{eqdemov} written in ENV \eqref{ENV1}, \eqref{vnq1} with $\alpha\neq 0$ and $\beta\neq 0$. Considering metric \eqref{mett}, connection \eqref{Gamma_000}, \eqref{Gamma_0ab}, and \eqref{Gamma_ijk}, the non-trivial components $E_{11}$, $E_{22}$, $E_{33}$, and $E_{31}$ of the field equation \eqref{eqdemov} are written in the Appendix \ref{app}. Besides the field equation, the covariant conservation of the source $\nabla_aT^{ab}=0$ given in \eqref{GRv1}, \eqref{GRv2}, and \eqref{GRv3} results in the time evolution of $\Omega_m$, $\eta$, and $q$, which are in the Appendix \ref{app}. The constraints $E_{00}$, $E_{01}$, and $E_{03}$ from the field equations \eqref{eqdemov} are also in the Appendix \ref{app}. Such that the dynamical system is completely defined in the Appendix \ref{app} and by the ENV first-order differential equations \eqref{deromegak}, \eqref{senv}. The initial conditions must always satisfy the constraints, and once they are initially satisfied, they must be maintained in time. That is the reason the constraints are used as a numerical check.

In this work, all the exact solutions analyzed for both gravitational theories satisfy $\Phi_3=0$; see the next section, \ref{sec3}. However, this non-diagonal shear variable is needed in order that the component $E_{ 31}$ of \eqref{eqdemov} for $\Phi_3$ or $\Phi_{3,1}$ in GR and QG, respectively, with the non-trivial dynamics for tilt $r\neq0$. In the same way, initial conditions must additionally satisfy $\Omega_K\neq 0$, $\eta\neq0$, $\eta\neq\pi$, and $\eta\neq\pi/2$, the EoS parameter $w\neq -1 $ for a non-trivial dynamics of the vorticity. If the matter density $\Omega_m= 0$, there is no tilt or vorticity since it is a property of the source. In the case of GR, this can be verified by inspection in the dynamical equation $E_{ 31}$ in \eqref{merg}.

The curvature invariants $R$, $R_{ab}R^{ab}$, and $R_{abcd}R^{abcd}$ in terms of the ENV \eqref{ENV1} and \eqref{vnq1} are written as
\begin{align}
 &R=\left\{-2\,\chi\, \left( -\left({\Phi_3}\right)^{2}-3\,\left({\Sigma_{+}}\right)^{2}-3\,\left({\Sigma_-}\right)^{2}+3\,\Omega_{{K}}\right.\right.\nonumber\\&\left.\left.-3\,Q_{{1}}-6 \right) \right\}/\left\{3\,\beta\,B\right\},\nonumber\\
&R_{ab}R^{ab}=\left\{ 2\,{\chi}^{2} \left( -2\,\Phi_3 \left(  \left( 3\,\Sigma_{+}\Sigma_--\Sigma_{-1} \right)\Phi_3\right.\right.\right.\nonumber\\&\left.\left.\left.+3\,\Sigma_-\Phi_{3 ,0} \right) \sqrt {3}+6\,\left({\Phi_3}\right)^{4}+ \left( 14\,\left({\Sigma_{+}}\right)^{2}+15\,\right.\right.\right.\nonumber\\&\left.\left.\left.-9\,\Omega_{{K}}+6\,Q_{{1}}-6\,\Sigma_{+1}+15\,\left({\Sigma_-}\right)^{2} \right)\left( {\Phi_3}\right)^{2}+6\,\Phi_{3 ,0} \right.\right.\nonumber\\&\left.\left.\Phi_3\left( \Sigma_{+}+1 \right) +18\,\left({\Sigma_{+}}\right)^{4}+ \left( 36\,\left({\Sigma_-}\right)^{2}-36\,\Omega_{{K}}\right.\right.\right.\nonumber\\&\left.\left. \left.+18\,Q_{{1}}+45 \right)\left( {\Sigma_{+}}\right)^{2}+18\,\Sigma_{+}\Sigma_{+1}+18\,\left({\Sigma_-}\right)^{4}\right.\right.\nonumber\\&\left.\left.+\left( 45+18\,Q_{{1}} \right) \left({\Sigma_-}\right)^{2}+18\,\Sigma_-\Sigma_{-1}+6\,\left({Q_{{1}}}\right)^{2}+6\, Q_{{1}}\right.\right.\nonumber\\&\left.\left. \left( 3-\Omega_{{K}} \right)+18+6\,\left({\Omega_{{K}}}\right)^{2}+\left({\Phi_{3 ,0}}\right)^{2}+3\,\left({\Sigma_{-1}}\right)^{2}\right.\right.\nonumber\\&\left.\left.-18\,\Omega_{{K}}+3\,\left({\Sigma_{+1}}\right)^{2} \right) \right\}/\left\{\left(3\,  \beta B \right)^{2}\right\},\nonumber\\
&R_{abcd}R^{abcd}=\left\{4\, \left[ -6\, \left( 2\,\Sigma_-\Phi_3\Sigma_{+}+ \left( 9\,\Sigma_--3\,\Sigma_{-1} \right) \right.\right.\right.\nonumber\\&\left.\left.\left.\Phi_3+12\,\Sigma_-\Phi_{3 ,0} \right) \Phi_3\sqrt {3}+27\,\left({\Sigma_{+}}\right)^{4}-12\,\left({\Sigma_{+}}\right)^{3}\right.\right.\nonumber\\&\left.\left.+\left( 36\,\left({\Phi_3}\right)^{2}+54\,\left({\Sigma_-}\right)^{2}-30\,\Omega_{{K}}+12\,Q_{{1}}-12\,\Sigma_{+1}\right.\right.\right.\nonumber\\&\left.\left.\left.+36 \right) \left({\Sigma_{+}}\right)^{2}+ \left( -6\,\left({\Phi_3}\right)^{2}+8\,\Phi_3\Phi_{3 ,0}+36\,\left({\Sigma_-}\right)^{2}\right.\right.\right.\nonumber\\&\left.\left.\left.+24\,\Sigma_-\Sigma_{-1}+24\,\Sigma_{+1} \right) \Sigma_{+}+11\,\left({\Phi_{3}}\right)^{4}+\left( {\Phi_3}\right)^{2}\right.\right.\nonumber\\&\left.\left. \left( 24\,\left({\Sigma_-}\right)^{2}-8\,\Omega_{{K}}+4\,Q_{{1}}-14\,\Sigma_{+1}+12 \right)+8\,\Phi_3\Phi_{{3,0}}\right.\right.\nonumber\\&\left.\left.+27\,\left({\Sigma_-}\right)^{4}+ \left( 36+12\,Q_{{1}}-6\,\Omega_{{K}}+12\,\Sigma_{+1} \right) \left({\Sigma_-}\right)^{2}\right.\right.\nonumber\\&\left.\left.+24\,\Sigma_-\Sigma_{-1}+6\,{\Sigma_{-1}}^{2}-6\,\Omega_{{K}}+3\,{\Omega_{{K}}}^{2}+3\,\left({Q_{{1}}}\right)^{2}\right.\right.\nonumber\\&\left.\left.+6\,Q_{{1}}+6\,\left({\Sigma_{+1}}\right)^{2}+6+2\,\left({\Phi_{{3,0}}}\right)^{2} \right] {\chi}^{2}\right\}/\left\{\left(3\, \beta B  \right)^{2}\right\}.\label{ec}
\end{align}
Note that $B \rightarrow 0$ is a physical singularity. 

It can be noted in the dynamical equations shown in the Appendix \ref{app} that the matter density $\Omega_m$ appears only as the product $B\Omega_m$. In this way, the matter density decouples from the dynamics as it approaches the physical singularity of $B\rightarrow 0$. This same independence of matter in the geometric evolution of the Universe was mentioned previously in \cite{Toporensky:2016kss} for the Bianchi I Universe without a tilted source.

\begin{figure*}[htpb]
      \begin{centering}
     \begin{tabular}{c c}
            \resizebox{\imsize}{!}{\includegraphics[width=0.4\textwidth]{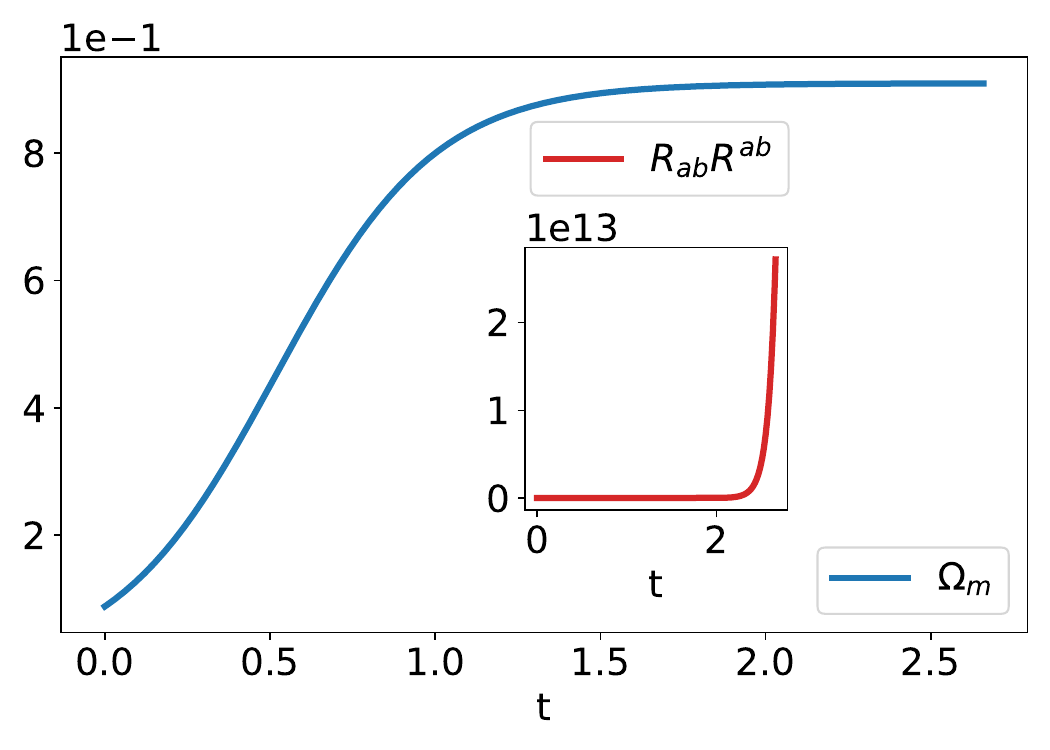}} &
        \resizebox{\imsize}{!}{\includegraphics[width=0.4\textwidth]{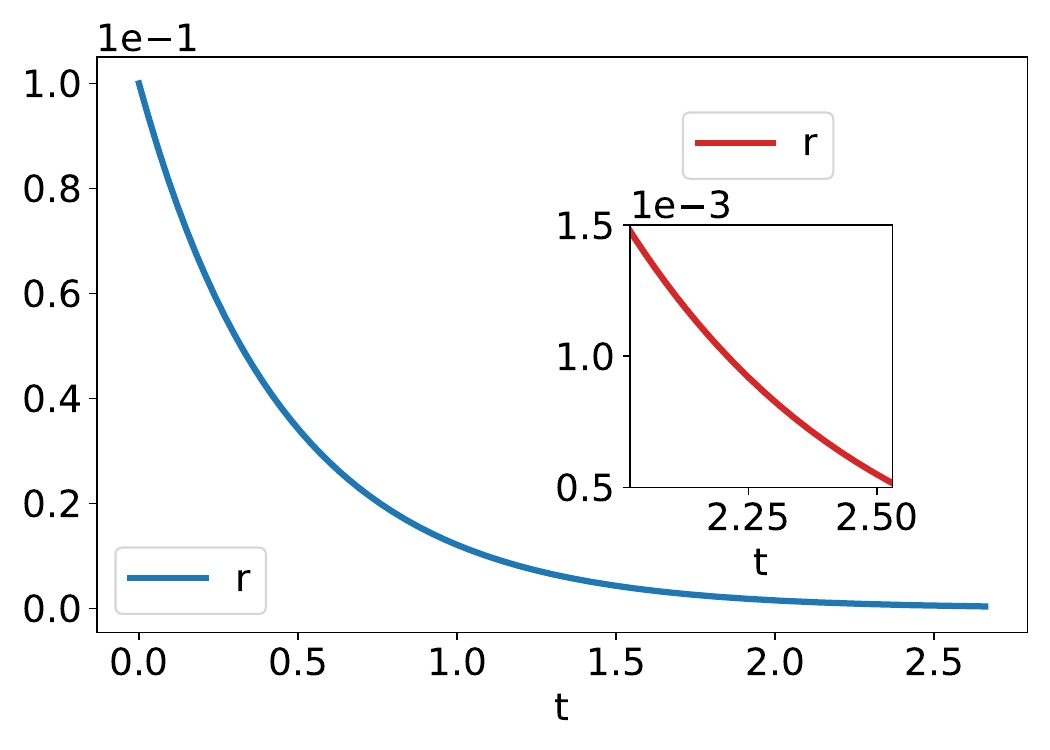}} \\
        (a)&(b)\\
            \resizebox{\imsize}{!}{\includegraphics[width=0.4\textwidth]{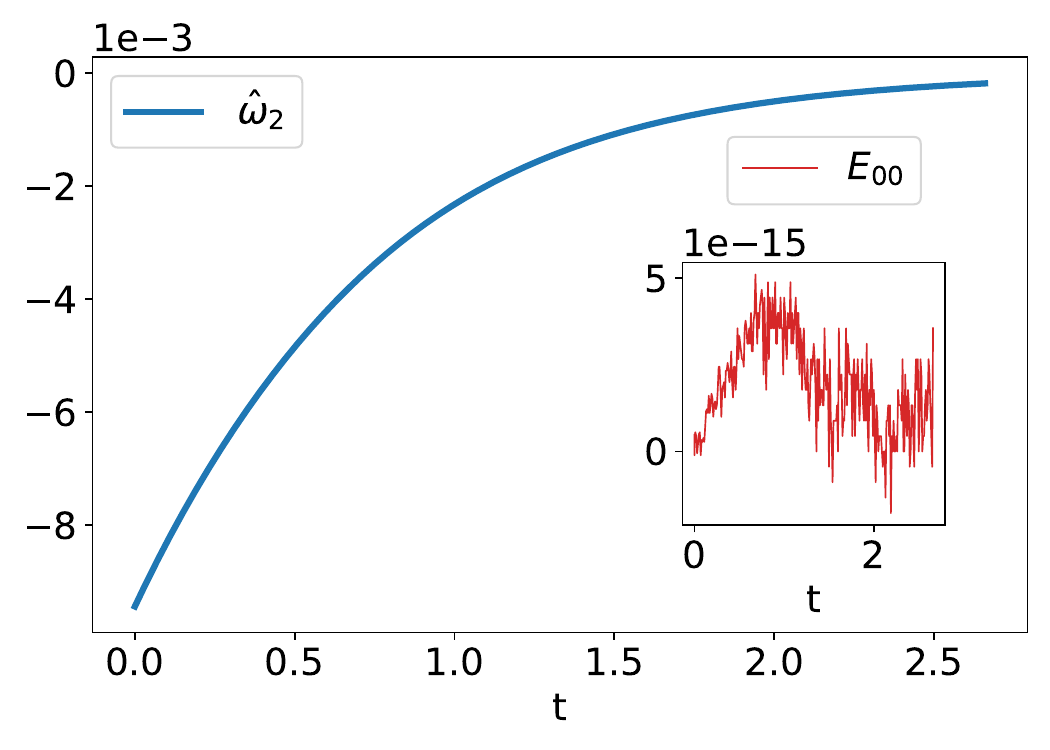}} &
            \resizebox{\imsize}{!}{\includegraphics[width=0.4\textwidth]{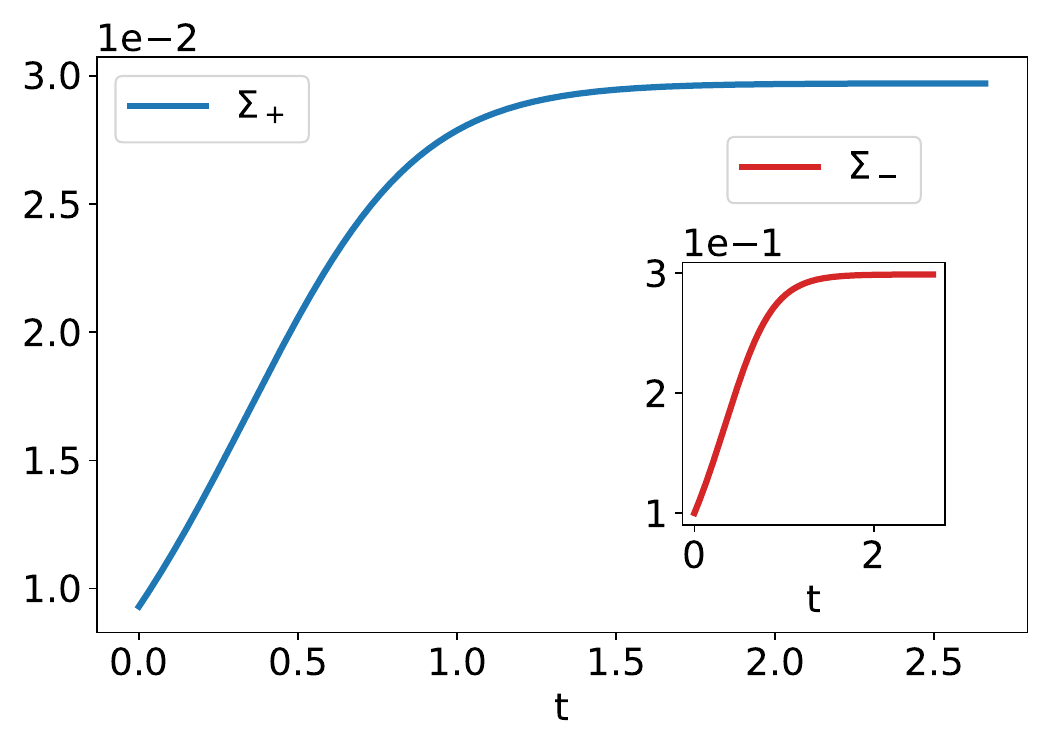}}\\
            (c) & (d)
           \end{tabular}
  \par\end{centering}\caption{According to GR it is plotted the numerical evolution to the past for an initial condition near Milne's exact solution. The initial condition is $H =-1.0$, $\Omega_K = 9.0e^{-1}$, $\Phi_3=-1.86e^{-3}$, $\Sigma_+=9.30e^{-3}$, $\Sigma_-=1.0e^{-1}$, and a source with $\Omega_m=8.81e^{-2}$, tilt $r=1.0e^{-1}$ with its direction $\eta=1.0e^{-1}$ for stiff matter with EoS parameter $w=1$. The solution is attracted to the kasner-like orbit, which satisfies \eqref{kw1} and \eqref{kw11}. a) The plot in blue shows that the matter density $\Omega_m$ approaches a constant of the value of $9.09612e^{-1}$. In the inset, it is plotted in red the increase of $R_{ab}R^{ab}$ following a divergence showing the presence of a curvature singularity. b) It is plotted in blue the tilt variable $r$ which approaches zero toward the singularity. Also, in the inset, in red, it shows a zoom of the plot $r$. c) The plot in blue shows that the vorticity approaches zero asymptotically. In the inset, it is plotted in red the numeric check for the constraint $E_{00}$ with fluctuations smaller than $10^{-14}$. d) The evolution of the diagonal shear components $\Sigma_+$ and $\Sigma_-$ is plotted, respectively, in blue and red, approaching the values $\Sigma_+\sim 2.97009e^{-2}$ and $\Sigma_-\sim 2.98918e^{-1}$}\label{f1}
\end{figure*} 

\begin{figure*}[htpb]
      \begin{centering}
     \begin{tabular}{c c}
            \resizebox{\imsize}{!}{\includegraphics[width=0.4\textwidth]{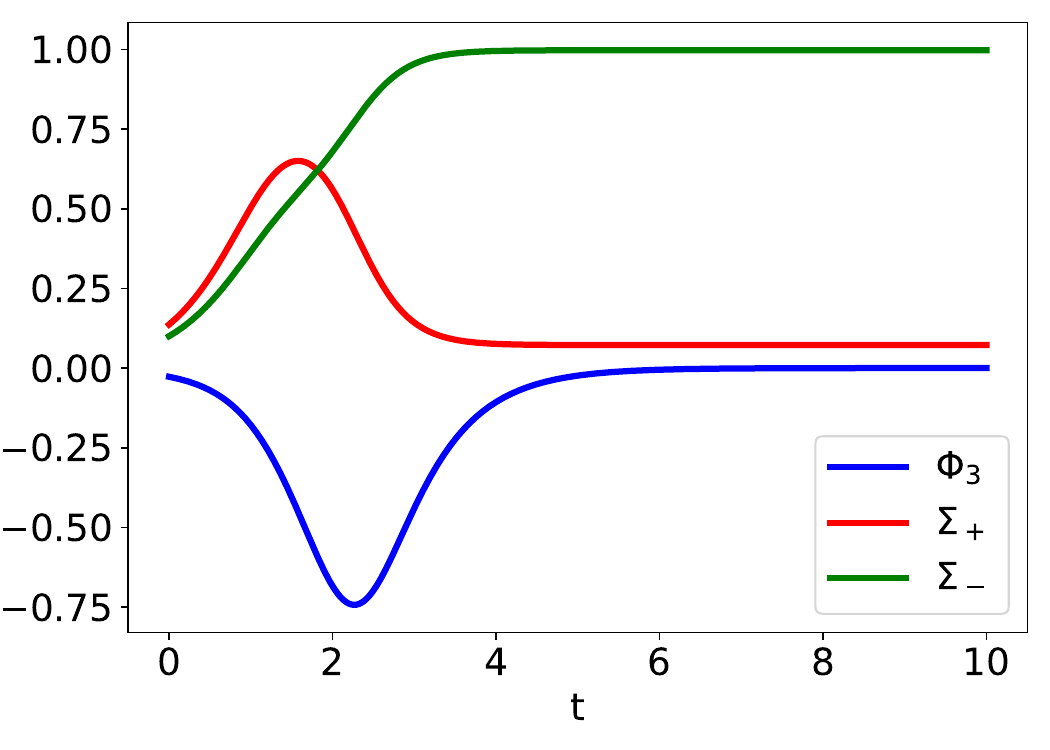}} &
            \resizebox{\imsize}{!}{\includegraphics[width=0.4\textwidth]{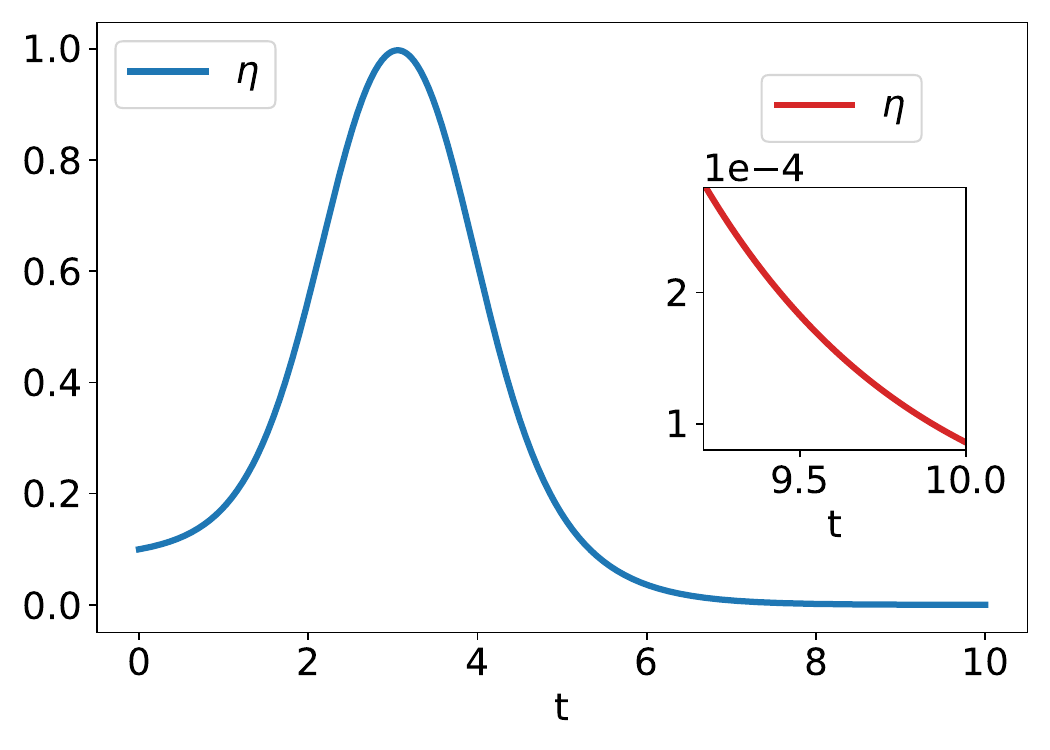}} \\
           \\
        (a) & (b)\\
            \resizebox{\imsize}{!}{\includegraphics[width=0.4\textwidth]{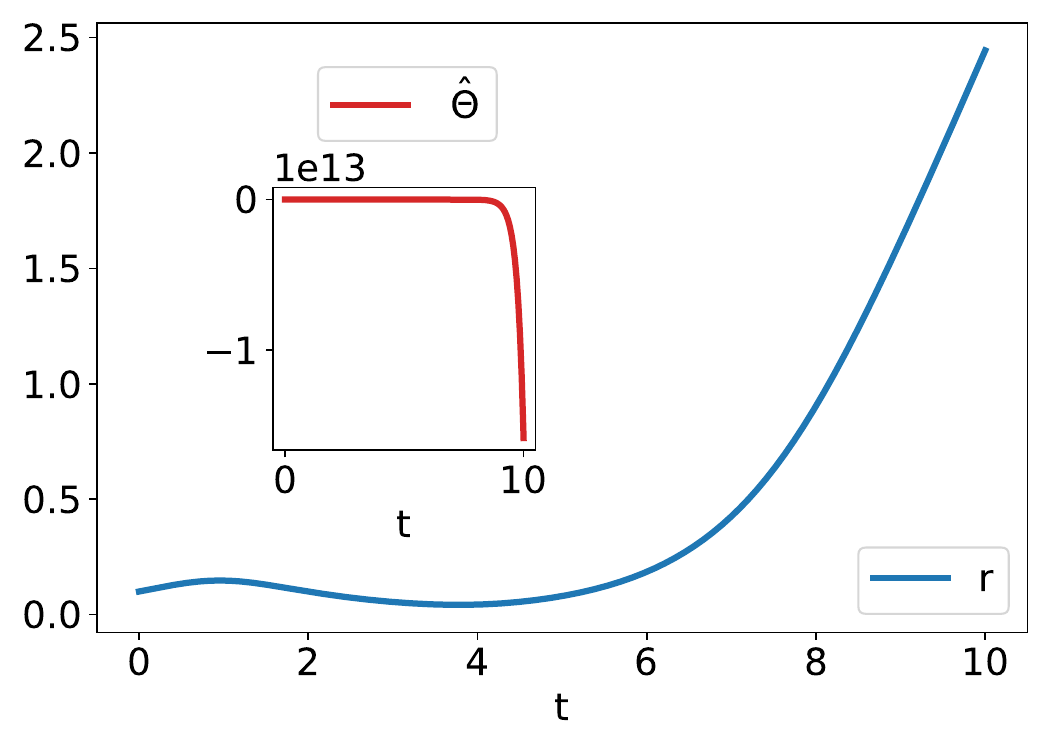}} &
            \resizebox{\imsize}{!}{\includegraphics[width=0.4\textwidth]{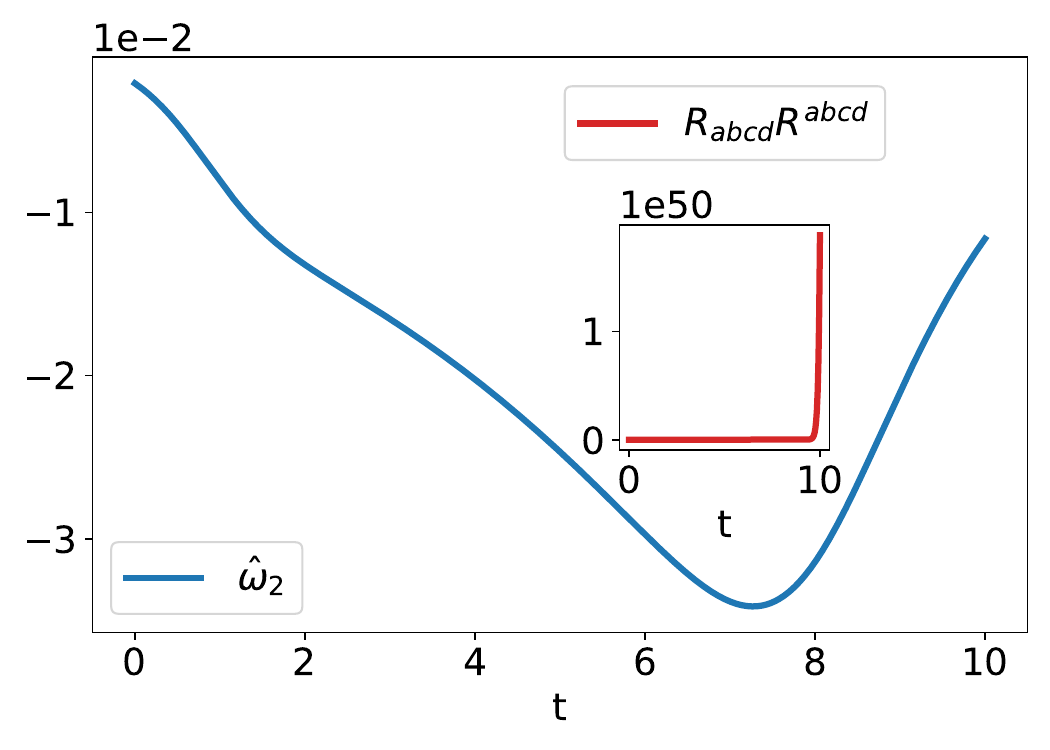}}\\
            (c) & (d)
           \end{tabular}
\par\end{centering}\caption{Now according to GR the orbit with initial condition $H =-6.66e^{-1}$, $\Omega_K = 1.0e^{-1}$, $\Omega_m=8.62e^{-1}$, $\Phi_3=-2.74e^{-2}$, $\Sigma_+=1.36e^{-1}$, $\Sigma_-=1.0e^{-1}$, $\eta=1.0e^{-1}$ and $r=1.0e^{-1}$ chosen near FLRW exact solution for dust fluid with $w=0$ is plotted. Again, the evolution is to the past. The solution is attracted to the Kasner orbit with $\phi=1.4987$. a) The graph shows the evolution of the shear components $\Phi_3$, $\Sigma_+$, and $\Sigma_-$, respectively, in blue, red, and green. Panel b) displays in blue the direction of tilt, $\eta$, which increases before it approaches zero. In the inset, in red, it is shown a zoom of the $\eta$ plot. c) Plotted in blue is the tilt variable $r$, which increases and decreases and then increases again as the orbit approaches the Kasner attractor. Also, in the inset, it is shown the matter contraction $\hat{\Theta}$ toward the past singularity. d) In blue, it is plotted, showing an increase of the vorticity in absolute value and then a decrease toward zero asymptotically. In the inset, plotted in red the increase of $R_{abcd}R^{abcd}$, which diverges, indicating the presence of a curvature singularity, as expected}\label{f2}
\end{figure*} 

\begin{figure*}[htpb]
      \begin{centering}
     \begin{tabular}{c c}
            \resizebox{\imsize}{!}{\includegraphics[width=0.4\textwidth]{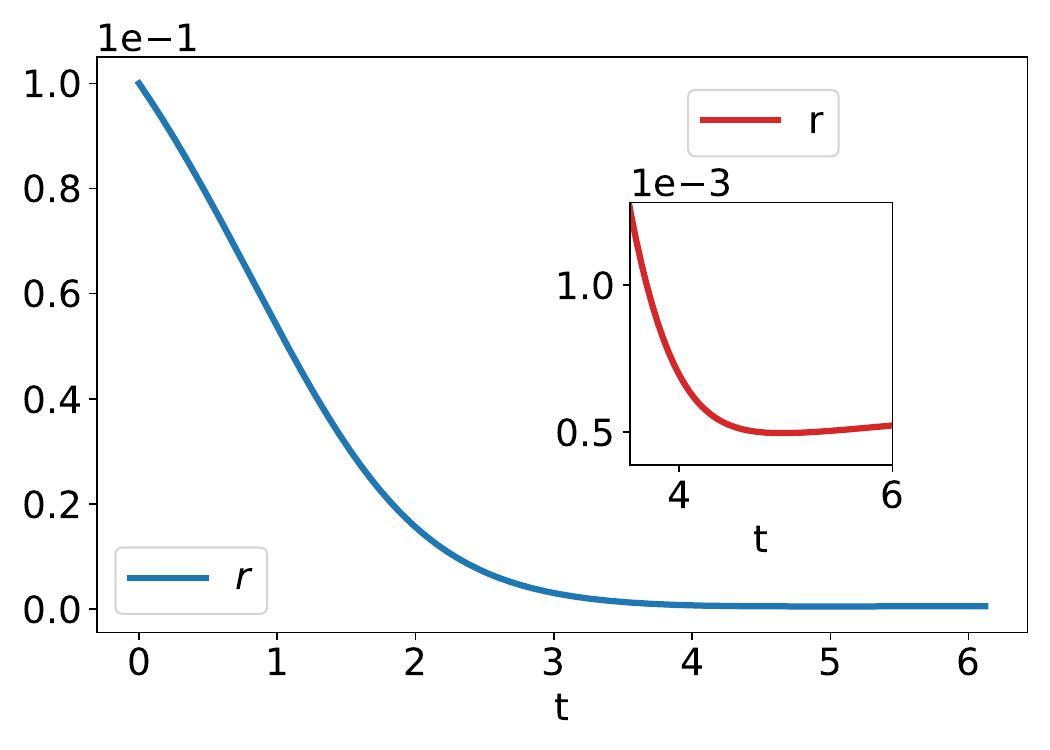}} &
        \resizebox{\imsize}{!}{\includegraphics[width=0.4\textwidth]{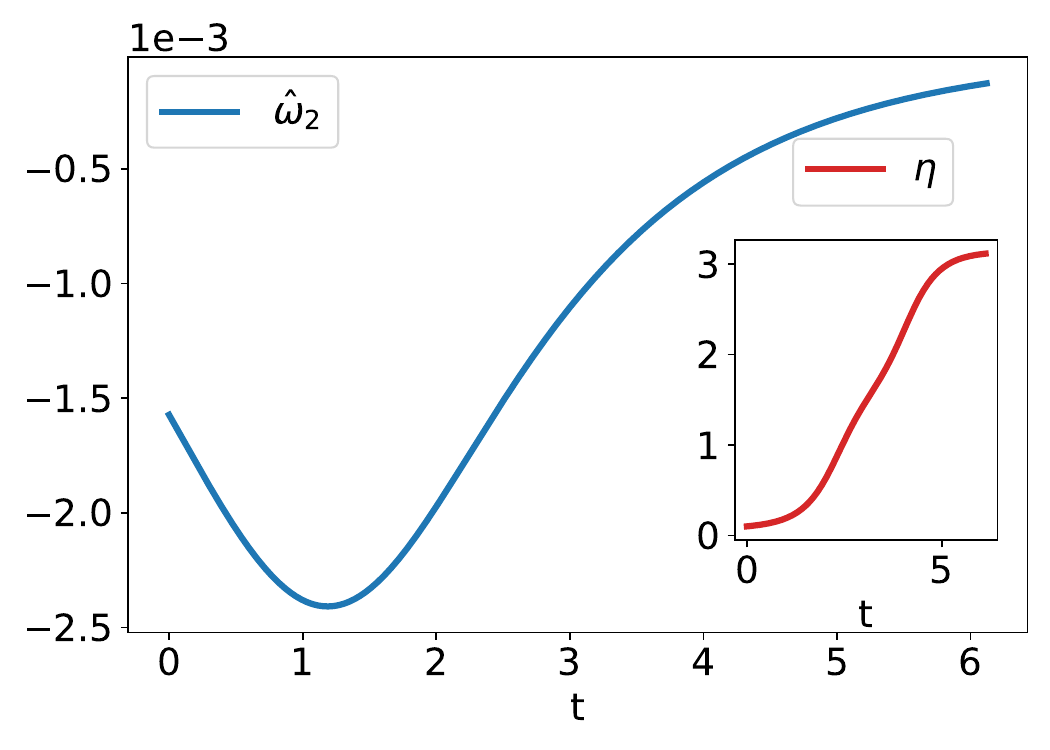}} \\
        (a) & (b)
           \end{tabular}
  \par\end{centering}\caption{The evolution to the past for GR shows for the initial condition $H =-5.0e^{-1}$, $\Omega_K = 0.1$, $\Omega_m=8.46e^{-1}$, $\Phi_3=-3.58e^{-2}$, $\Sigma_+=1.78e^{-1}$, $\Sigma_-=1.0e^{-1}$, $\eta=1.0e^{-1}$, and $r=1.0e^{-1}$ near FLRW exact solution for radiation with $w=1/3$ that the solution is attracted to the Kasner orbit with $\phi=1.6047$. a) The graph in blue shows the tilt variable $r$ approaches zero toward the singularity. In the inset, it is shown in red a zoom of the plot $r$. b) The vorticity in blue increases in absolute value and then approaches zero. In the inset, it is plotted in red the direction of the tilt $\eta$ approaching $\pi$}\label{f3}
\end{figure*} 

\section{Fixed-points and their stability}\label{sec3}

In this section, we will consider the stability of some selected exact solutions for Einstein-Hilbert and quadratic gravity models. In expansion-normalized variables, exact solutions are described by fixed points of the dynamical system.

The dynamical equations are linearized near the static solutions, and the eigenvalues for each fixed point are shown. The past or future evolution near the solutions is discussed.

\subsection{Einstein-Hilbert gravity}\label{subsec31}

\subsubsection{FLRW without spatial curvature solution}

The exact solution of FLRW without spatial curvature for GR is an exact solution with zero tilt $r=0$
\begin{align}
 &H =\frac{2}{3\,(1+w)}\exp\left [ -\frac{3}{2}\left ( 1+w \right )t \right ], & \Omega_{m}= 1.
\end{align}
The other variables are null. See, for instance, the textbooks \cite{Weinberg:1972kfs,Stephanibook}.

The Ricci and Riemann scalar products and the Ricci scalar increase and then diverge as they evolve towards the past and tend to zero as they evolve towards the future. For radiation fluid, the EoS parameter is $w=1/3$ and the Ricci scalar is zero.

It is an asymptotic fixed point $t \rightarrow \infty$, and the eigenvalues together with the degeneracies are
\[3\,w+1,\, 3\,w-1,\, 3\,w,\, -{3}/{2}\,(1+w),\, \left[{-3}/{2}\,(1-w)\right]_3,\,0.\]
The null eigenvalue corresponds to the variable $\eta$.
The eigenvalues show that for the values of the EoS parameter $-1< w<-{1}/{3}$, the FLRW orbit is an attractor to the future.

\subsubsection{Milne solution}

Milne Universe is an exact solution for both GR and QG, also non-tilted
\begin{align}
    &H=\exp(-t), & \Omega_{K}= 1,\nonumber\\&\eta=0, &r=0,
\end{align}
with the other variables being null. Milne solution is Minko\-wski space, the Riemann tensor is zero \cite{Mukhanov:2005sc}.

It is also an asymptotic fixed point when $t\rightarrow \infty $, the eigenvalues and the degeneracies are 
\[3\,w-1,\, -(3\,w+1),\, [-1]_2,\, [-2]_3,\,0.\] 
The null eigenvalue corresponds to the variable $\eta$. The eigenvalues show that for values of the EoS parameter $-{1}/{3}< w<{1}/{3}$, the Milne orbit is an attractor to the future. Our numeric results show that, for $1/3\leq w\leq1$, the Milne Universe is also a future attractor, and tilt $r$ increases toward future evolution. This result was previously mentioned in \cite{Coley:2008zz}.

\subsubsection{Kasner solution}\label{subsubsubsecks}

The Kasner solution is an exact vacuum solution with zero tilt $r=0$ for both GR and QG theories \cite{Barrow:2006xb,Toporensky:2016kss}
\begin{align}
    &H=({1}/{3})\exp\left ( -3\,t \right ),\label{kw1}
\end{align}
and $\Sigma_+$ and $\Sigma_-$ constants that must satisfy the Einstein-Hilbert $00$ field equation
\begin{align}  
&\Sigma_{+}^2+\Sigma_{-}^2=1,\label{k}
\end{align}
which connect the Kasner circle with the parameter $\phi$ given in radians as 
\begin{align}    
      &\Sigma_{+} = \cos{\phi}, &\Sigma_{-}= \sin{\phi}.
\end{align}
The other variables are null.

The $R_{abcd}R^{abcd}$ curvature scalar increases and then diverges as it evolves toward the past and tends to zero as it evolves towards the future. However, the Ricci tensor is identically zero.

It is also an asymptotic fixed point when $t \to \infty$, with eigenvalues and degeneracies  
\begin{align*}
    &\sqrt{3}\sin{\phi}-3\cos{\phi},\, 4+4\cos{\phi},\, 2\cos{\phi}+3\,w-1,\, 3,\,-3,\\&3\,(1-w),\, [0]_2.
\end{align*}
One of the null eigenvalues corresponds to the variables $\Sigma_{+}$ and $\Sigma_{-}$, and it is connected to the Kasner circle for the parameter $\phi$. The other null eigenvalue corresponds to the direction of the tilt $\eta$.

The Kasner solution is not a future attractor, for that to happen, all the eigenvalues must have negative real part, which is not the case. Nevertheless, orbits are attracted to Kasner to the past if all the real parts are positive, for example, when the parameter $\phi=\pi/2$. The only negative eigenvalue $-3$ corresponds to the Hubble parameter $H$, which increases toward the past singularity and then diverges.

The tilt variable $r$ is associated with the eigenvalue \[2\cos{\phi}+3\,w-1.\] For instance, for the past attractor orbit with the parameter $\phi=\pi/2$, the variable $r$ increases for $w<1/3$ and decreases for $w>1/3$.

\subsubsection{Stiff matter}\label{subsubssm}

The exact solution for stiff matter $p=w\rho$, with EoS parameter $w=1$, is a solution with zero tilt $r=0$ for GR given by Eq. \eqref{kw1}, and $\Omega_{m}$, $\Sigma_{+}$, and $\Sigma_{-}$ constants which must satisfy Einstein $00$ field equation
\begin{align}  
&\Sigma_{+}^2+\Sigma_{-}^2+\Omega_m=1.\label{kw11}
\end{align}
The other variables are null. Similar to the previously discussed Kasner vacuum solution \ref{subsubsubsecks}, the constraint \eqref{kw11} connects $\Sigma_{+}$ and $\Sigma_{-}$ to the Kasner circle with radius $(1-\Omega_m)^{1/2}$ for the parameter $\phi$ in radians by $\Sigma _-=\sin{\phi}$, and $ \Sigma_+=(\sin^2{\phi}+1-\Omega_m)^{1/2}$.

Past evolution leads to the curvature singularity. 

This solution is also an asymptotic fixed point when $t\rightarrow\infty$ and has the following eigenvalues with multiplicities 
\[2\,(\Sigma_++1),\, 4\,(\Sigma_++1),\, \left[-3\,\Sigma_++\sqrt{3}\,\Sigma_-\right]_2,\, 3,\, -3,\, [0]_2.\]
The null eigenvalues correspond to the variables $\Omega_{m}$, $\Sigma_{+}$, and $\Sigma_{-}$, and it is connected to the Kasner circle for the parameter $\phi$ and radius $(1-\Omega_m)^{1/2}$. These eigenvalues show that for the stiff matter, the orbit with the exact solution \eqref{kw1} and \eqref{kw11} is a past attractor for $-1<\Sigma_+<\sqrt{3}/3\,\Sigma_-$. However, like in the Kasner exact solution, the only negative eigenvalue $-3$ corresponds to the Hubble parameter $H$, which increases toward the past and then diverges at the singularity.

\subsection{Quadratic gravity}\label{subsec32}

\subsubsection{Isotropic singularity asymptotic solution}

The isotropic singularity is a vacuum $T_{ab}=0$ asymptotic $t\to -\infty$ solution for $B \rightarrow 0$ for QG theory. The non-null ENV are:
\begin{align}
    & Q_{1}= -2,& Q_{2} = 4.
\end{align}
With $\eta=0$ and $r = 0$.

Curvature invariants diverge as $B \rightarrow 0$, while the Ricci scalar approaches zero, which is the reason this solution is also called false radiation since it is a vacuum solution for QG. 

This solution has the following eigenvalues with degeneracies
\[ 3\,w-1,\, -3\,w+1,\, \,4,\,5,\,[1]_3, [3]_4,\, [2]_4,\,0.\]
The null eigenvalue is associated to the variable $\eta$.

The eigenvalues show that for the EoS parameter $w>1/3$, the past $\Omega_m$ increases toward the singularity and then diverges, while the tilt variable $r$ tends to zero. However, the opposite situation occurs for $w<1/3$: the variable $\Omega_m$ tends to zero toward the singularity while the tilt variable $r$ increases.

\subsubsection{Milne solution}

The non-null ENV for the exact Milne solution for the QG are:
\begin{align}
    &B =({\chi }/{3\,\beta})\exp(2\,t),&\Omega_K = 1,\nonumber\\ & Q_{1} = -1,&Q_{2} = 1,
\end{align}
with $\eta=0$ and $r = 0$.

Again, it is a fixed point when $t \to -\infty$, with eigenvalues and multiplicities 
\[3\,w-1,\, -(3\,w+1),\, -1,\, [-2]_2,\, [1]_2,\,[2]_3,\,[0]_5.\]
The eigenvalues show that the Milne orbit is not an attractor to the past or to the future for QG. 

\subsubsection{Kasner solution}
The exact Kasner solution for QG theory is
\begin{align}
    &B=({3\,\chi}/{\beta})\exp{(6\,t)},& \Omega_{K} = 0, \nonumber\\& \Omega_{m} = 0,&Q_1=-3, \nonumber\\&Q_2=9, &\Phi_{3}= 0,\nonumber\\
       &\Phi_{3 ,0}=0,
 &\Phi_{3,1}=0,\nonumber\\&\Sigma_{+} = \cos{\phi},&\Sigma_{+1}=-3\cos{\phi},\nonumber\\&\Sigma_{+2}=9\cos{\phi},&\Sigma_{-}= \sin{\phi},\nonumber\\&\Sigma_{-1}=-3\sin{\phi},&\Sigma_{-2}=9\sin{\phi},\nonumber\\
 &\eta=0,&r=0.
\end{align}
With $\phi$ a constant parameter given in radians.

It is also an asymptotic $t \to -\infty$, fixed point, with eigenvalues  
\begin{align*}
    &-3\,(w+1),\, 9,\, [6]_4,\, [0]_2,\,  4+4\cos{\phi},\,\\& 2\,\cos{\phi}+3\,w-1,\,\lambda_0,\, \lambda_1,\, \lambda_2.
  \end{align*}
Where the eigenvalues $\lambda_i$ are obtained from the characteristic equation 
\[{\lambda_i}^3+a{\lambda_i}^2+b\lambda_i+c=0\]
where
\begin{align*}
a=&-\sin{\phi} \sqrt {3}+3\,\cos{\phi} -12,\\
b=&\,6\,\sin{\phi}  \left( \cos{\phi} +2 \right) \sqrt {3}-6\, \left( \cos{\phi}  \right) ^{2}-36\,\cos{\phi}\\& +33,\\
c=&\left( 24\, \left( \cos{\phi}  \right) ^{2}\sin{\phi} -33\,\sin{\phi}  \right) \sqrt {3}+81\,\cos{\phi},
\end{align*}
whose solution gives the eigenvalues
\begin{align*}
    \lambda_0=&\,1/6\,\sqrt [3]{ \xi }-6\, \left(1/3\,b-1/9\,{a}^{2}\right)/\left(\sqrt [3]{ \xi }\right)-1/3\,a,\\
   \lambda_1= &-1/12\,\sqrt [3]{ \xi }+3\,\left(1/3\,b-1/9\,{a}^{2}\right)/\left(\sqrt [3]{ \xi }\right)-1/3\,a\\&+1/2\,i\sqrt {3} \left[1/6\,\sqrt [3]{ \xi }+6\,\left(1/3\,b-1/9\,{a}^{2}\right)/\left(\sqrt [3]{ \xi }\right) \right],\\
    \lambda_2=&-1/12\,\sqrt [3]{ \xi }+3\,\left(1/3\,b-1/9\,{a}^{2}\right)/\left(\sqrt [3]{ \xi }\right)-1/3\,a\\&-1/2\,i\sqrt {3} \left[1/6\,\sqrt [3]{ \xi }+6\,\left(1/3\,b-1/9\,{a}^{2}\right)/\left(\sqrt [3]{ \xi }\right) \right].
\end{align*}
The terms $\Delta$ and $\xi$ are defined to simplify the notation as
\begin{align*}
    \Delta=&\,12\,{a}^{3}c-3\,{a}^{2}{b}^{2}-54\,abc+12\,{b}^{3}+81\,{c}^{2},\\
     \xi =&-8\,{a}^{3}+36\,ab+12\,\sqrt{\Delta}-108\,c.
\end{align*}

Particular choices among Kasner solutions can be an attractors to the past if all of eigenvalues have positive real part. For example, when $\phi = 1.5707 \sim \pi/2$, the Kasner eigenvalues and the degeneracies become
\begin{align*}
   &4,\,[6]_4,\, [0]_2,\, 9,\,  -9.99\times10^{-1}+3\,w,\,7.73+5.0\times10^{-10} i,\\& 1.73+5.66\times 10^{-10} i,\, 4.26-1.16\times10^{-9} i,\,-3\,(w+1).
\end{align*}

\begin{figure*}[htpb]
      \begin{centering}
     \begin{tabular}{c c}
            \resizebox{\imsize}{!}{\includegraphics[width=0.4\textwidth]{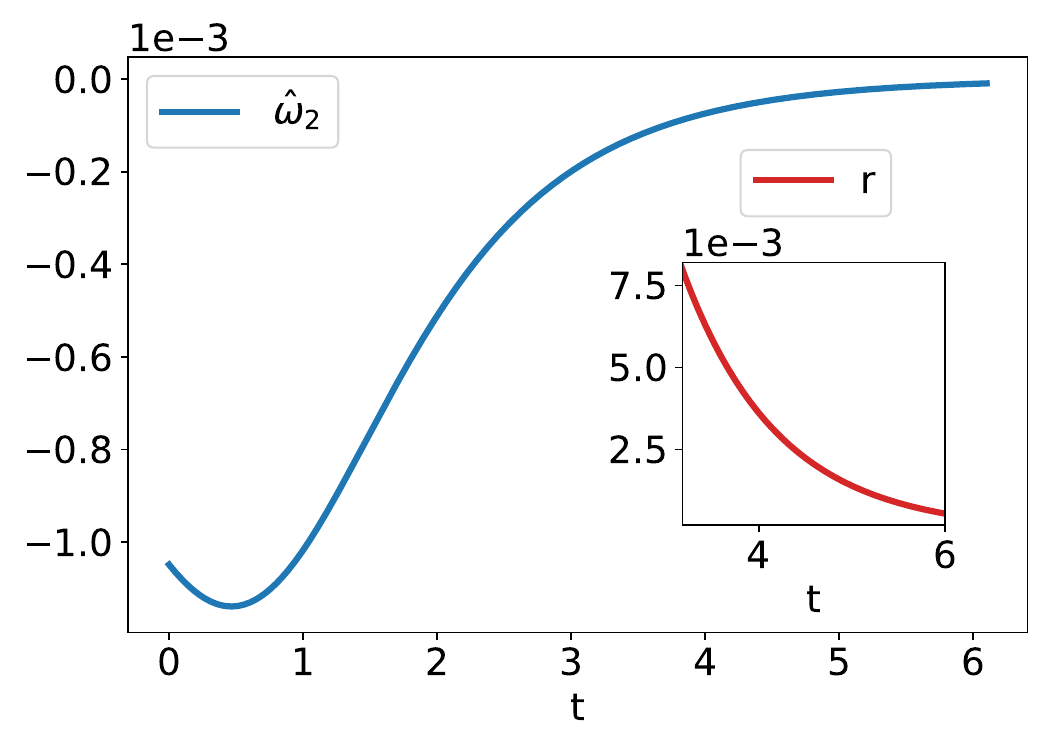}} &
        \resizebox{\imsize}{!}{\includegraphics[width=0.4\textwidth]{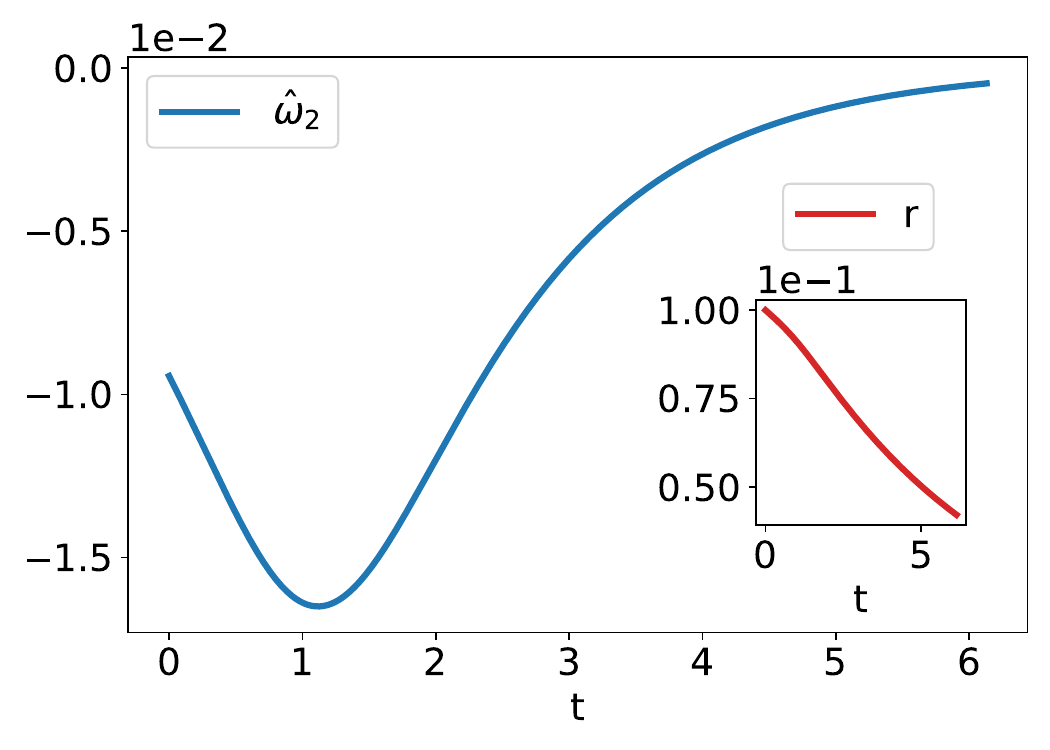}} \\ (a) & (b)
           \end{tabular}
  \par\end{centering}\caption{a) It is plotted, in blue, the numeric evolution to the past of the vorticity for an initial condition near Kasner exact solution for radiation fluid with $w=1/3$ and $\phi=0.4$ for GR. The initial condition is $H =-3.33e^{-1}$, $\Omega_K = 1.0e^{-1}$, $\Omega_m=7.78e^{-1}$, $\Phi_3=-3.29e^{-2}$, $\Sigma_+=1.64e^{-1}$, $\Sigma_-=2.89e^{-1}$, $\eta=1.0e^{-1}$, and $r=1.0e^{-1}$. This solution is attracted to Kasner orbit with $\phi=1.0930$. The graph shows that the vorticity has a small increase in absolute value, followed by a decrease that approaches zero. In the inset, it is plotted in red the decrease of the tilt variable $r$. b) According to GR, in blue, it is plotted the numeric evolution to the past of the vorticity for the initial condition $H =-1.0$, $\Omega_K = 9.0e^{-1}$, $\Omega_m=8.87e^{-2}$, $\Phi_3=-1.25e^{-3}$, $\Sigma_+=6.24e^{-3}$, $\Sigma_-=1.0e^{-1}$, $\eta=1.0e^{-1}$, and $r=1.0e^{-1}$, which is near Milne's exact solution for radiation fluid with $w=1/3$. This solution is past attracted to a Kasner orbit with $\phi=1.4927$. The plot shows the vorticity increase in absolute value before decreasing and approaching zero. In the inset, it is shown in red the decrease of the tilted parameter $r$ toward the singularity}\label{f4}
\end{figure*} 

\begin{figure*}[htpb]
      \begin{centering}
     \begin{tabular}{c c}
            \resizebox{\imsize}{!}{\includegraphics[width=0.4\textwidth]{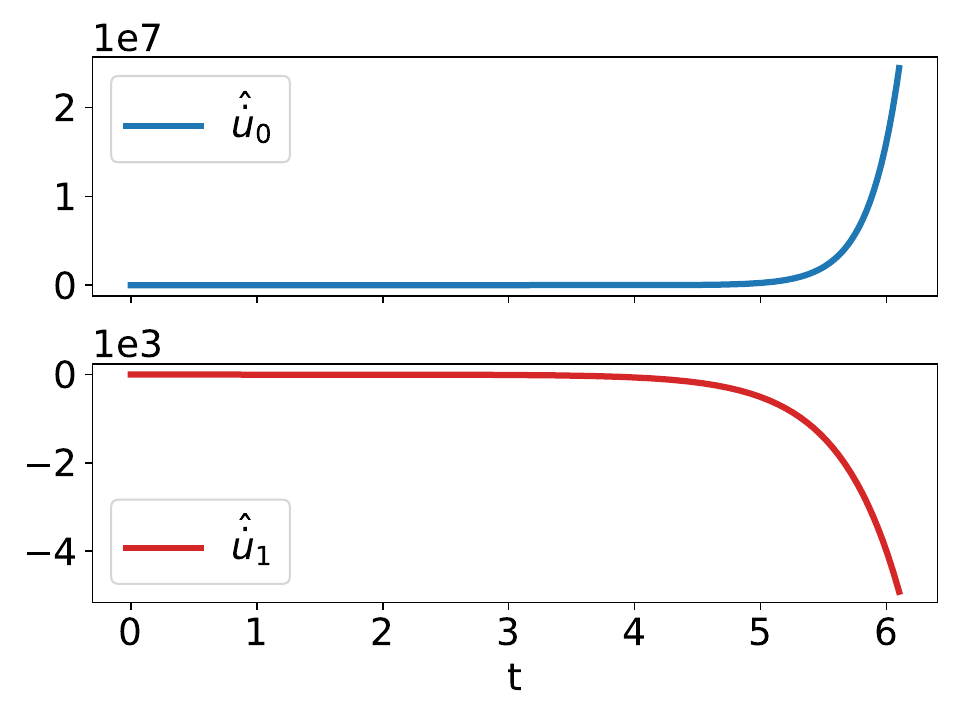}} &
            \resizebox{\imsize}{!}{\includegraphics[width=0.4\textwidth]{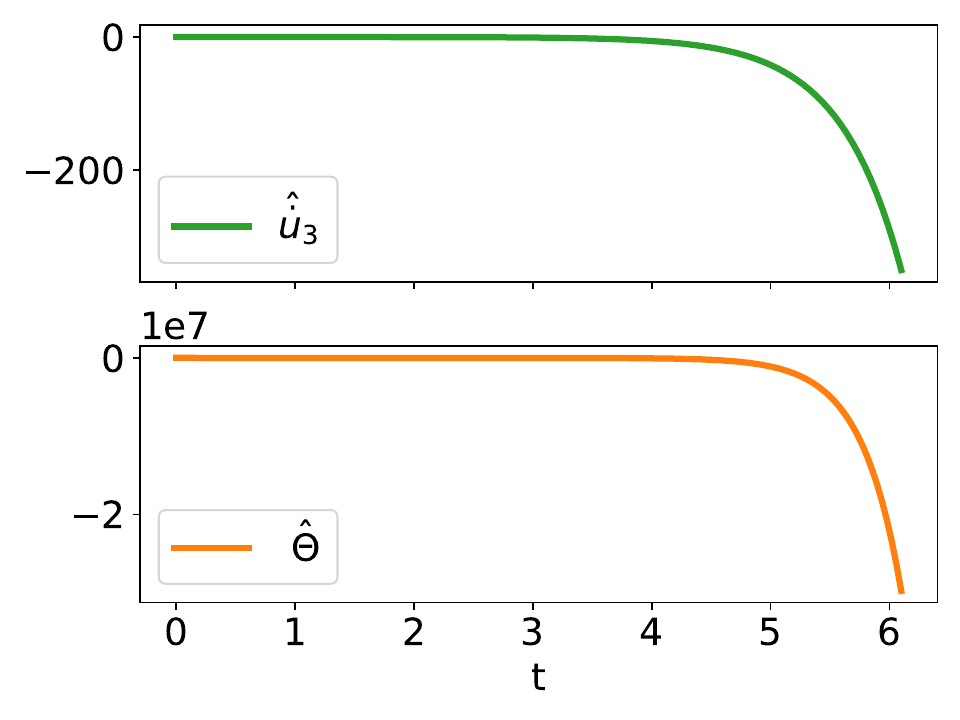}} \\ (a) & (b)\\
            \resizebox{\imsize}{!}{\includegraphics[width=0.4\textwidth]{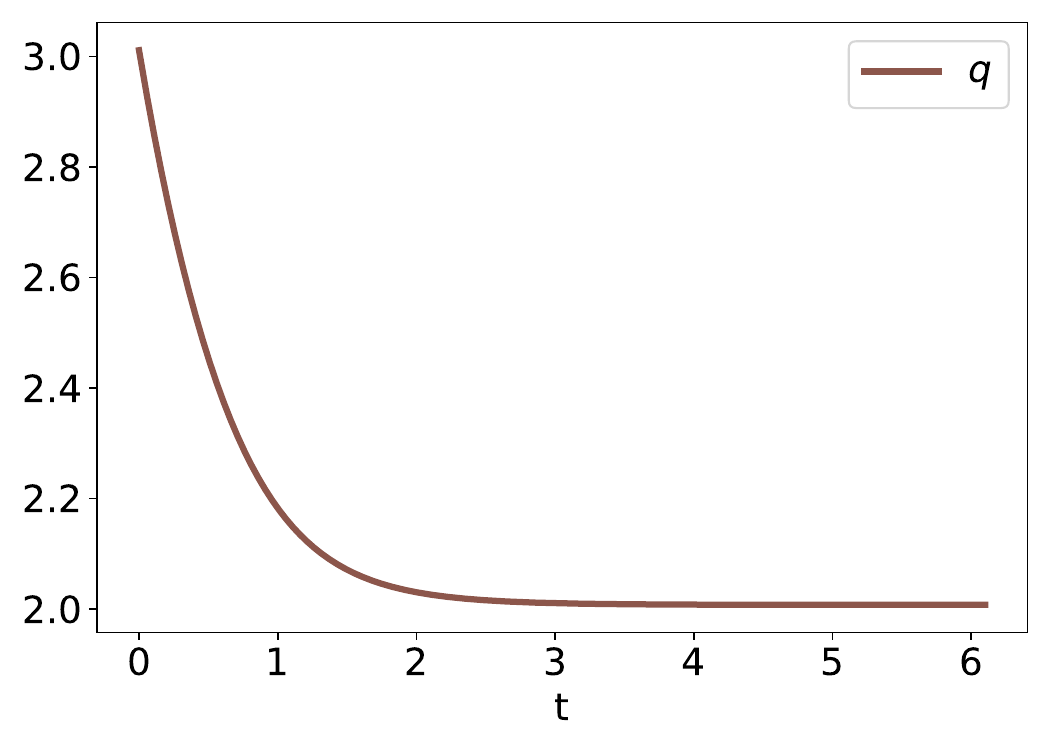}} \\ (c) 
           \end{tabular}
  \par\end{centering}\caption{The numerical past evolution of the kinematic variables are plotted for the same initial condition as in Figure \ref{f4}a, which is near Kasner solution for radiation fluid with $w=1/3$ according to GR. The matter acceleration components $\hat{\dot{u}}_0$ and $\hat{\dot{u}}_1$ shown in a) are respectively in blue and red. While the remaining non-null component, $\hat{\dot{u}}_3$, and the matter contraction $\hat{\Theta}$ are respectively displayed in b) in green and orange. It can be seen that matter contraction diverges at the singularity. c) The graph in brown shows that the deceleration parameter $q$ approaches $q=2$ as the solution is attracted to the Kasner orbit}\label{f5}
\end{figure*} 

\begin{figure*}[htpb]
      \begin{centering}
     \begin{tabular}{c c}
            \resizebox{\imsize}{!}{\includegraphics[width=0.4\textwidth]{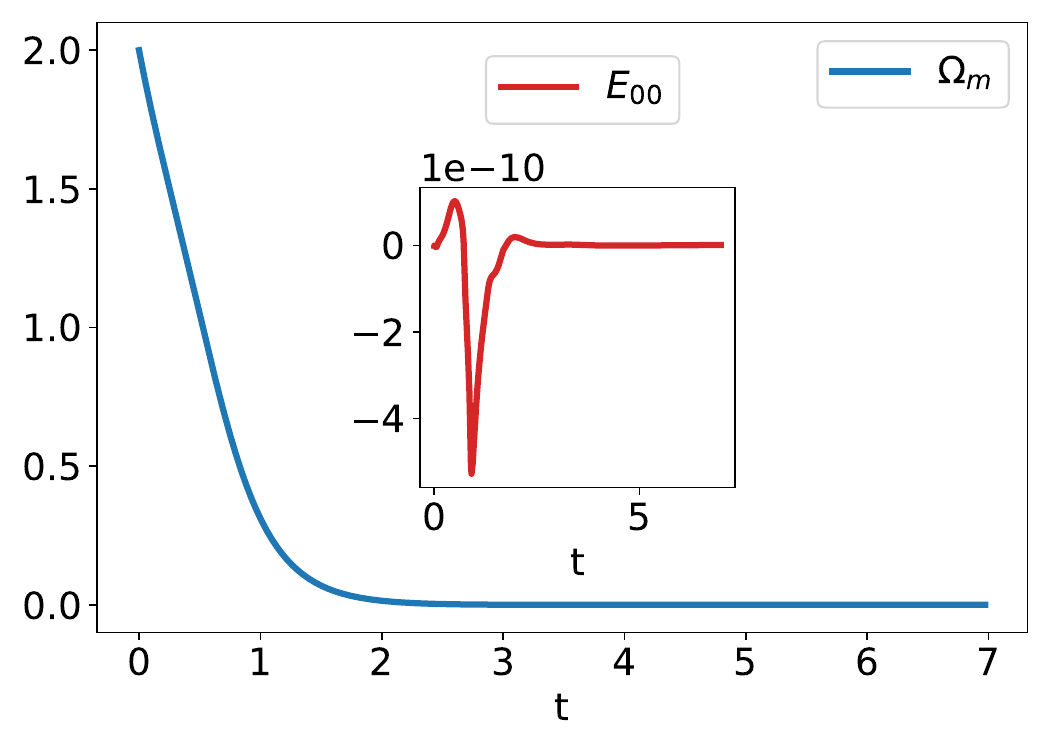}} &
            \resizebox{\imsize}{!}{\includegraphics[width=0.4\textwidth]{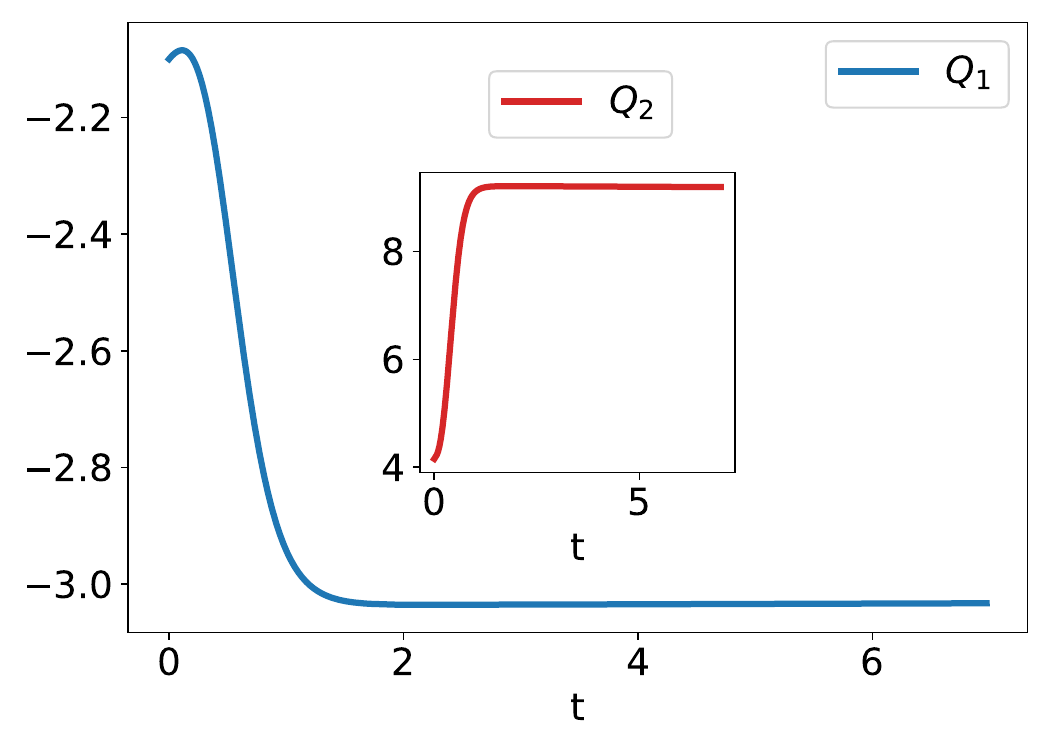}} \\
           \\
        (a) & (b)\\
            \resizebox{\imsize}{!}{\includegraphics[width=0.4\textwidth]{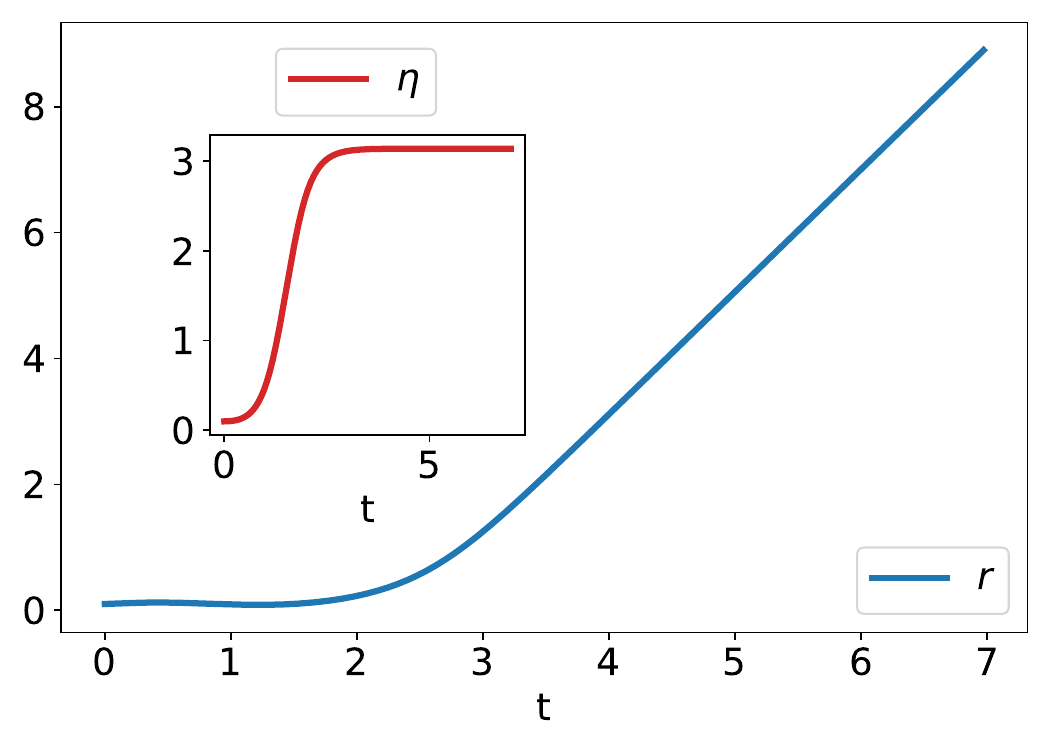}} &
            \resizebox{\imsize}{!}{\includegraphics[width=0.4\textwidth]{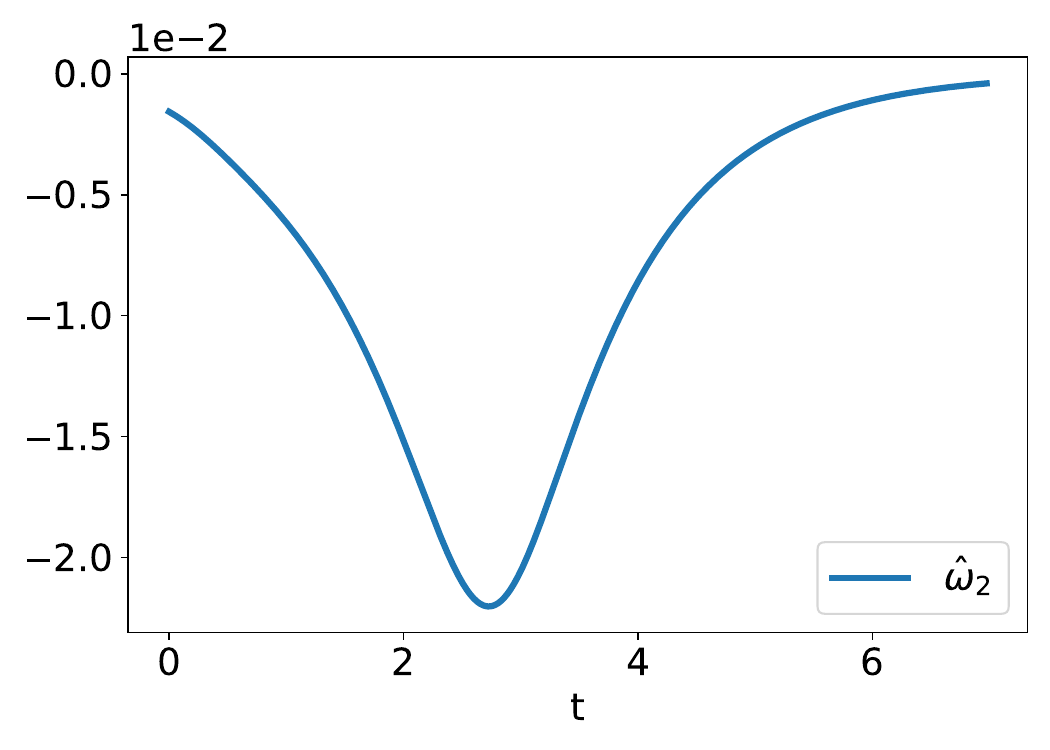}}\\
            (c) & (d)\\
            \resizebox{\imsize}{!}{\includegraphics[width=0.4\textwidth]{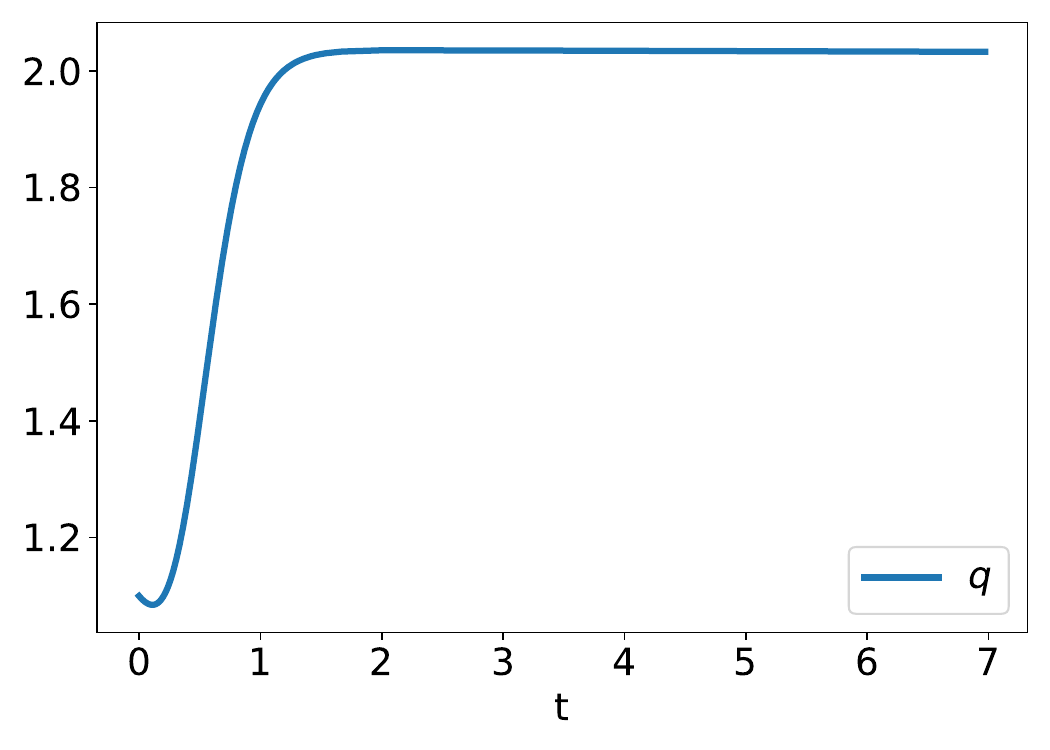}} \\ (e)
           \end{tabular}
  \par\end{centering}\caption{It is plotted the orbit with initial condition $B=8.0e^{-1}$, $\Omega_K=1.0e^{-1}$, $\Phi_3=1.0e^{-1}$, $\Phi_{3,0}=1.0e^{-1}$, $\Phi_{3,1}=-2.19$, $Q_1=2.10$, $Q_2=4.15$, $\Sigma_+=1.0e^{-1}$, $\Sigma_{+1}=1.0e^{-1}$, $\Sigma_{+2}=1.16e^{1}$, $\Sigma_-=1.0e^{-1}$, $\Sigma_{-1}=1.0e^{-1}$, $\Sigma_{-2}=1.0e^{-1}$ with matter source $\Omega_m=2.0$, tilt $r=1.0e^{-1}$ and its direction $\eta=1.0e^{-1}$ chosen to be near isotropic singularity for the dust fluid $w=0$. This orbit is attracted to Kasner orbit with $\phi= 2$. Here the evolution for QG is to the past, with $\chi=1.2$ and $\beta=2$. a) The matter density decreases to zero as the singularity is approached. In the inset, in red, is shown the numerical check for the constraint $E_{00}$ with fluctuations smaller than $10^{-9}$. b) It is plotted in blue the ENV $Q_1$ and in the inset, in red, the ENV $Q_2$, tending to the values $Q_1=-3$ and $Q_2=9$, which are the values of the Kasner orbit for these variables. c) The graph in blue shows the increase of the tilt variable $r$. In red, in the inset, it is shown that the direction of the tilt $\eta$ approaches the angle $\pi$. d) It plots the increase in absolute value of the vorticity following its approach to zero toward the singularity. e) It is plotted, in blue, the deceleration parameter $q\rightarrow2$ as the solution is attracted to the Kasner orbit}\label{f6}
\end{figure*} 

\begin{figure*}[htpb]
      \begin{centering}
     \begin{tabular}{c c}
            \resizebox{\imsize}{!}{\includegraphics[width=0.4\textwidth]{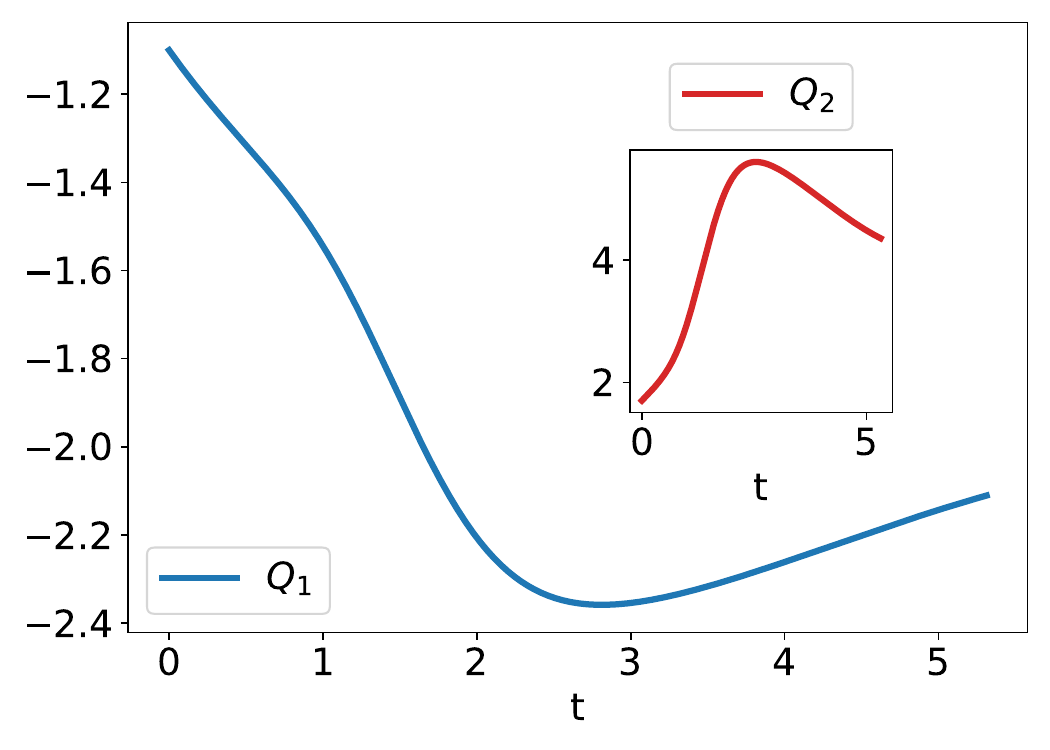}} &
            \resizebox{\imsize}{!}{\includegraphics[width=0.4\textwidth]{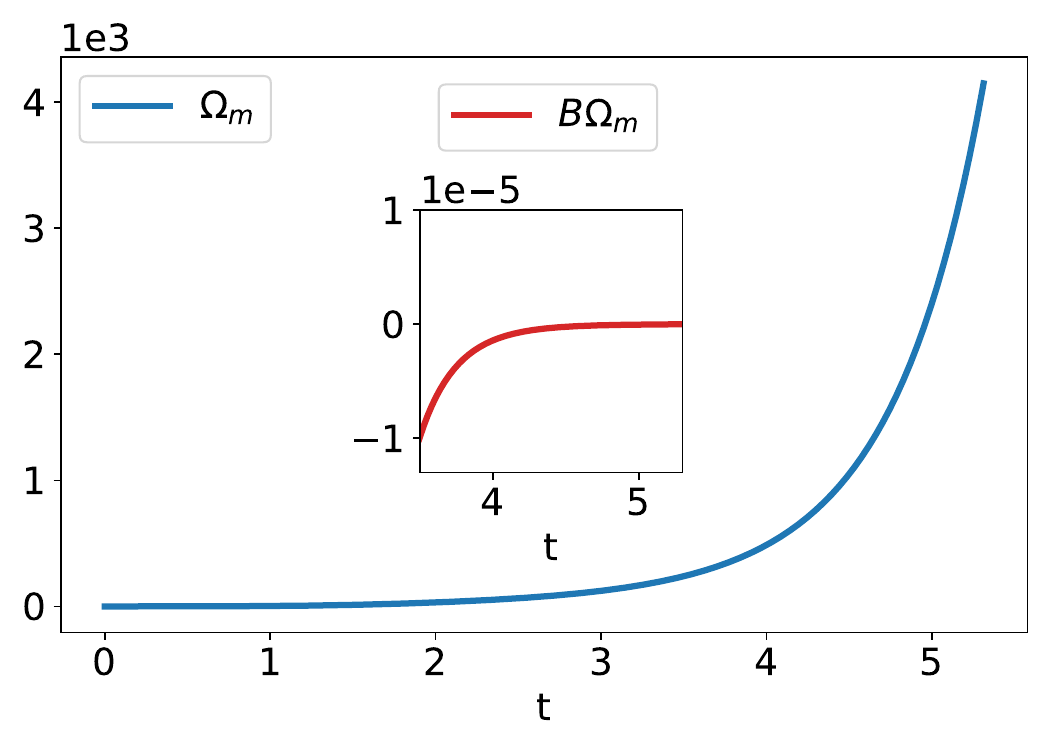}} \\
           \\
        (a) & (b)\\
            \resizebox{\imsize}{!}{\includegraphics[width=0.4\textwidth]{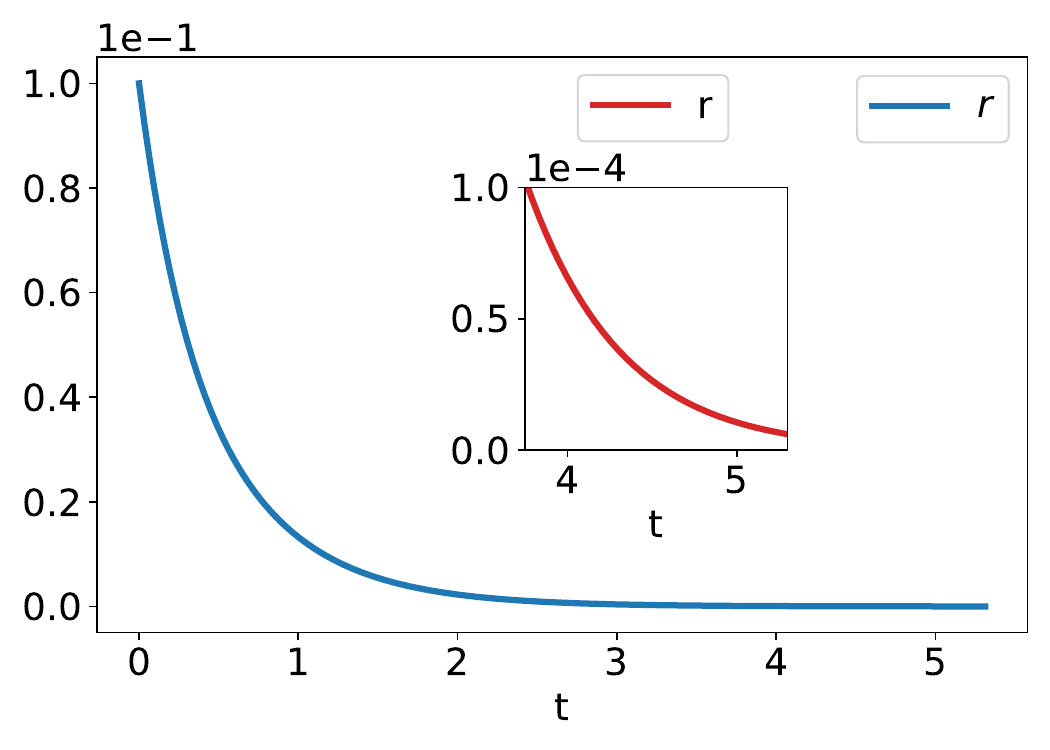}} &
            \resizebox{\imsize}{!}{\includegraphics[width=0.4\textwidth]{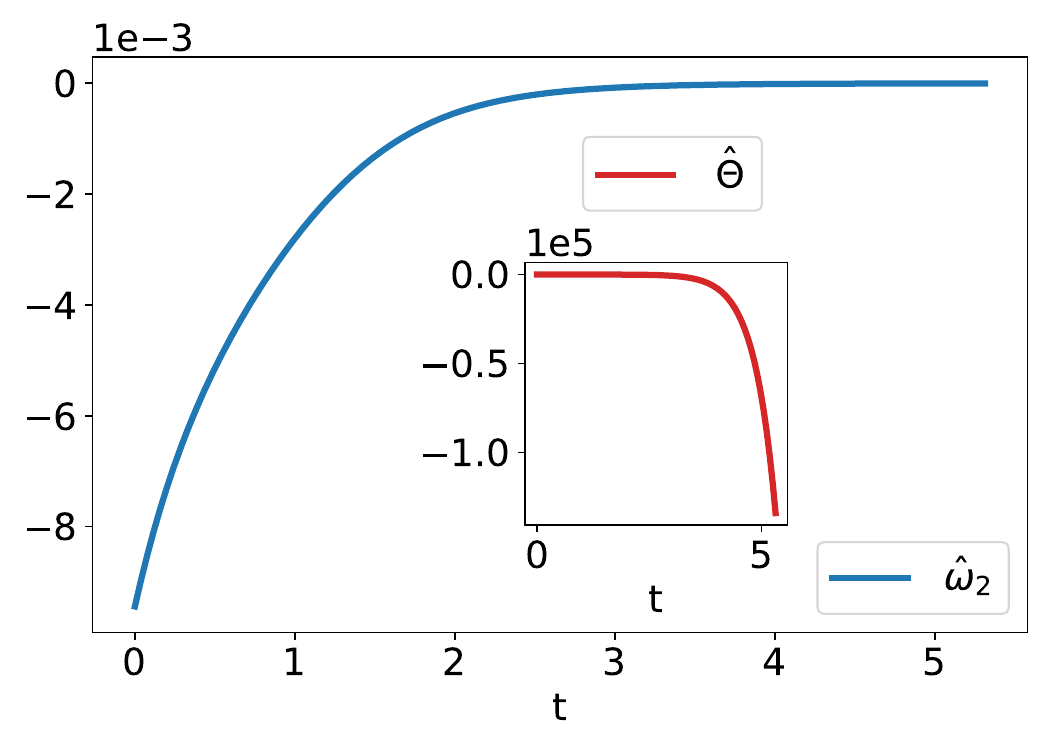}}\\
            (c) & (d)\\
             \resizebox{\imsize}{!}{\includegraphics[width=0.4\textwidth]{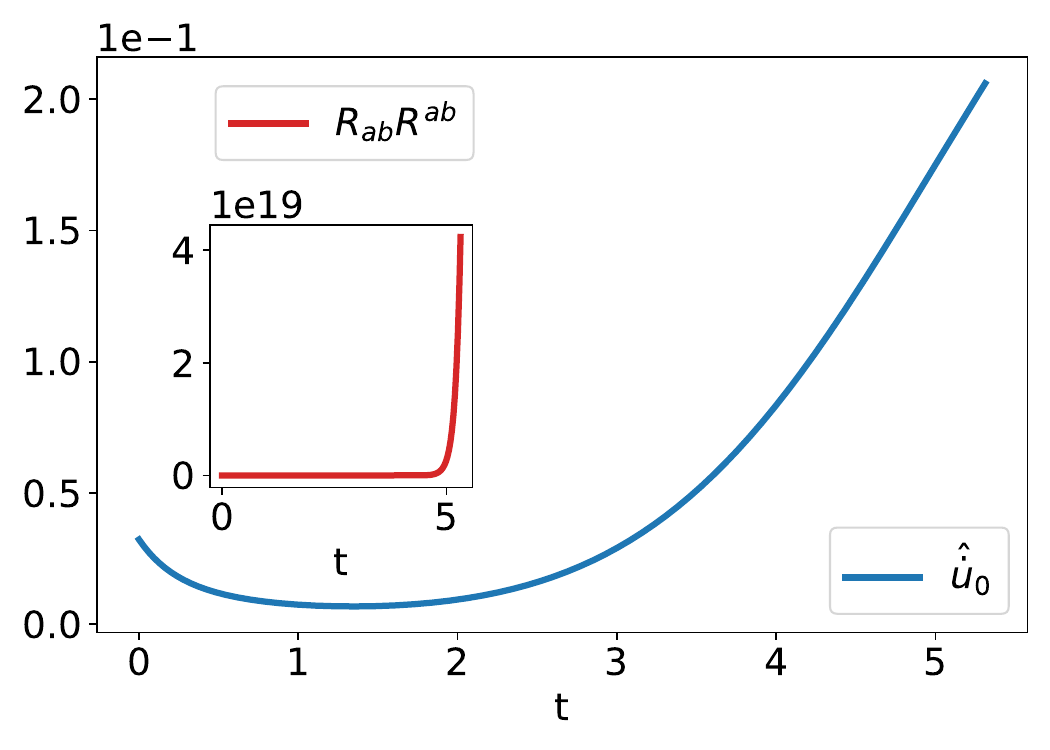}}&
             \resizebox{\imsize}{!}{\includegraphics[width=0.4\textwidth]{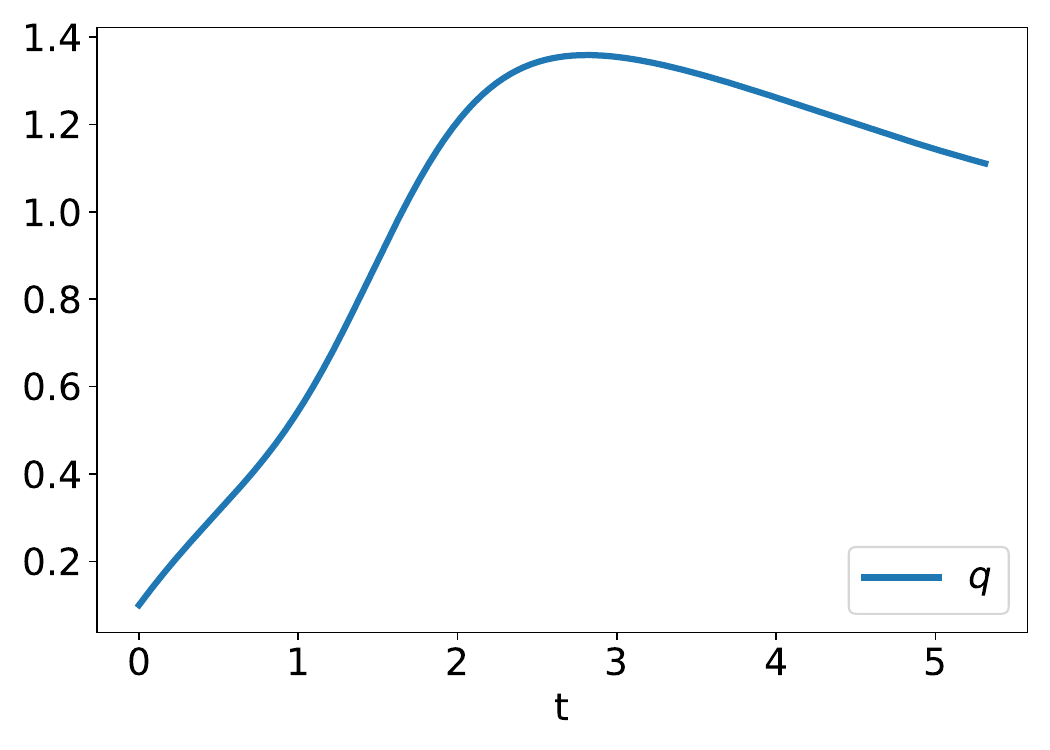}}\\
             (e) & (f)
           \end{tabular}
  \par\end{centering}\caption{QG shows the orbit with initial condition $B=2.0e^{-1}$, $\Omega_K=9.0e^{-1}$, $\Omega_m=1.0e^{-1}$, $\Phi_3=1.0e^{-1}$, $\Phi_{3,0}=1.0e^{-1}$, $\Phi_{3,1}=-1.66$, $Q_1=-1.10$, $Q_2=1.70$, $\Sigma_+=1.0e^{-1}$, $\Sigma_{+1}=1.0e^{-1}$, $\Sigma_{+2}=-1.50$, $\Sigma_-=1.0e^{-1}$, $\Sigma_{-1}=1.0e^{-1}$, $\Sigma_{-2}=1.0e^{-1}$, $\eta=1.0e^{-1}$ and $r=1.0e^{-1}$ near Milne's exact solution for stiff fluid, $\chi=1.2$ and $\beta=2$. The numeric evolution is to the past and the solution is attracted to the isotropic singularity. a) It is plotted in blue, and in the inset, in red, the ENVs $Q_1$ and $Q_2$ approach $-2$ and $-4$, respectively, which are the values of the isotropic singularity orbit for these variables. b) The matter density increases and is plotted in blue. In the inset, it is shown, in red, that the product $B\Omega_m$ tends to zero as $B \rightarrow 0$ towards the singularity. As expected, it shows the matter decouples from the dynamic when $B\rightarrow0$. c) The graph shows in blue the tilt variable $r$ tending to zero. In the inset, in red, is shown a zoom of the plot $r$. d) The vorticity approaches zero and is plotted in blue. In the inset, in red, the graph shows the matter contraction $\hat{\Theta}$. e) The component of the matter acceleration $\hat{\dot{u}}_0$ is plotted in blue. Also, in the inset, it is shown in red the divergence of the scalar product $R_{ab}R^{ab}$, showing the presence of the curvature singularity. f) It is plotted, in blue, the deceleration parameter approaches $q=1$ when the solution is attracted to the isotropic singularity}\label{f7}
\end{figure*} 

\begin{figure*}[htpb]
      \begin{centering}
     \begin{tabular}{c c}
           \resizebox{\imsize}{!}{\includegraphics[width=0.4\textwidth]{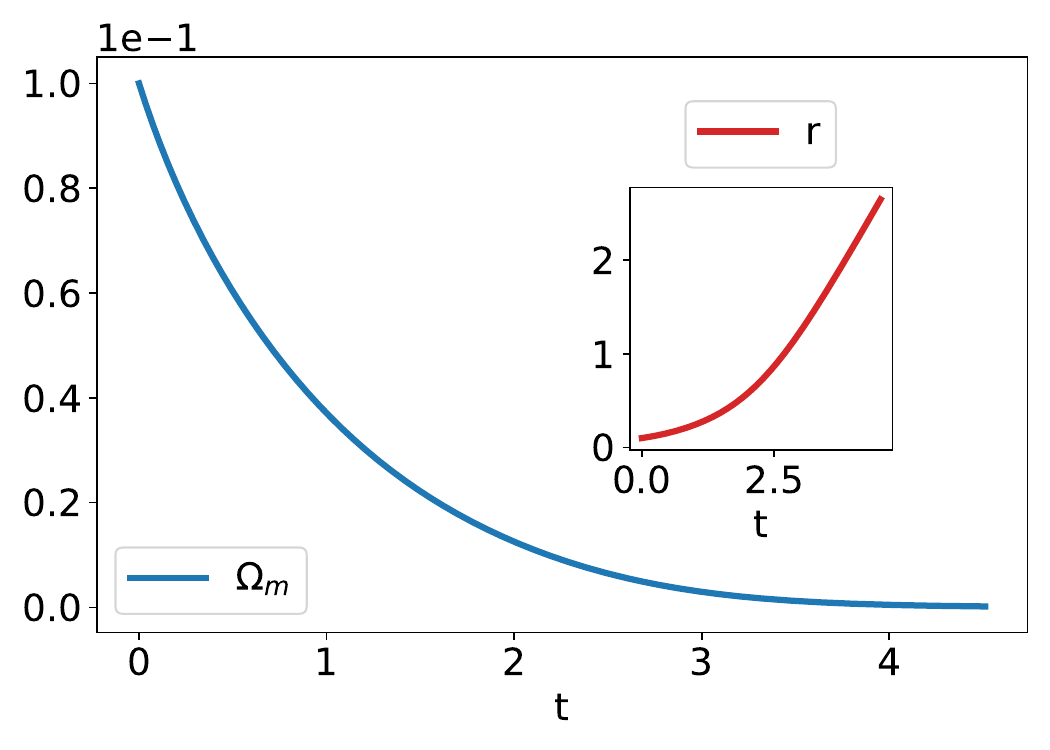}} &
            \resizebox{\imsize}{!}{\includegraphics[width=0.4\textwidth]{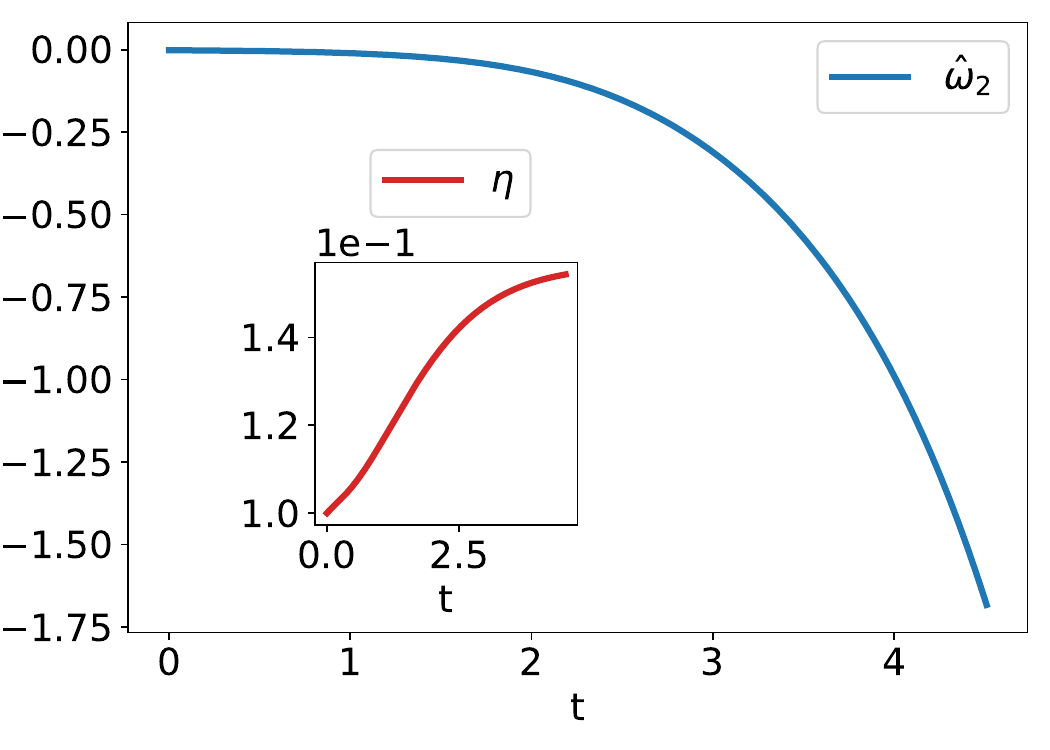}} \\
            \\
            (a)&(b)
           \end{tabular}
  \par\end{centering}\caption{The graph shows the backward time evolution according to QG for the orbit with initial conditions $B=8.0e^{-1}$, $\Omega_K=1.0e^{-1}$, $\Omega_m=1.0e^{-1}$, $\Phi_3=1.0e^{-1}$, $\Phi_{3,0}=1.0e^{-1}$, $\Phi_{3,1}=-1.62 e^{-2}$, $Q_1=-2.10$, $Q_2=3.82$, $\Sigma_+=1.0e^{-1}$, $\Sigma_{+1}=1.0e^{-1}$, $\Sigma_{+2}=7.86e^{-1}$, $\Sigma_-=1.0e^{-1}$, $\Sigma_{-1}=1.0e^{-1}$, $\Sigma_{-2}=1.0e^{-1}$, $\eta=1.0e^{-1}$, and $r=1.0e^{-1}$ near to asymptotic isotropic singularity solution for dust fluid, $w=0$, $\beta=2$, and $\chi=1.2$. This orbit is attracted to the isotropic singularity orbit. a) The plot in blue shows the decrease of the matter density towards zero. In red in the inset, it is shown the past evolution increases of the tilt variable $r$. b) It is plotted in blue the increase of the vorticity toward negative values. In the inset, it is plotted, in red, the direction of the tilt, $\eta$}\label{f8}
\end{figure*} 

\begin{figure*}[htpb]
      \begin{centering}
     \begin{tabular}{c c}
            \resizebox{\imsize}{!}{\includegraphics[width=0.4\textwidth]{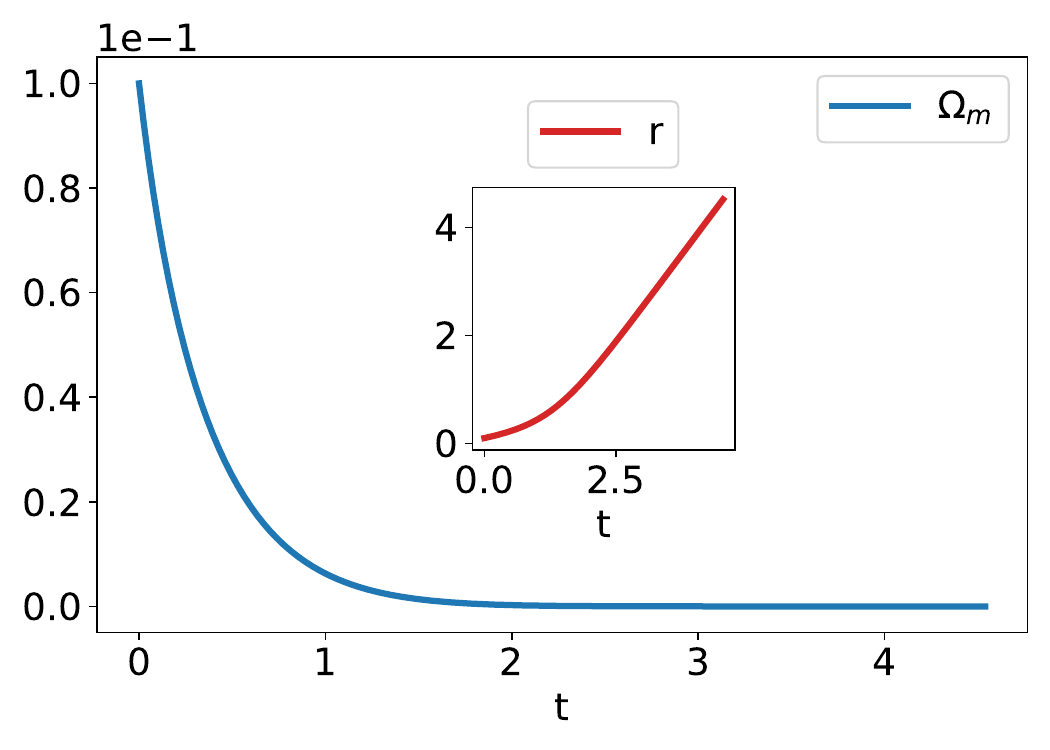}} &
            \resizebox{\imsize}{!}{\includegraphics[width=0.4\textwidth]{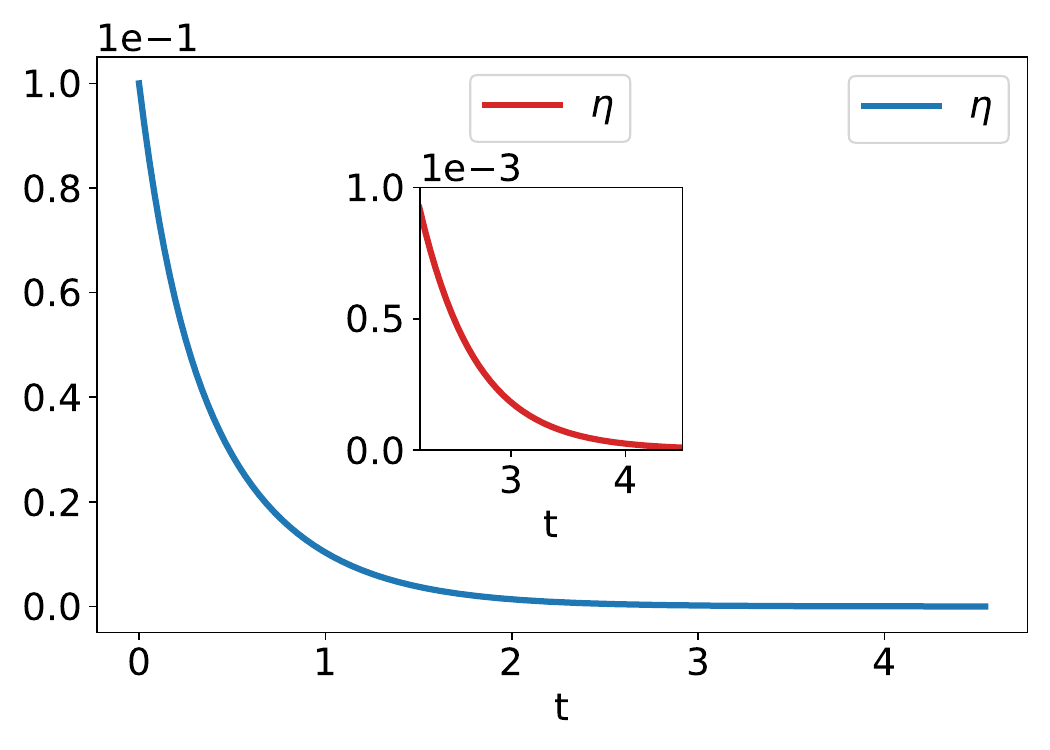}} \\
           \\
        (a) & (b)\\
                    \resizebox{\imsize}{!}{\includegraphics[width=0.4\textwidth]{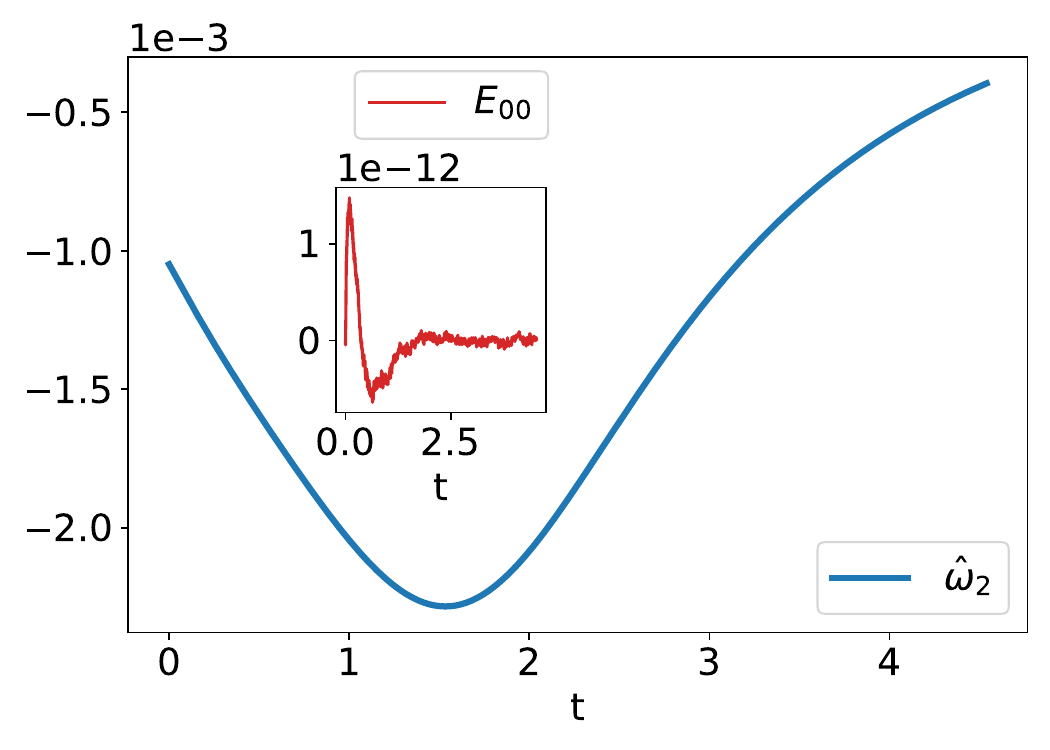}} &\\(c)
           \end{tabular}
  \par\end{centering}\caption{It is plotted the orbit with initial condition $B=1.80$, $\Omega_K=1.0e^{-1}$, $\Omega_m=1.0e^{-1}$, $\Phi_3=1.0e^{-1}$, $\Phi_{3,0}=1.0e^{-1}$, $\Phi_{3,1}=-2.28$ $Q_1=-3.10$, $Q_2=8.16$, $\Sigma_+=-5.16e^{-1}$, $\Sigma_{+1}=1.14$, $\Sigma_{+2}=8.15$, $\Sigma_-=8.09e^{-1}$, $\Sigma_{-1}=-2.82$, $\Sigma_{-2}=8.08$, $\eta=1.0e^{-1}$, and $r=1.0e^{-1}$ near to Kasner exact solution with $\phi=2$ for dust fluid, $w=0$. Again, the dynamics from QG is toward the past, and it is setting $\beta=2$ and $\chi=1.2$. This solution is attracted to isotropic singularity. a) The graph in blue shows the matter density decreasing towards zero. Also, in the inset, in red, the increase of the tilt variable $r$ is plotted. b) In blue, it is plotted that the direction of the tilt, $\eta$, tends to zero. In the inset, it is shown, in red, a zoom of the plot $\eta$. c) It is plotted, in blue, the vorticity increase in absolute value followed by a zero approach as $\eta \rightarrow 0$. In the inset, it is shown, in red, the numerical check for the constraint $E_{00}$ with fluctuations smaller than $10^{-11}$}\label{f9}
\end{figure*} 

\begin{figure*}[htpb]
      \begin{centering}
     \begin{tabular}{c c}
            \resizebox{\imsize}{!}{\includegraphics[width=0.4\textwidth]{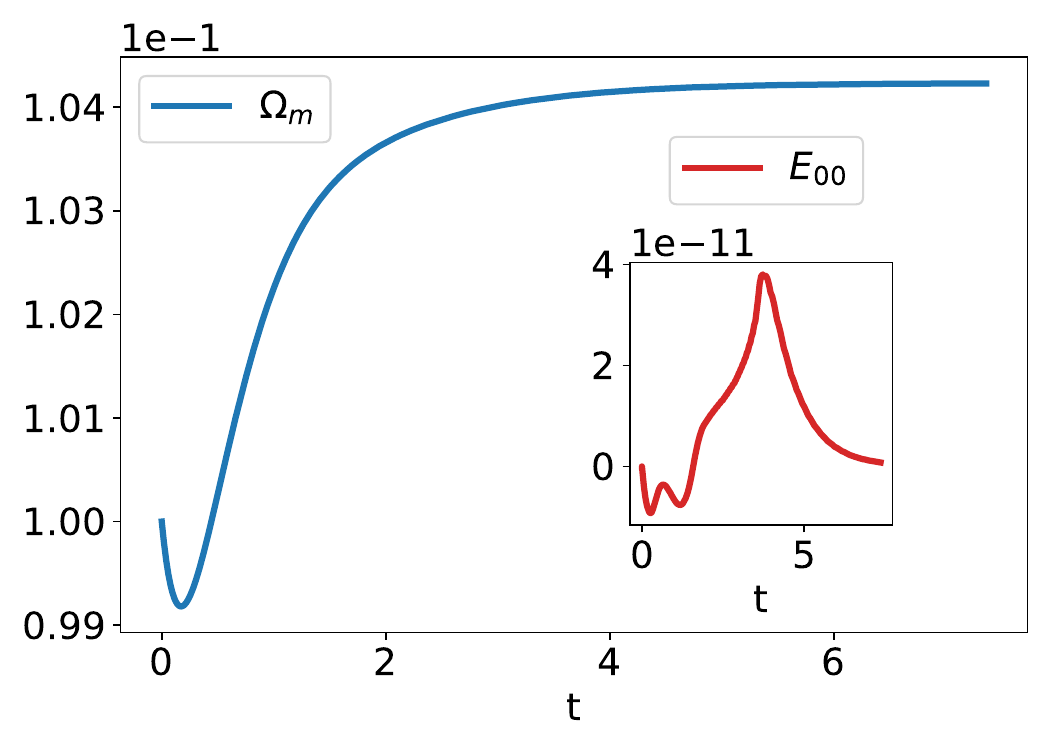}} &
            \resizebox{\imsize}{!}{\includegraphics[width=0.4\textwidth]{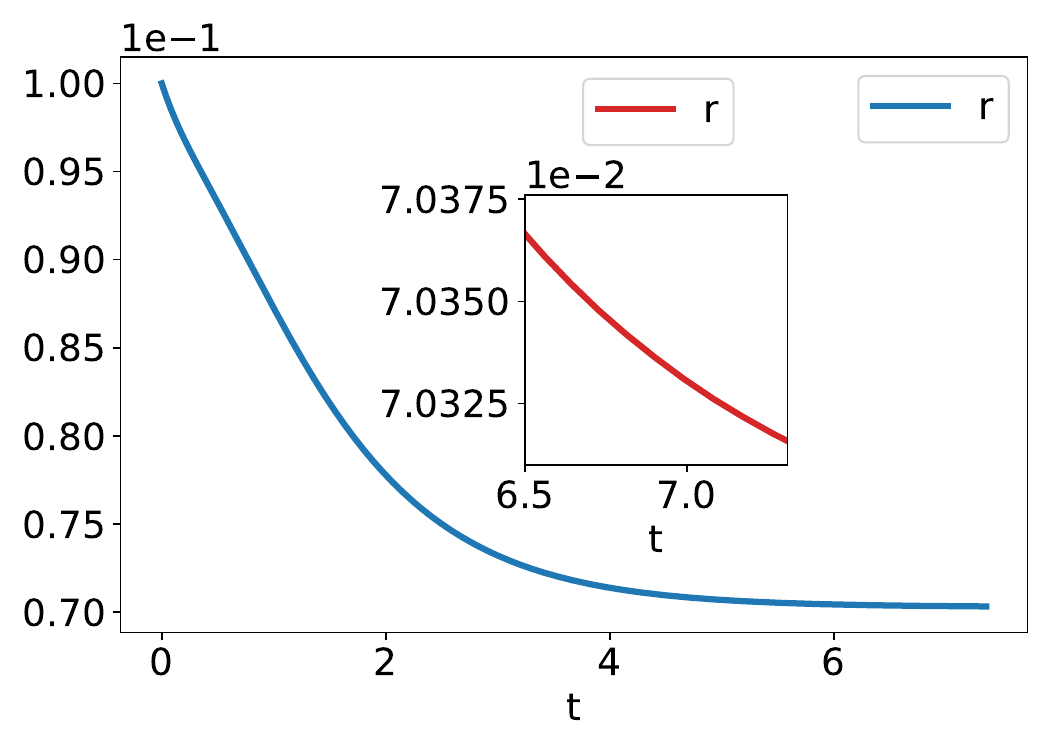}} \\
           \\
        (a) & (b)\\
            \resizebox{\imsize}{!}{\includegraphics[width=0.4\textwidth]{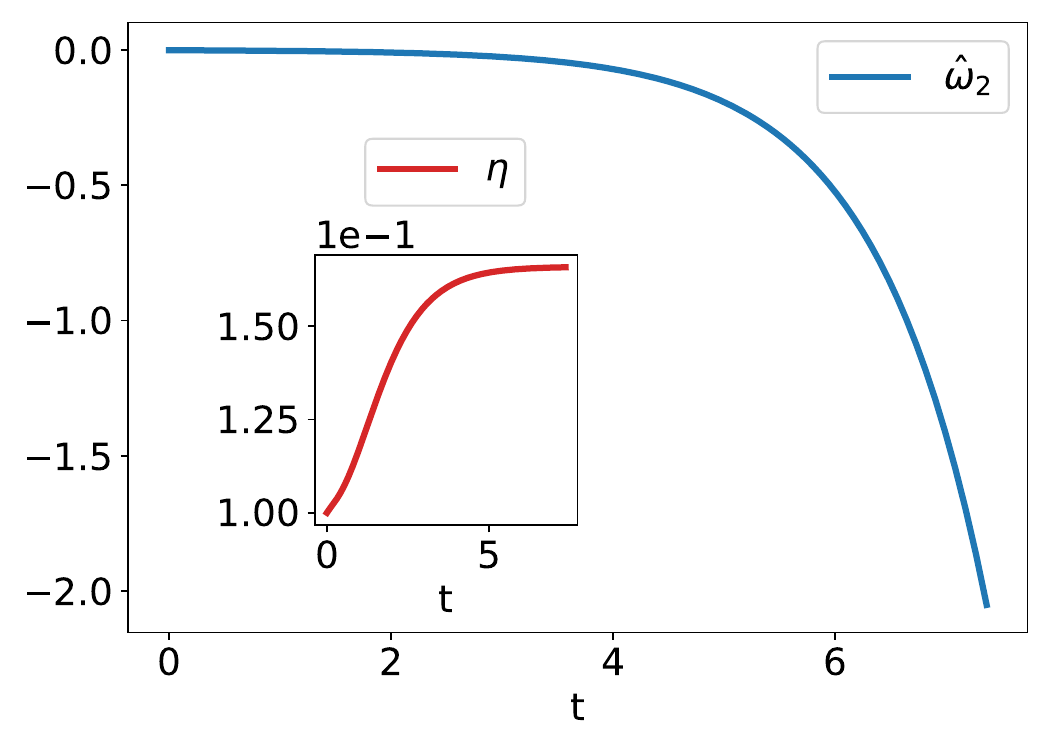}} &
            \resizebox{\imsize}{!}{\includegraphics[width=0.4\textwidth]{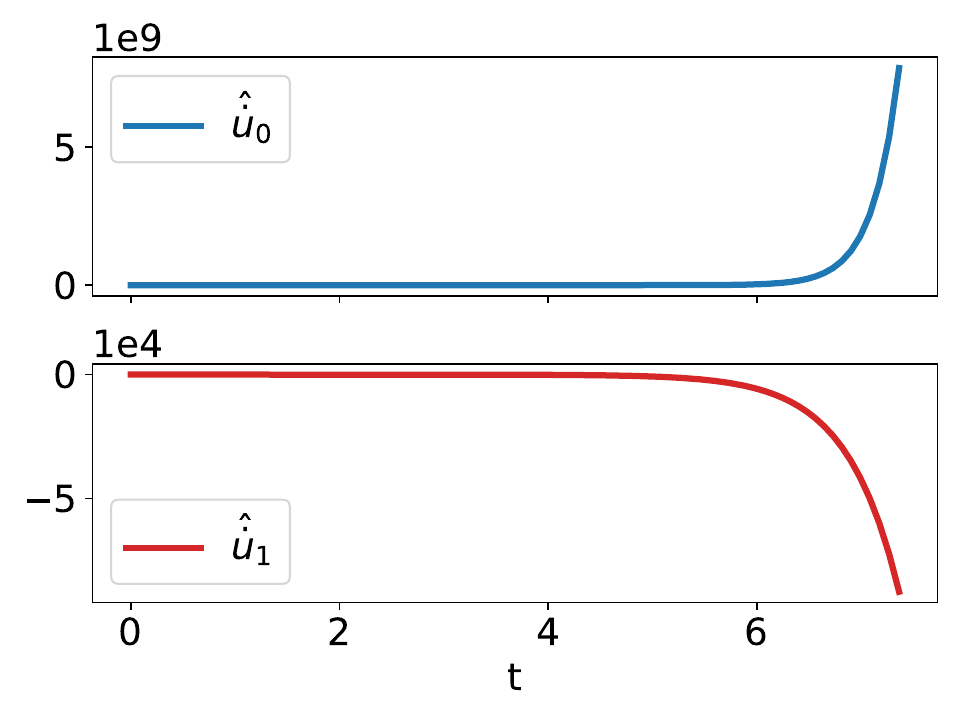}} \\
           \\
        (c) & (d)\\
                    \resizebox{\imsize}{!}{\includegraphics[width=0.4\textwidth]{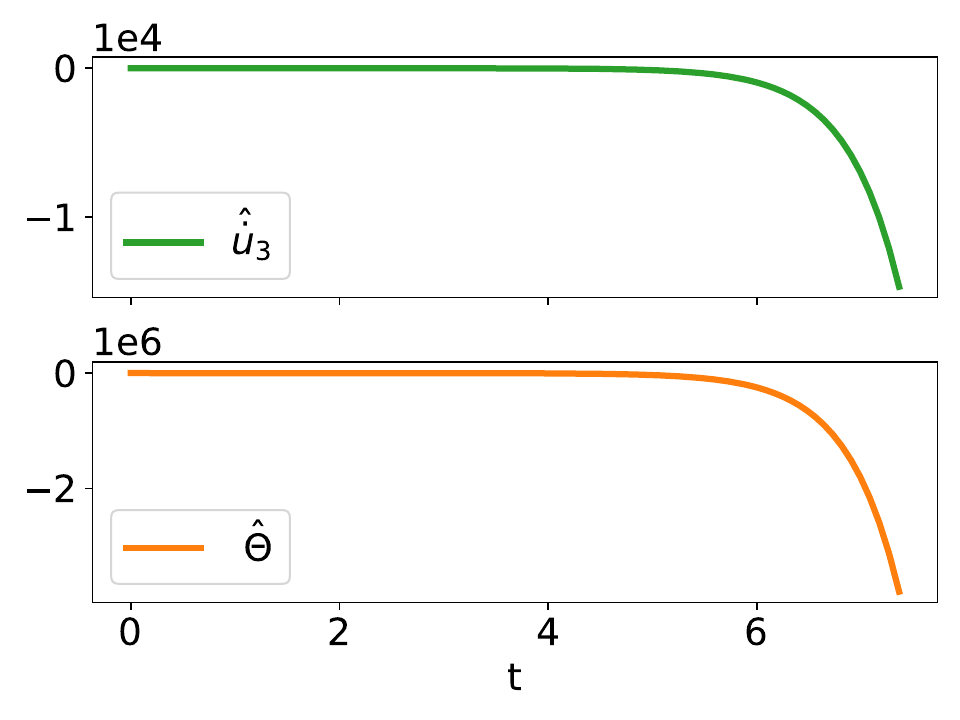}} &\\
                    (e)
           \end{tabular}
  \par\end{centering}\caption{According to QG the graph shows the past evolution for the initial condition $B=8.0e^{-1}$, $\Omega_K=1.0e^{-1}$, $\Omega_m=1.0e^{-1}$, $\Phi_3=1.0e^{-1}$, $\Phi_{3,0}=1.0e^{-1}$, $\Phi_{3,1}=-5.43e^{-2}$ $Q_1=-2.10$, $Q_2=3.82$, $\Sigma_+=1.0e^{-1}$, $\Sigma_{+1}=1.0e^{-1}$, $\Sigma_{+2}=9.76e^{-1}$, $\Sigma_-=1.0e^{-1}$, $\Sigma_{-1}=1.0e^{-1}$, $\Sigma_{-2}=1.0e^{-1}$, $\eta=1.0e^{-1}$, and $r=1.0e^{-1}$ near to isotropic singularity for radiation fluid $w=1/3$, with $\beta=2$ and $\chi=1.2$. a) It is plotted in blue the matter density behavior before it approaches a constant. In the inset, in red, it is shown the numerical check for the constraint $E_{00}$ with fluctuations smaller than $10^{-10}$. Panel b) shows in blue that the tilt variable $r$ decreases towards the singularity. The inset shows a zoom, in red, of the plot $r$. c) The graph shows the increase in vorticity in blue. In the inset, it is shown in red the past evolution of the direction of the tilt, $\eta$. d) The matter acceleration components $\hat{\dot{u}}_0$ and $\hat{\dot{u}}_1$ are plotted in blue and in red, respectively, showing their divergence towards the singularity. e) It is plotted in green the divergence of the remaining non-null matter acceleration component $\hat{\dot{u}}_3$. Also, it is shown in orange the matter contraction $\hat{\Theta}$}\label{f10}
\end{figure*} 

\begin{figure*}[htpb]
      \begin{centering}
     \begin{tabular}{c c}
            \resizebox{\imsize}{!}{\includegraphics[width=0.4\textwidth]{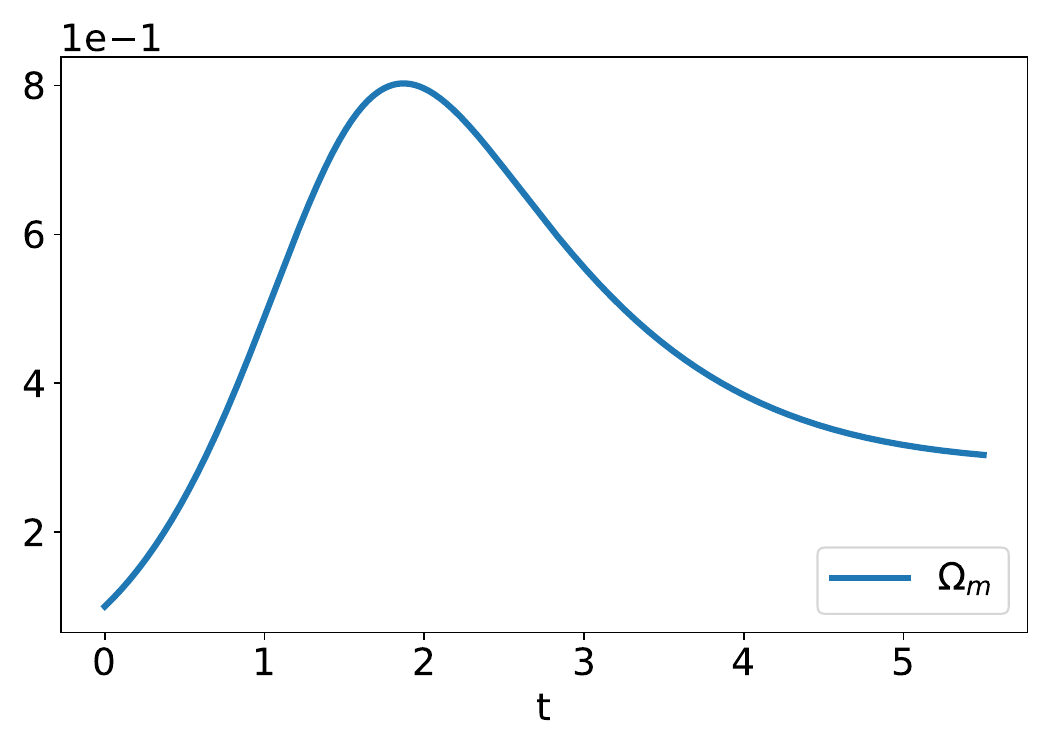}} &
            \resizebox{\imsize}{!}{\includegraphics[width=0.4\textwidth]{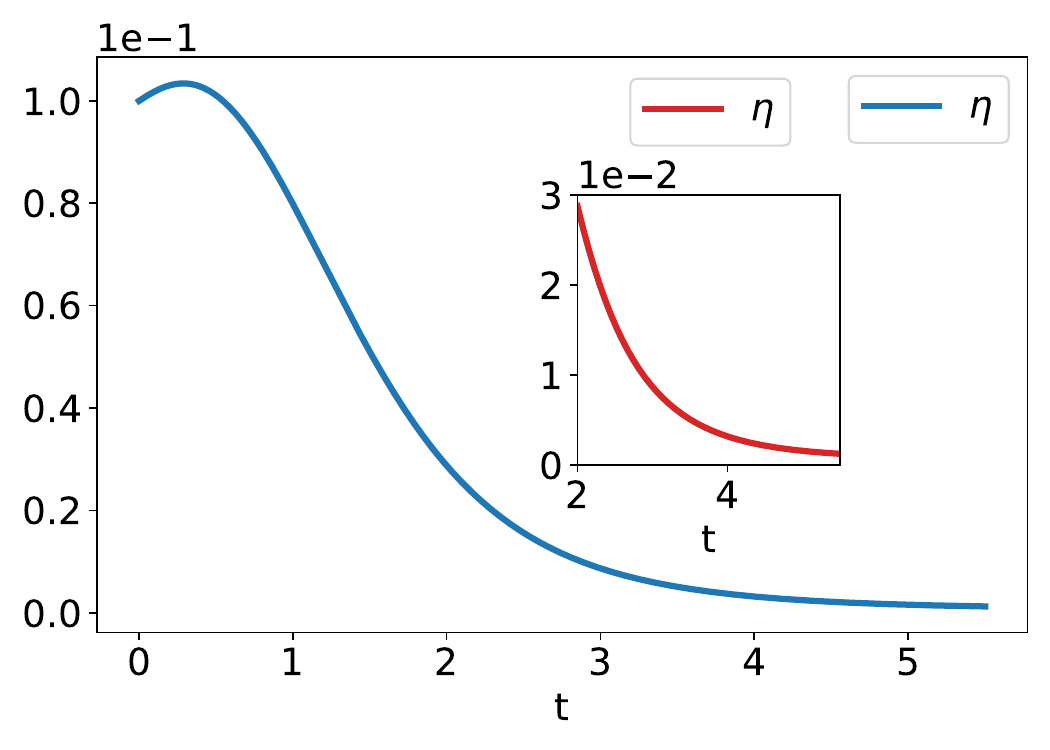}} \\
           \\
        (a) & (b)\\
            \resizebox{\imsize}{!}{\includegraphics[width=0.4\textwidth]{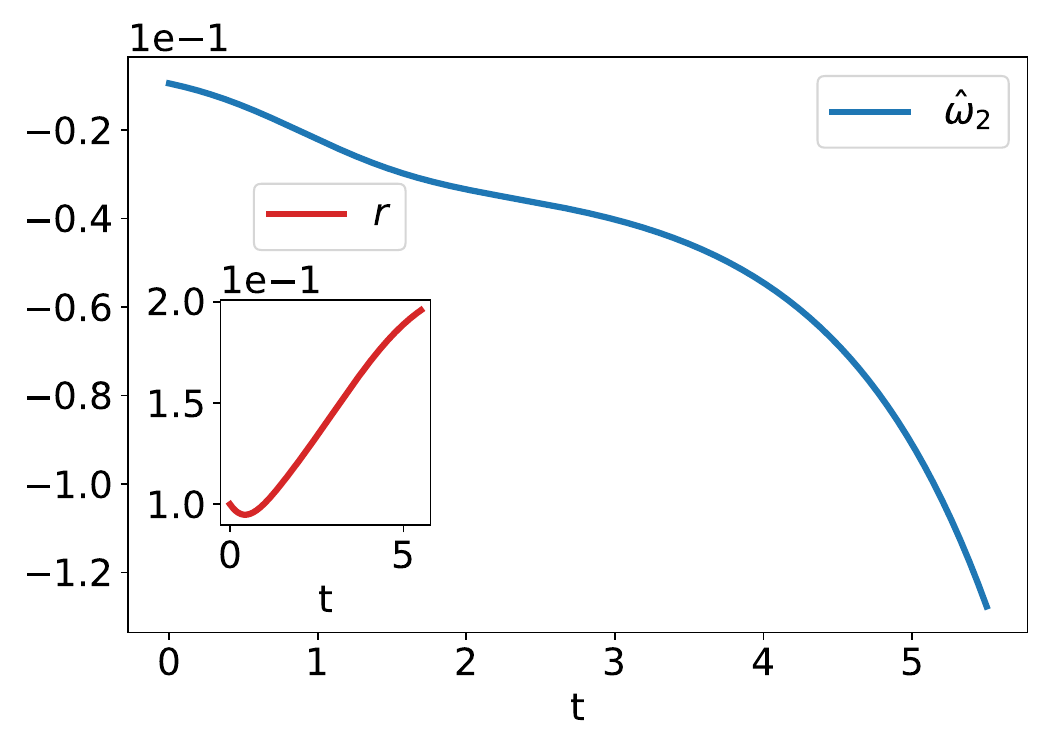}} &
            \resizebox{\imsize}{!}{\includegraphics[width=0.4\textwidth]{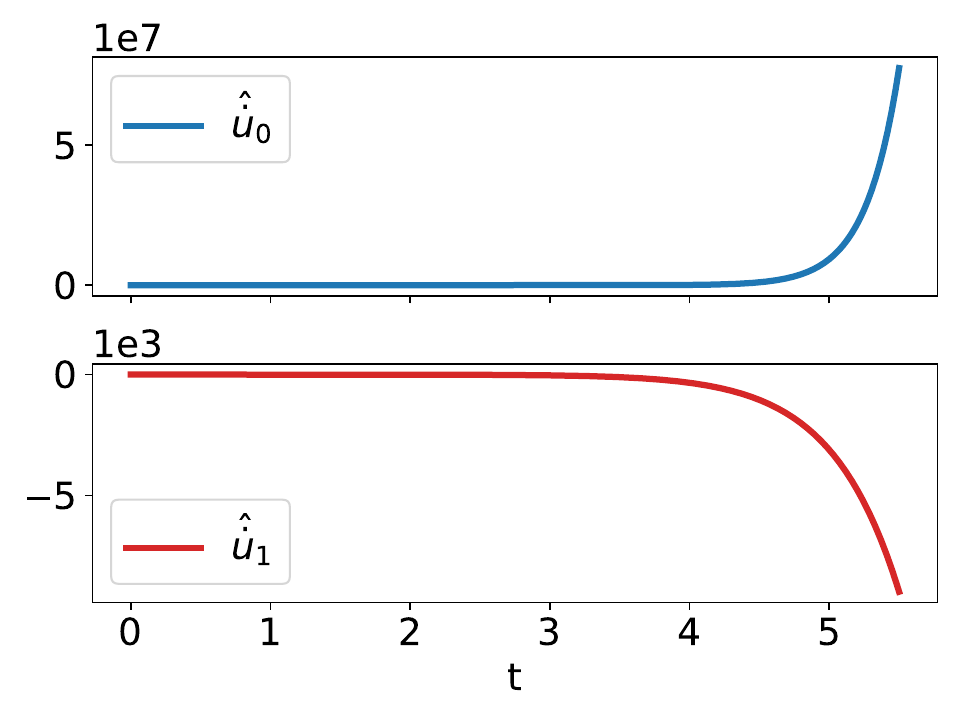}} \\
           \\
        (c) & (d)\\
        \resizebox{\imsize}{!}{\includegraphics[width=0.4\textwidth]{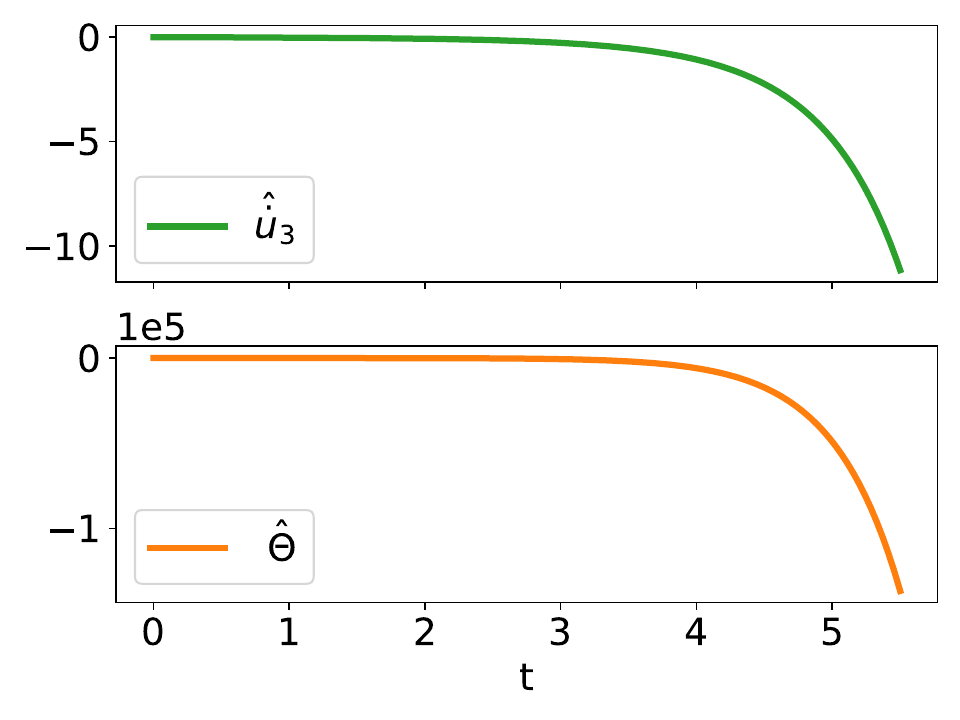}}&\\
        (e)
           \end{tabular}
  \par\end{centering}\caption{The evolution to the past for QG shows the orbit with initial condition $B=2.0e^{-1}$, $\Omega_K=9.0e^{-1}$, $\Omega_m=1.0e^{-1}$, $\Phi_3=1.0e^{-1}$, $\Phi_{3,0}=1.0e^{-1}$, $\Phi_{3,1}=-1.65$ $Q_1=-1.10$, $Q_2=1.70$, $\Sigma_+=1.0e^{-1}$, $\Sigma_{+1}=1.0e^{-1}$, $\Sigma_{+2}=-1.54$, $\Sigma_-=1.0e^{-1}$, $\Sigma_{-1}=1.0e^{-1}$, $\Sigma_{-2}=1.0e^{-1}$, $\eta=1.0e^{-1}$, and $r=1.0e^{-1}$ near Milne's solution for radiation fluid, $w=1/3$, $\beta=2$, and $\chi=1.2$. This solution is attracted to isotropic singularity orbit. a) It is shown in blue that the matter density has a brief increase, followed by a decrease, and then approaches a constant value. b) The direction of the tilt, $\eta$, is plotted in blue. In the inset, it is shown in red a zoom of the plot $\eta$. c) The graph in blue shows the increase of the vorticity in absolute value. In the inset, in red, it is shown the increase of the tilt variable $r$. d) It is plotted the increase following a divergence in the matter acceleration components $\hat{\dot{u}}_0$ and $\hat{\dot{u}}_1$, in blue and red, respectively. e) It is plotted in green the increase of the remaining non-null matter acceleration component $\hat{\dot{u}}_3$. Also, it is plotted in orange the matter contraction $\hat{\Theta}$ divergence at the singularity, as expected}\label{f11}
\end{figure*} 

\begin{figure*}[htpb]
      \begin{centering}
     \begin{tabular}{c c}
            \resizebox{\imsize}{!}{\includegraphics[width=0.4\textwidth]{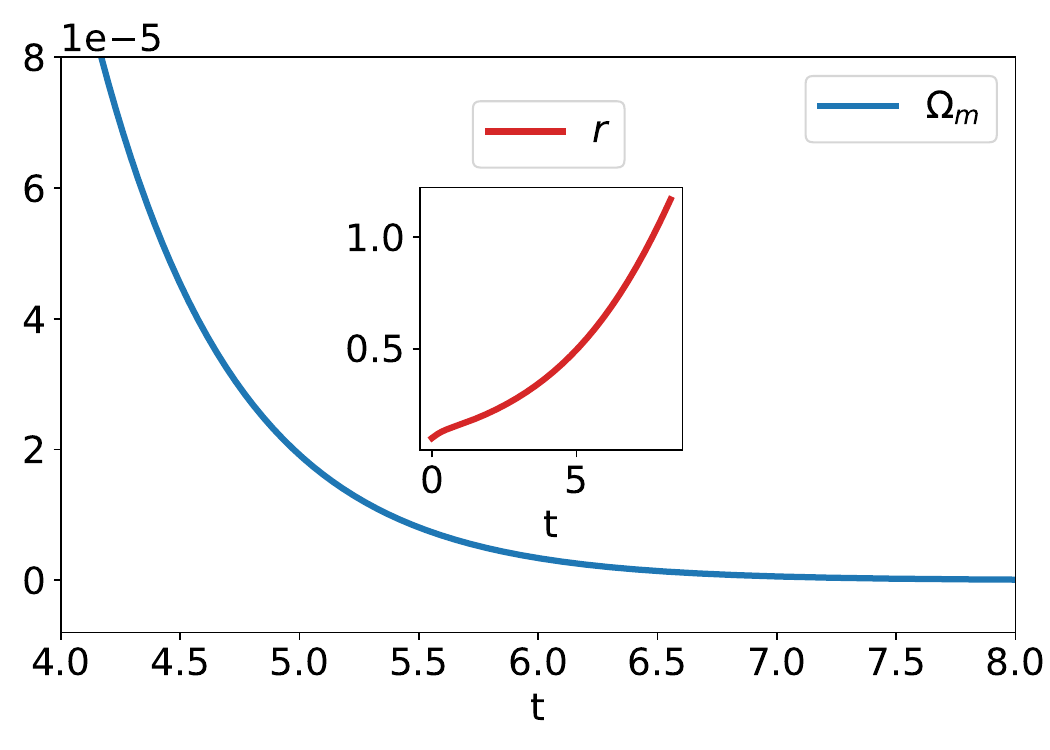}} &
            \resizebox{\imsize}{!}{\includegraphics[width=0.4\textwidth]{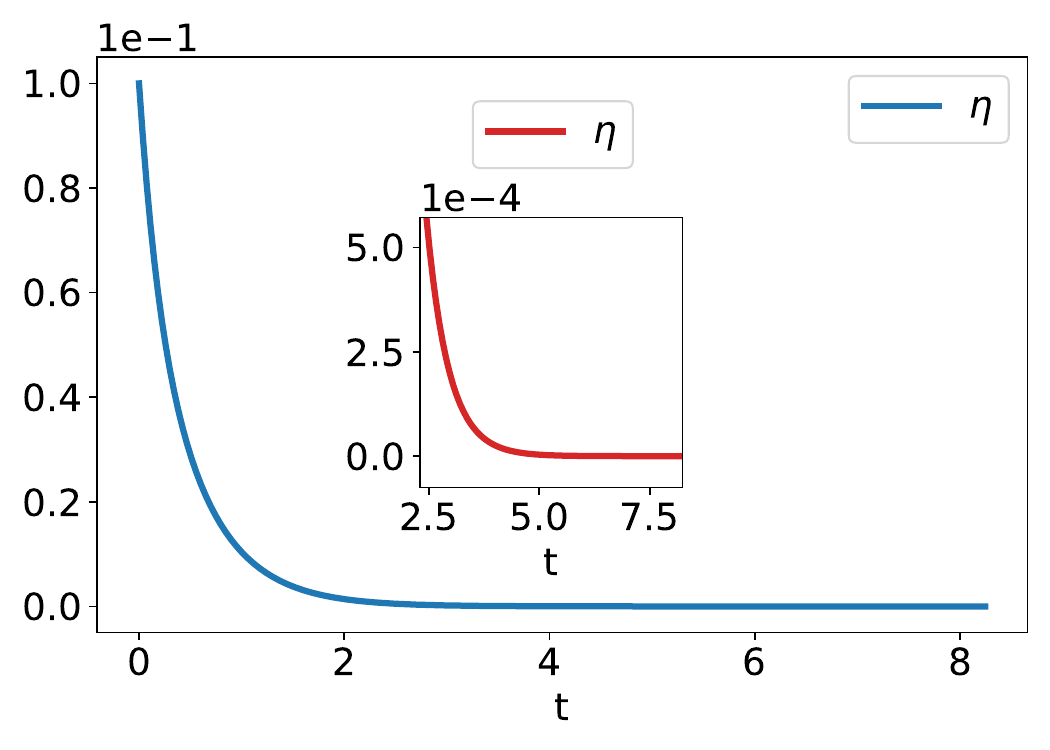}} \\
           \\
        (a) & (b)\\
            \resizebox{\imsize}{!}{\includegraphics[width=0.4\textwidth]{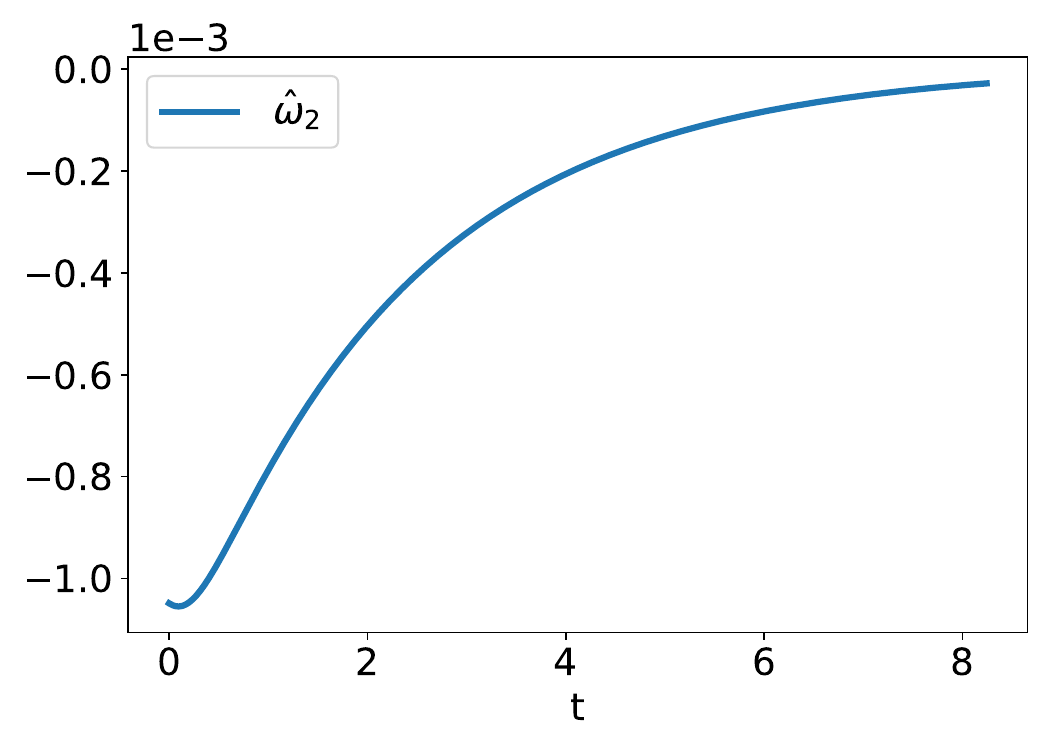}} &
            \resizebox{\imsize}{!}{\includegraphics[width=0.4\textwidth]{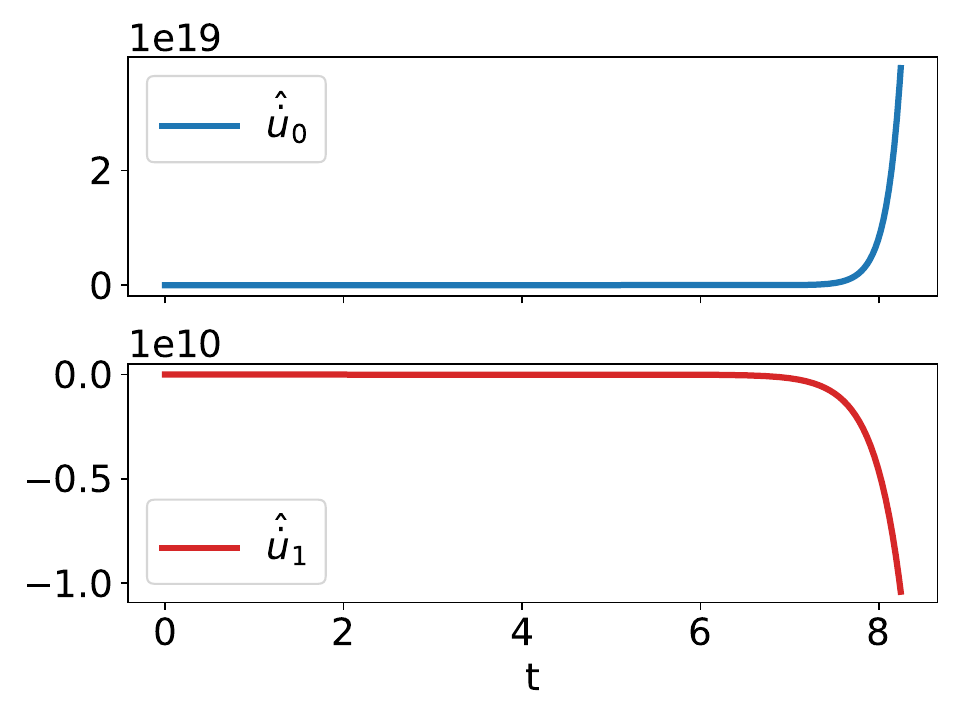}} \\
           \\
        (c) & (d)\\
        \resizebox{\imsize}{!}{\includegraphics[width=0.4\textwidth]{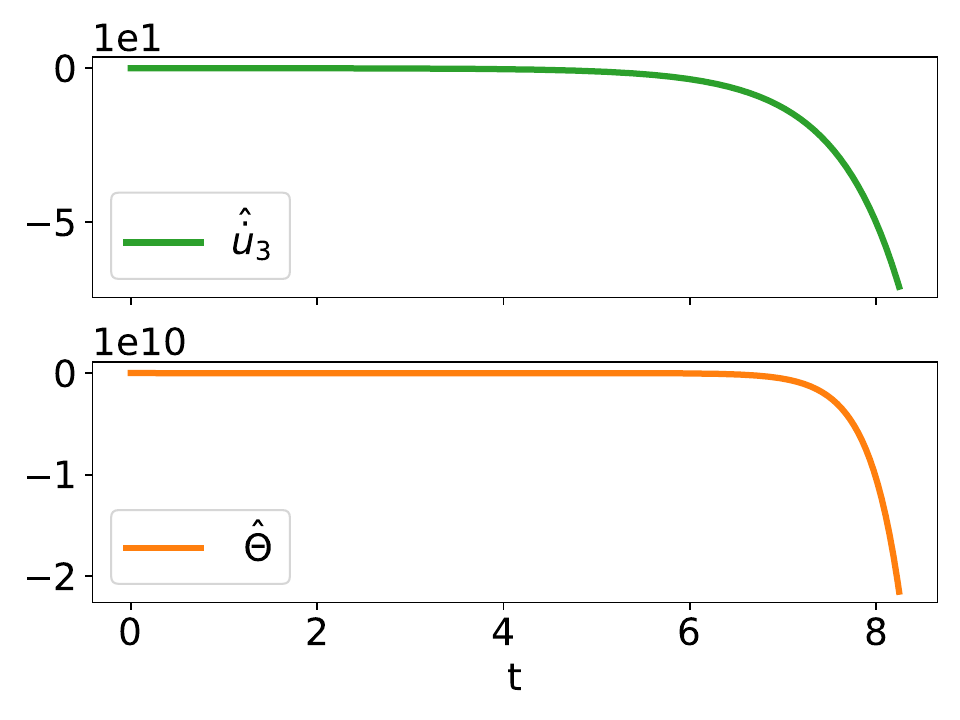}}&\\
        (e)
           \end{tabular}
  \par\end{centering}\caption{It is plotted the orbit with initial condition $B=1.80$, $\Omega_K=1.0e^{-1}$, $\Omega_m=1.0e^{-1}$, $\Phi_3=1.0e^{-1}$, $\Phi_{3,0}=1.0e^{-1}$, $\Phi_{3,1}=-2.36$ $Q_1=-3.10$, $Q_2=8.16$, $\Sigma_+=-5.16e^{-1}$, $\Sigma_{+1}=1.14$, $\Sigma_{+2}=8.58$, $\Sigma_-=8.09e^{-1}$, $\Sigma_{-1}=-2.82$, $\Sigma_{-2}=8.08$, $\eta=1.0e^{-1}$, and $r=1.0e^{-1}$ near to Kasner solution with $\phi=2$ for radiation fluid, $w=1/3$, $\beta=2$, and $\chi=1.2$. Here the evolution is to the past according to QG, and this solution is attracted to isotropic singularity. a) The matter density approaches zero is plotted in blue. In the inset, it is shown, in red, the increase of the tilted variable $r$. b) The graph shows in blue that the direction of the tilt, $\eta$, approaches zero. In the inset, it is shown, in red, a zoom of the plot $\eta$. c) The graph of the vorticity approaching zero is shown in blue. d) It is plotted the divergence of the matter acceleration components $\hat{\dot{u}}_0$ and $\hat{\dot{u}}_1$ in blue and red, respectively. e) The remaining non-null matter acceleration component $\hat{\dot{u}}_3$ evolves without a divergence and is plotted in green. It is plotted in orange the divergence of the matter contraction $\hat{\Theta}$ at the singularity}\label{f12}
\end{figure*} 

\section{Numerical results}\label{sec4}

The rest of the article describes the numerical results for initial conditions near to exact solutions mentioned in the above section. This present work is going to investigate only time evolution to the past, toward the singularity. As already mentioned, the dynamical system is defined in Section \ref{subsec21} for GR and in the Appendix \ref{app}, and by \eqref{deromegak}, \eqref{senv} for QG, see Section \ref{subsec22}. The initial condition must satisfy the constraints; for GR, see Section \ref{subsec21}, and for QG, see Section \ref{subsec22} and Appendix \ref{app}. Since the constraints are maintained in time, they are used to check the numerical results.

\subsection{Einstein-Hilbert gravity}\label{subsec41}

Numerical evolution to the past shows that according to GR, initial conditions near all exact solutions analyzed hitherto, namely FLRW, Milne, and Kasner do approach Kasner solution with $\Omega_m \rightarrow 0$ if $w<1$. This Kasner attractor falls in two cases either with $\eta\rightarrow \pi$ when $\phi>\pi/2$, or either with $\eta \rightarrow 0$ when $\phi<\pi/2$. As mentioned in Subsection \ref{subsec21}, the vorticity vector $\hat{\omega}_2$ in \eqref{cininc} tends to zero when tilt assumes the values $r=0$ and/or $\eta=0$ or $\eta=\pi$. Such that Kasner is a past attractor with zero vorticity. When $w=1$ for stiff fluid, initial conditions near all exact solutions mentioned above do approach Kasner-like orbits with exact solution \eqref{kw1} and \eqref{kw11} for GR. The tilt variable $r$ decreases and approaches zero. The matter density $\Omega_m$, the shear components $\Sigma_+$, and $\Sigma_-$ approach constant values, which must satisfy Einstein $00$ field equation \eqref{kw11}. This Kasner-like orbit falls in two cases: with $\eta\rightarrow \pi$ when $\Sigma_+<0$, or with $\eta \rightarrow 0$ when $\Sigma_+>0$. Thus, this orbit is also a past attractor with zero vorticity. In addition, the matter acceleration evolves without divergence.

Vorticity $|\hat{\omega}^2|$ increases together with $r$ in absolute value, and then $|\hat{\omega}^2|\rightarrow 0$ decreases to zero asymptotically while tilt $r$ increases together with the matter acceleration $\hat{\dot{u}}^a$. Radiation fluid also shows this same behavior for vorticity and matter acceleration, despite the tilt variable $r$ decreasing. According to \eqref{cininc}, this behavior occurs for several initial conditions, even when the tilt $r$ decreases as long as $H$ increases sufficiently. Also, this matter substance expansion increases and then diverges towards negative values, which show an infinite contraction of the Universe as the singularity is approached. The numerical experiments confirm the boundless increase of curvature invariants, as expected.

It must be mentioned that although asymptotically the solution approaches Kasner solution with $R_{ab}=0$, it is not exact Kasner such that the scalar $R_{ab}R^{ab}$ has a denominator, $B^2\rightarrow 0$, Eq. \eqref{ec}, that goes to zero while at the same time the numerator is not exactly zero. That is the reason that this curvature scalar also diverges numerically toward the singularity. 

It should be noted that, according to Eq. \eqref{decpar}, the deceleration parameter evolution to the past approaches $q\rightarrow2$ for the Kasner attractor toward the singularity.

The dynamical system is defined in Subsection \ref{subsec21} for GR. The numerical solutions are analyzed for initial conditions near to exact solutions of FLRW, Milne, and Kasner for stiff matter, dust, and radiation fluid. In FLRW case, the matter energy-momentum source is assumed tilted, while for vacuum solutions, Milne and Kasner, a small amount of tilted source is considered. As mentioned in Subsect. \ref{subsec21}, the tilt is described by the energy-momentum tensor \eqref{tem}, where the metric is \eqref{mett} and the time-like vector is \eqref{timeve}, with normalization $\hat{u}^a\hat{u}^bg_{ab}=-1$. Also, in GR, the initial conditions for $\Omega_m$, $\Phi_3$, and $\Sigma_+$ are given by the constraints $E_{00}$, $E_{03}$, and $E_{01}$ in Eq. \eqref{rgvinc2}, respectively. The constraints are maintained in time and used to check the numerical results. For GR, it is verified that the constraints fluctuations are always smaller than $10^{-11}$. 

\subsubsection{Stiff fluid}

Initial conditions that are near FLRW, Milne, and Kasner solutions for stiff matter $p=w\rho$, with the EoS parameter $w=1$ are past attracted to the Kasner-like orbit with exact solution \eqref{kw1} and \eqref{kw11} for GR. The matter density, the shear components $\Sigma_+$, and $\Sigma_-$ approach constant values, which satisfy the Einstein ${00}$ field equation \eqref{kw11}. As previously mentioned, this Kasner-like orbit falls in two cases: with $\eta\rightarrow \pi$ when $\Sigma_+<0$, or with $\eta \rightarrow 0$ when $\Sigma_+>0$. Thus, this orbit is also a past attractor with zero vorticity. Furthermore, the curvature invariants and the matter contraction diverge toward the singularity, showing both the presence of a curvature singularity and a divergent contracting Universe $\hat{\Theta}\rightarrow -\infty$. However, the matter acceleration evolves without divergence.

Figure \ref{f1} depicts an orbit with initial condition near Milne's exact solution with a small amount of tilted source for stiff matter. Here, the numeric time evolution is for the past. The solution is attracted to the Kasner-like orbit with solution \eqref{kw1} and \eqref{kw11}. In Figure \ref{f1}a, it is shown in blue that the matter density approaches a constant. In Figure \ref{f1}b, it is shown in blue that the tilt variable $r$ approaches zero. In the inset of Figure \ref{f1}b, a zoom of the variable $r$ is shown in red in the plot. Also, in red in the inset of Figure \ref{f1}a, it is shown the divergence of the curvature scalar $R_{ab}R^{ab}$. It can be seen in blue in Figure \ref{f1}c that the vorticity decreases to zero toward the singularity. Also, the inset of Figure \ref{f1}c shows the numerical check for the constraint $E_{00}$ in Eq. \eqref{kw11} with fluctuations smaller than $10^{-14}$. Figure \ref{f1}d shows the evolution and approaches to constants of the diagonal shear components $\Sigma_+$ and $\Sigma_-$ plotted, respectively, in blue and red.

Even orbits which are out of the linear stability range are attracted to these Kasner-like orbits \eqref{kw1} and \eqref{kw11}, indicating that this is a non linear attractor. 

\subsubsection{Dust fluid}

As mentioned in Subsect. \ref{subsec21}, for dust Universe, the EoS parameter is $w=0$ and the matter acceleration, shown in \eqref{cininc}, is zero. 

Past evolution for initial conditions near FLRW, Milne, and Kasner solutions for dust fluid is addressed. 

In Figure \ref{f2}, it is shown the past numerical evolution of an initial condition near FLRW solution for a source with $\Omega_m=8.62e^{-1}$ and tilt $r=1.0e^{-1}$ and tilt direction $\eta=1.0e^{-1}$ for dust fluid with the EoS parameter $w=0$. This solution is attracted to a Kasner orbit with $\phi=1.4987$; see Figure \ref{f2}a, where the graph of the shear components is plotted. In blue, it is shown the non-diagonal shear component $\Phi_3$. It is plotted in red the diagonal shear component $\Sigma_+$ and in green the component $\Sigma_-$. 

As mentioned at the beginning of Subsect. \ref{subsec41}, the Kasner attractor falls in the case with $\eta \rightarrow 0$ when $\phi<\pi/2$. However, plotted in blue in Figure \ref{f2}b is an increase in $\eta$ before it approaches zero when the Kasner orbit is attracted. In the inset of Figure \ref{f2}b, it is shown in red a zoom of the $\eta$ plot.

Numerical results for orbits near Milne and Kasner solutions show that the tilt variable $r$ increases. Nevertheless, for initial conditions near to FLRW solution, it is shown plotted in blue in Figure \ref{f2}c, at the beginning $r$ increases and decreases and then increases again as the orbit approaches the Kasner attractor. 

Also, as already mentioned in Sect. \ref{subsec21}, the vorticity tends to zero when the direction of the tilt assumes the value $\eta=0$. However, vorticity $|\hat{\omega}^2|$ increases with the tilt variable $r$ in absolute value, and then $|\hat{\omega}^2|\rightarrow 0$ decreases to zero asymptotically. The numerical past evolution of the vorticity is plotted in blue in Figure \ref{f2}d.

In the inset of Figure \ref{f2}c, in red, it shows the matter contraction toward singularity. Also, the product $R_{abcd}R^{abcd}$ increases and then diverges, showing the presence of the curvature singularity. This numerical evolution is shown in the inset, in red, of Figure \ref{f2}d.

\subsubsection{Radiation fluid}

Here we address the last case analyzed for GR, orbits to the past near FLRW, Milne, and Kasner solutions for radiation fluid with EoS parameter $w=1/3$.

The numerical results show that an initial condition near to FLRW solution with initial conditions shown in Figure \ref{f3} is past attracted to the Kasner orbit with $\phi=1.6047$, such that as $\phi>\pi/2$ the direction of the tilt $\eta \rightarrow \pi$. The eigenvalue found in Subsect. \ref{subsec31} for the tilt $r$ is small in absolute value and negative, $-6.78\times 10^{-2}$. However, as it is small in modulo, numerically, $r$ decreases toward singularity. In Figure \ref{f3}a it is plotted in blue the tilt variable $r$. In the inset of Figure \ref{f3}a shows in red a zoom of the plot $r$. Vorticity increases in absolute value before decreasing towards the singularity, as plotted in blue in Figure \ref{f3}b. The direction of the tilt $\eta$ is plotted in red in the inset of Figure \ref{f3}b.

A near Kasner orbit for a small tilted source with $\phi=0.4$ and initial conditions shown in Figure \ref{f4}a approaches to the past the Kasner attractor with $\phi=1.0930$. In this case, $\phi<\pi/2$ and the tilt direction $\eta \rightarrow 0$. Also, the eigenvalue found in Subsect. \ref{subsec31} for the tilt $r$ is positive. In the inset of Figure \ref{f4}a, it is plotted in red the decrease of $r$ to the past in accordance with the sign of the eigenvalue.

An initial condition near Milne's exact solution is past attracted to a Kasner orbit with $\phi=1.4927$, such that $\phi<\pi/2$ with tilt direction $\eta \rightarrow 0$. The eigenvalue for $r$ is positive, thus the tilt variable $r$ decreases for evolution toward the past. In the inset, in red, in Figure \ref{f4}b, it is plotted the tilt $r$. 

Figs. \ref{f4}a and \ref{f4}b show that for orbits near Kasner and Milne's solutions, respectively, the vorticity increases in absolute value before decreasing toward the singularity.

We observe that for all orbits analyzed for the radiation fluid, there is an increase, followed by a divergence in the matter acceleration. The matter acceleration diverges toward the singularity as shown in Figs. \ref{f5}a and \ref{f5}b for initial conditions near to Kasner solution. Also, the matter contraction $\hat{\Theta}$, \eqref{cininc}, is displayed in the lower graph shown in Figure \ref{f5}b.

Fig. \ref{f5}c shows the deceleration parameter $q$ for initial conditions near to Kasner solution for radiation fluid. According to Eq. \eqref{decpar}, the deceleration parameter approaches $q=2$ as the orbit is past attracted to the Kasner solution.

\subsection{Quadratic gravity} \label{subsec42}

Numerical time evolution to the past for QG shows that orbits with initial conditions near asymptotic isotropic singularity solution and Milne and Kasner exact solutions do approach Kasner with $\Omega_m \rightarrow 0$ and $w\leq 1$ for larger value of the initial $\Omega_m$. In both cases, a small amount of tilt is considered.

We must mention here that, as in GR, this Kasner attractor falls in two cases: $\eta\rightarrow\pi$ when $\phi>\pi/2$, or $\eta\rightarrow 0$ when $\phi<\pi/2$. Again, the Kasner orbit is a past attractor with zero vorticity and deceleration parameter approaching $q=2$. As mentioned in Subsection \ref{subsec41}, the tilt $r$ increases together with the matter acceleration $\hat{\dot{u}}^a$. Vorticity $|\hat{\omega}^2|$ increases in absolute value with $r$, and then $|\hat{\omega}^2|\rightarrow 0$ decreases to zero asymptotically. However, as in GR, for radiation, this is true even though $r$ decreases towards zero. 

As an example, Figure \ref{f6} plots the orbit with initial condition near isotropic singularity for dust fluid, but with $\Omega_m=2$. This solution is past attracted to the Kasner orbit with $\phi=2$. It is plotted in Figure \ref{f6}a, in blue, the matter density approaches zero. In the inset of Figure \ref{f6}a, it is shown, in red, the numerical check for the constraint $E_{00}$ with small fluctuation on $10^{-9}$. Figure \ref{f6}b shows plotted in blue, and in the inset, in red, the ENVs $Q_1$ and $Q_2$ tending to the values $Q_1=-3$ and $Q_2=9$, which are the Kasner values for these variables. The tilt variable $r$ increases toward the singularity, as plotted in blue in Figure \ref{f6}c. However, as expected, for $\phi>\pi/2$, the direction of tilt, $\eta$, approaches $\pi$. In the inset of Figure \ref{f6}c, it is shown, in red, the plot of $\eta$. According to Eq. \eqref{cininc}, in the graph shown in Figure \ref{f6}d, the vorticity increases in absolute value with $r$ and then approaches zero as a consequence of $\eta\rightarrow\pi$ at the singularity. In Figure \ref{f6}e, the deceleration parameter $q$ is plotted in blue, tending to the value $q=2$.

In the case of $w=1$ for stiff matter, initial conditions near all solutions mentioned above approach Kasner. The matter density $\Omega_m$ and the shear components $\Sigma_+$ and $\Sigma_-$ approach a constant, as in GR. However, as previously mentioned in Subsect. \ref{subsec22}, in QG, the matter decouples from dynamics when $B\rightarrow 0$. Such that the Kasner orbit satisfies the quadratic gravity $E_{00}$ field equation \eqref{k}. This is a difference between GR and QG; in GR, it can be seen, from the dynamic system shown in Subsect. \ref{subsec21}, that matter is not decoupled from the dynamics towards past evolution, and for stiff matter, the Einstein $00$ field equation satisfies \eqref{kw11}. 

The dynamic system to QG is defined in the Appendix \ref{app}, and by \eqref{deromegak}, \eqref{senv}, see Sect. \ref{subsec22}. The initial condition must satisfy the constraints, see Appendix \ref{app}. In QG, the initial conditions for $Q_2$, $\Sigma_{+2}$, and $\Phi_{3,1}$ are determined by the constraints $E_{00}$, $E_{01}$, and $E_{03}$, respectively. Since the constraints are maintained in time, they are used to check the numerical results. It is verified that constraints numerical fluctuations are always smaller than $10^{-9}.$ 

Solutions near the asymptotic isotropic singularity solution, Milne and Kasner exact solutions for the stiff matter, dust, and radiation fluid; are past attracted to the isotropic singularity. Also, for these vacuum solutions, a small amount of a tilted source is considered. The isotropic singularity orbit, as a past attractor, may have an increase in the vorticity. Note that, as $Q_1=-2$ and from Eq. \eqref{decpar}, the isotropic singularity past attractor has the deceleration parameter $q\rightarrow1$.

\subsubsection{Stiff fluid}

Numerical time evolution to the past for orbits with initial conditions near asymptotic isotropic singularity solution, Milne and Kasner exact solutions with a small amount of source tilted for stiff matter $p=w\rho$, with EoS parameter $w=1$, is attracted to the isotropic singularity orbit. The matter density, $\Omega_m$, increases. Oppositely, the vorticity approaches zero together with the tilt variable $r$; see \eqref{cininc}. The matter contraction diverges, showing a Universe contraction. The matter acceleration evolves without a divergence for initial conditions near isotropic singularity and Milne solutions. However, there is an increase in the matter acceleration for initial condition near Kasner exact solution. Also, the curvature scalars $R_{abcd}R^{abcd}$ and $R_{ab}R^{ab}$ increase and then diverge at the singularity.

Figure \ref{f7} shows the past evolution of the orbit with an initial condition near Milne's solution for stiff fluid, $w=1$. In Figure \ref{f7}a, the ENVs $Q_1$ and $Q_2$ are plotted in blue and red, respectively, with $Q_2$ plotted in the inbox. Their approach the values $Q_1=-2$ and $Q_2=-4$, which correspond to the asymptotic isotropic singularity solution; see Subsect. \ref{subsec32}. Figure\ref{f7}b shows that matter density increases as expected from the eigenvalues of the isotropic singularity shown in Subsect. \ref{subsec32}. In the inset of Figure \ref{f7}b, it is shown, in red, that the product $B\Omega_m$ tends to zero. The matter enters the dynamic only with this product; see the Appendix \ref{app}. Also, this product tends to zero as $t\rightarrow-\infty$ because $B\rightarrow0$, thus all the other variables are past attracted to the isotropic singularity. Figure\ref{f7}c shows the decrease of the tilt variable $r$ tending to zero. In the inset, it is plotted in red, a zoom of the plot $r$. Also, it agrees with the eigenvalues shown in Subsect. \ref{subsec32}. In Figure \ref{f7}d, it can be seen that the vorticity decreases as a consequence of the $r$ decreases. In the inset of Figure \ref{f7}d, it is shown in red the matter contraction. In the inset of Figure \ref{f7}e, it is shown in red the $R_{ab}R^{ab}$ increase, followed by a divergence. The matter acceleration evolves without diverging; see Figure \ref{f7}e for the plot in blue of the matter acceleration component $\hat{\dot{u}}_0$. In Figure \ref{f7}f, it is shown, in blue, that the deceleration parameter approaches the value $q=1$, which corresponds to Eq. \eqref{decpar} for the solution attracted to the isotropic singularity.

\subsubsection{Dust fluid}

Also, numerical time evolution to the past for the orbit with initial condition near all solutions considered here with a small amount of source tilt for dust fluid, $w=0$, is attracted to the isotropic singularity orbit. The matter density tends to zero while the tilt $r$ increases. This numerical result can be seen in Figs. \ref{f8}a and \ref{f9}a for the orbit with initial condition near to asymptotic isotropic singularity solution and Kasner exact solution, respectively. The matter density is plotted in blue, and in the inset, in red, the tilt $r$. The matter contraction diverges. The matter acceleration is zero for dust fluid, \eqref{cininc}. The products $R_{abcd}R^{abcd}$ and $R_{ab}R^{ab}$ increase and then diverge at the singularity.

Figure \ref{f8} shows the past evolution toward the singularity for the orbit with initial condition near to asymptotic isotropic singularity solution for dust fluid, $w=0$. It is plotted, in red, in Figure \ref{f8}b the direction of the tilt $\eta$. It is plotted in blue in Figure \ref{f8}b, the vorticity increases in absolute value.

Numerical time evolution to the past for an initial condition near Milne's solution for dust fluid shows the direction of the tilt $\eta$ evolves to values different from $0$ and $\pi$, and the vorticity increases in absolute value.

Figure \ref{f9} shows the past evolution towards the singularity for an initial condition near to Kasner solution for the dust fluid, $w=0$. The direction of the tilt $\eta$ tends to zero, plotted in blue in Figure \ref{f9}b. In the inset of Figure \ref{f9}b, it is shown in red a zoom of the plot $\eta$. In Figure \ref{f9}c, the increase in vorticity in absolute value together with $r$ and then approaches zero when $\eta \rightarrow 0$ is plotted in blue. In the inset of Figure \ref{f9}c, it is shown in red the numerical check for the constraint $E_{00}$ with fluctuations smaller than $10^{-11}$.

\subsubsection{Radiation fluid}

The numerical time evolution shows that orbits with initial conditions near all solutions mentioned in Subsection. \ref{subsec32} for radiation fluid $w=1/3$ is also past attracted to the orbit of the isotropic singularity. The scalar products $R_{abcd}R^{abcd}$ and $R_{ab}R^{ab}$ increase and then diverge at the singularity.

Figure \ref{f10} plots the numerical past evolution for an initial condition near asymptotic isotropic singularity solution for radiation fluid, $w=1/3$. This solution is attracted to the isotropic singularity orbit. It is plotted in blue in Figure \ref{f10}a the matter density evolution and its approaches to a constant value. In the inset of Figure \ref{f10}a, it is shown in red the numerical check for the constraint $E_{00}$ with fluctuation smaller than $10^{-10}$. In Figure \ref{f10}b, it is shown in blue that the tilt variable $r$ decreases toward the singularity. In the inset of Figure \ref{f10}b, it is shown in red a zoom of the plot $r$. It is plotted in the inset of Figure \ref{f10}c in red the numerical evolution of the direction of the tilt, $\eta$. The vorticity increase in absolute values is plotted in blue in Figure \ref{f10}c. The matter acceleration non-null components $\hat{\dot{u}}_0$, $\hat{\dot{u}}_1$, and $\hat{\dot{u}}_3$ increase followed by a divergence are plotted in blue, red, and green, respectively, in Figs. \ref{f10}d and \ref{f10}e. The matter contraction is plotted in orange in Figure \ref{f10}e.

It is plotted in Figure \ref{f11} the numerical time evolution for the past to the orbit with an initial condition near Milne's solution for radiation fluid. This solution is also attracted to the isotropic singularity orbit. Figure \ref{f11}a shows in blue the graph of the matter density, which tends to a constant value. The direction of the tilt, $\eta$, is shown in blue in Figure \ref{f11}b. In the inset of Figure \ref{f11}b, it is shown in red a zoom of the plot of $\eta$. The tilt variable $r$ increases is shown plotted in red in the inset of Figure \ref{f11}c. It is shown in blue in Figure \ref{f11}c the increase of the vorticity in absolute values. The matter acceleration non-null components $\hat{\dot{u}}_0$, $\hat{\dot{u}}_1$, and $\hat{\dot{u}}_3$ increase are plotted in blue, red, and green, respectively, in Figs. \ref{f11}d and \ref{f11}e. The matter contraction, showing a Universe contraction, is plotted in orange in Figure \ref{f11}e.

Figure \ref{f12} shows the numerical time evolution toward the past to the orbit with an initial condition near the Kasner exact solution for radiation fluid, $w=1/3$ and $\phi=2$. Again, this solution is attracted to the isotropic singularity orbit. Figure \ref{f12}a shows in blue the matter density decrease. In the inset of Figure \ref{f12}a, it is plotted in red the increase in the tilted variable $r$. It is plotted in blue in Figure \ref{f12}b, the direction of the tilt $\eta$ approaching zero. In the inset, it is shown in red a zoom of the plot $\eta$. The vorticity tending to zero together with $\eta \rightarrow 0$ is plotted in blue in Figure \ref{f12}c. The non-null matter acceleration components $\hat{\dot{u}}_0$ and $\hat{\dot{u}}_1$ increases followed by a divergence are plotted in Figure \ref{f12}d in blue and in red, respectively. It is plotted in green in Figure \ref{f12}e, where the remaining non-null matter acceleration component $\hat{\dot{u}}_3$ evolves without a divergence. The divergence of the matter contraction at the singularity is shown in orange in Figure \ref{f12}e.

\section{Conclusions}\label{summary}

In this work, we investigate the behavior of the kinematic variables for a tilted source, which is a non-perfect fluid with energy and momentum flux \cite{King:1972td}, for Bianchi V cosmological model. If the tilt is non-zero, this source substance has accelerated non-geodesic motion with vorticity and a different expansion than the geometric one \cite{King:1972td,ellisking,Coley_2006,Lim_2006,Coley:2008zz,Coley_2009}. However, a different type of singularity may occur, in which the kinematic variables diverge for tilted Bianchi models \cite{Coley_2006,Lim_2006,Coley:2008zz,Coley_2009}. On the other hand, when tilt is zero, the source is the usual perfect fluid present in any GR textbook, with the fluid following geodesic, vorticity-free world lines \cite{Weinberg:1972kfs,Stephanibook}. The dynamics for time evolution to the past for the Einstein-Hilbert GR and quadratic gravity QG are numerically obtained. Backward evolution was chosen because we wanted to understand the singularity. 

The tilted source is described by the energy-momentum tensor \eqref{tem}, where the metric is \eqref{mett} and the time-like vector $\hat{u}^a$ is \eqref{timeve}, with normalization $\hat{u}^a\hat{u}^bg_{ab}=-1$. The connection is defined by \eqref{Gamma_000}, \eqref{Gamma_0ab} with \eqref{sig1}, \eqref{Gamma_ijk} and \eqref{deromegak}. We started with the full dynamical system for GR as defined by \eqref{deromegak}, \eqref{GRv1}-\eqref{merg}, while the dynamical system for QG is defined in the Appendix \ref{app} and by \eqref{deromegak}, \eqref{senv}. The constraints shown in \eqref{rgvinc2} and in the Appendix \ref{app}, respectively, for GR and QG, must be satisfied by the initial conditions. Once these constraints are initially satisfied, they must be maintained during time evolution and are used in this work as a numeric check. It was verified that all the constraints for GR show numerical fluctuations, always smaller than $10^{-11}$, while for QG they fluctuate numerically at $|E_{00}|,\,|E_{01}|,\,|E_{03}|<10^{-9}$.

According to linear stability analysis shown in Subsection \ref{subsec31} for the asymptotic fixed point when $t \rightarrow \infty$, the FLRW orbit is a future attractor for EoS parameter $-1<w<-1/3$. While for $-1/3<w<1/3$, Milne's orbit is a future attractor. In addition, numerical time evolution to the future for solutions with initial conditions near FLRW, Milne, and Kasner for $1/3< w\leq1$ are attracted to Milne's orbit, and the tilt increases, which agrees with Coley, Hervik, and Lim \cite{Coley:2008zz} as well as Krishnan, Mondol, and Sheikh-Jabbari \cite{Krishnan:2022qbv,Krishnan:2022uar}. 

Our numerical experiments confirm that time evolution to the past in GR for tilted Bianchi V orbits with initial conditions near FLRW, Milne, and Kasner shows that solutions are attracted to Kasner orbit with $\Omega_m \rightarrow 0$ and $w<1$, according to Figures \ref{f2}-\ref{f5}. For GR, the tilt variable $r$ increases or decreases for orbits when the eigenvalue $2\cos{\phi}+3\,w-1$ is negative or positive, respectively. This Kasner attractor falls into two cases with $\eta\rightarrow\pi$ when $\phi>\pi/2$; otherwise, when $\phi<\pi/2$, $\eta\rightarrow 0$. As mentioned in Subsect. \ref{subsec21}, the vorticity component $\omega_2$ in \eqref{cininc} tends to zero when the tilt angle assumes the values $\eta=0$ or $\eta=\pi$. For this reason, according to GR Kasner is a past attractor with zero vorticity. However, the vorticity together with the tilt $r$ show an increase and then an asymptotic decrease to zero at the singularity. Radiation fluid also shows this same behavior for the vorticity despite the tilt variable $r$ decreasing, as shown in Figure \ref{f4}. As expected, the curvature scalars do diverge for backward evolution.

The matter expansion assumes ever decreasing negative values $\hat{\Theta}\rightarrow -\infty$, showing an infinite contraction of the Universe as the geometrical singularity is approached, which is expected. Furthermore, in GR, for initial conditions near Kasner exact solution when $w=1/3$, the matter acceleration increases and then diverges, as shown in Figure \ref{f5}. According to \eqref{cininc}, this behavior occurs for several initial conditions, even when the tilt $r$ decreases as long as $H$ increases sufficiently.

Backward evolution shows that, according to GR, initial conditions near all the solutions already mentioned in Subsect. \ref{subsec21} for stiff matter EoS parameter $w=1$ approach Kasner-like orbits with exact solution \eqref{kw1} and \eqref{kw11}. As an example, it is plotted in Figure \ref{f1} the orbit with an initial condition near Milne's exact solution for stiff matter. Figure \ref{f1} shows that the matter density $\Omega_m$ and the diagonal shear components $\Sigma_+$ and $\Sigma_-$ approach constant values, while the tilt variable $r$, together with the vorticity, all approach zero and the matter acceleration evolves without a divergence. Even orbits which are out of the linear stability range are attracted to these Kasner-like orbits \eqref{kw1} and \eqref{kw11}, so that this is a non linear attractor. 

For QG, it is already known that there are two attractors to the past, Kasner attractor and false radiation attractor, also called isotropic singularity attractor. However, there is a difference between GR and QG with stiff matter substance with EoS $w=1$. While in GR the Kasner-like exact solution with stiff matter $w=1$ is a past attractor, in QG this solution does not exist, even asymptotically. This happens because $\Omega_m$ only appears in the specific combination $B\Omega_m$, which goes to zero at the physical singularity $B\rightarrow 0$ when $t\rightarrow-\infty$.

In QG, the linear stability for the isotropic singularity attractor to the past shows that for the EoS parameter $w<1/3$, the variable $\Omega_m$ decreases and tends to zero toward the singularity while the variable $r$ increases. Nevertheless, the opposite situation occurs for $w>1/3$: $\Omega_m$ increases while the tilt variable $r$ approaches zero. Nevertheless, according to the above-mentioned decoupling, the increase in $\Omega_m$ as compared to the decrease in $B$ is insufficient to modify the dynamics at the singularity. The isotropic singularity is maintained.

According to Eq. \eqref{decpar}, it is also pointed out that orbits past-attracted to Kasner and to isotropic singularity have deceleration parameters approaching $q=2$ and $q=1$, respectively. This is shown in Figures \ref{f5}-\ref{f7} for GR and QG.

According to QG and the set of assumptions herein, initially smaller $\Omega_m$ solutions fall into the false radiation attractor, while initially larger values of $\Omega_m$ solutions fall into the Kasner attractor. Kasner past attractor shows a similar behavior as in GR, which allows kinematic singularity with zero vorticity, together with geometric singularity as in Figure \ref{f6}. While the isotropic singularity attractor allows a kinematic singularity with the divergence of all kinematic variables, tilt, acceleration, and vorticity, together with the geometric singularity, as shown in Figures \ref{f7}-\ref{f12}.

We end the article by remarking that for all initial conditions and configurations analyzed in GR, for a Universe with a singularity with initially any small amount of vorticity, this vorticity will always grow. We were not able to find a single Universe in which this infinitesimally small vorticity at the singularity decreases. While in QG, we found universes with infinite vorticity at the singularity, and this vorticity decreases to finite values as a consequence of time evolution only. The decrease of the vorticity in QG occurs for the false radiation singularity, a solution which does not exist in GR. As is well known, tilted substance can also present non zero accelerations, i.e., non geodesic motion. Both GR and QG allow infinite accelerations at the curvature singularity, which decrease to finite values to the future due to time evolution only.

According to all conditions supposed in this article, in this sense QG is a stronger theory than GR because in QG, an initial curvature singularity together with a divergence in acceleration, tilt, matter expansion, and vorticity subject only to time evolution is sufficient to obtain a perfect fluid source to the future. While in GR, any small amount of initial vorticity at the singularity will increase. 

This work addresses the time evolution to the past so that radiation is not decoupled and CMBR does not exists in this regime, especially as the singularity is approached. The influence of cosmological tilt implications on the CMBR occurs for future expansion and is not taken into account in this work. 

Anisotropic models and CMBR is analyzed for instance by the Maartens Ellis Stoeger famous article \cite{Maartens_1995} which is a generalization of the Ehlers Geren Sachs theorem \cite{Ehlers1968}.

\begin{acknowledgement}

\section*{Acknowledgements}

W. P. F. de Medeiros gratefully acknowledges the financial assistance provided by the Brazilian agency \textit{Coordena\c c\~ao de Aperfeiçoamento de Pessoal de N\'ivel Superior} (CAPES) project number $88887.803891/2023-00$ for their financial support. D. A. Sales acknowledges the \textit{Conselho Nacional de Desenvolvimento Cient\'ifico e Tecnol\'ogico} (CNPq) and the \textit{Funda\c c\~ao de Amparo \`a Pesquisa do Estado do Rio Grande do Sul} (FAPERGS). 
\end{acknowledgement}

\appendix

\section{Appendix}\label{appa}
The key point in the procedure is a change of base which we describe in this Appendix. 

We begin by discussing the structure constants shown in \eqref{sc1}. Of course they must satisfy the Jacobi identity
\begin{equation}
{C^{m}_{il}C^{l}_{jk}+C^{m}_{jl}C^{l}_{ki}+C^{m}_{kl}C^{l}_{ij}=0}. \label{cjab}
\end{equation}
The structure constants are written in their symmetric and anti symmetric parts as $\frac{1}{2}\epsilon ^{ijl}C^{k}_{ij}=N^{kl}$ and $N^{kl}=n^{kl}+\epsilon ^{kli}a_i$  with $n_{ij}=n_{ji}$
        \begin{equation}
        {C^{k}_{ij}=\epsilon _{lij}n^{kl}+\delta ^{k}_ja_i-\delta ^{k}_ia_j}. \label{cestrutura}
    \end{equation}
Considering \eqref{cjab} and the \eqref{cestrutura} the Jacobi identity is given by 
\[
n^{ij}a_i=0, 
\]
where the symmetric part $n^{ij}$ can always be diagonalized and orthonormalized. In the same way, the vector $a_i$ can be chosen as \cite{Stephanibook,Stephani:2003tm}
\begin{equation}
    {a_i=(b,0,0)}.\label{chosen}
\end{equation}

It can be shown that the following line element is the most general line element for a spatially homogeneous spacetime
\begin{equation}
 {ds^2=-\frac{dt^2}{H^2}+h_{ij}(t)\rho ^i\otimes \rho ^j},
\end{equation}
where $\rho^j$ is the appropriate base for the particular Bianchi type. The shift is set to zero by choosing the time vector $\partial_t$ orthogonal to the homogeneous surfaces. 
While the lapse function $1/N^2$ can always be set to $1$ by choosing the time coordinate. The spatial componetes $h_{ij}$ are arbitrary functions of time. 

To obtain the specific shear used in this work we start with the following spatial metric components
\begin{equation}
 {h_{ij}(t)=\begin{pmatrix}
e^{-2\upsilon}+e^{-2\zeta } &0 &e^{-\varphi -\upsilon} \\
0& e^{-2\gamma} & 0\\
 e^{-\varphi -\upsilon} & 0& e^{-2\varphi }
\end{pmatrix}},\label{hiii}
\end{equation}
where the time dependence is contained in the metric. Following \cite{Stephanibook,Stephani:2003tm}, the anisotropic Bianchi V 
model has the following base 
\begin{align}
&\rho^0=dt,\,\rho ^1=dx,\,\rho ^2=e^xdy,\,\rho ^3=e^xdz,\nonumber\\
 &\eta_0=\partial_t,\,\eta _1=\partial_x,\,\eta _2=e^{-x}\partial_y,\,\eta _3=e^{-x}\partial_z ,\label{eta}
\end{align}
with $\rho^a\eta_b=\delta^a_b$ and structure constants
\begin{align}
&C^2_{12}=C^3_{13}=1,&C^2_{21}=C^3_{31}=-1.
\end{align}
These structure constants correspond to $n^{ij}=0$ and $a_i=(1,0,0)$ in \eqref{cestrutura} and \eqref{chosen}, see for instance \cite{Stephanibook,Stephani:2003tm}.

Through a base transformation 
\begin{align}
    &\eta_a={M_a}^be_b
\end{align} with
\begin{align}
    {M_a}^b&=\begin{pmatrix}
    1&0&0&0\\
0&e^{-\zeta} &0  & e^{-\upsilon}\\
0&0 &e^{-\gamma}  &0  \\
0&0 &0  & e^{-\varphi } \\
\end{pmatrix}\label{mtra},
\end{align}
the base $\left\{\eta_a\right\}$ is transformed to a new base $\left\{e_a\right\}$. Here the matrix ${M_a}^b$ must be invertible and cover all the space, except for zero measure sets.

For the base transformation matrix \eqref{mtra} the dual $\left\{\omega^a\right\}$ and its new base $\left\{e_a\right\}$ are
\begin{align}
 &\omega^0=\rho^0,\,\omega ^1=e^{-\zeta}\rho ^1,\,\omega ^2=e^{-\gamma}\rho ^2,\,\omega ^3=e^{-\upsilon}\rho^1+e^{-\varphi }\rho ^3,\nonumber
\\
&e_0=\eta_0,\,e_1=e^{\zeta}\eta_1-e^{\zeta-\upsilon+\varphi }\eta_3,\,e_2=e^{\gamma}\eta_2,\,e_3=e^{\varphi }\eta_3,\label{basenew}
\end{align}
and $\omega^ae_b=\delta^a_b$. The new base is time-dependent, and the metric becomes a tetrad in the spatial part
\begin{equation}
 {ds^2=-\frac{dt^2}{H^2}+g_{ij}\omega ^i\otimes \omega ^j}
\end{equation}
\begin{equation}
 {g_{ij}=\begin{pmatrix}
1 & 0 &0 \\
0 &1 &0 \\
0 & 0 &1
\end{pmatrix}},
\end{equation}
as in \eqref{mett}.

Since the base is non-coordinate, the connection is determined by metricity and zero torsion given by \eqref{mtr_torc}
\begin{align*}
   &\nabla _cg_{ab}=0 \implies e_c^d\partial_d g_{ab}-\Gamma^d_{ac}g_{db}-\Gamma^d_{bc}g_{ad}=0,\nonumber\\
   &\nabla _ae_{b}-\nabla _be_{a}=[e _{a},e_{b}]\implies (\Gamma^d_{ba}-\Gamma^d_{ab})e_d=[e _{a},e_{b}], 
\end{align*}
\begin{align*}
\Gamma_{a\,bc}&=\frac{1}{2}\left(-e_a^f\partial_fg_{bc} +e_{b}^f\partial_fg_{ac}+e_c^f\partial_f g_{ab} \right. \nonumber\\ 
&\left. +D_{acb}-D_{bca}-D_{cba}\right).
\end{align*}

The commutation relations for the basis vectors $\left\{e_a\right\}$ is given by
\begin{equation}
    {[e _{a},e_{b}]=e_a^f\partial_fe_b^d-e_b^f\partial_fe^d_a=D^{c}_{ab}e_{c}^d},\label{comd}
\end{equation}
where $D_{cab}=-D_{cba}$, and usually the $d$ index is omitted in the last part. The pure spatial part the commutator results in the structure constants, which now must be time dependent 
\begin{equation}
    {[e _{j},e_{k}]=e_j^f\partial_fe_k^d-e_k^f\partial_f e_j^d=-C^{i}_{jk}e_{i}^d}.\label{comd1}
\end{equation}
According to the above, the spatial structure constants for this new base \eqref{basenew} become
\begin{align}
&C^2_{12}=C^3_{13}=e^{\zeta},&C^2_{21}=C^3_{31}=-e^{\zeta},\label{cee}
\end{align}
from this equation we can immediately read out $b(t)=e^{\zeta(t)}$ from \eqref{ce}.

The spatial derivatives of the metric \eqref{mett} are zero and do not contribute to the connection 
\[
\Gamma _{i\,jk}=\frac{1}{2}(C_{jki}+C_{kji}-C_{ikj}), 
\]
which according to the above structure constants results in
\begin{align*}
&\Gamma _{2\,12}=\Gamma _{3\,13}=e^\zeta, & \Gamma _{1\,22}=\Gamma _{1\,33}=-e^\zeta,
\end{align*}
while there's only one component which comes from metricity part of the connection $\partial_0g_{00}-2\Gamma_{0\,00}=0$, as in \eqref{Gamma_000}
\begin{equation}
    {\Gamma ^0_{00}=-{\dot{H}}/{H}}. \label{H00_apendice}
\end{equation}

With respect to the base \eqref{basenew}, the commutators $[e_0,e_i]$ 
\begin{align}
&D_{110}=-\dot{\zeta},\;\; D_{220}=-\dot{\gamma},\;\;
D_{330}=-\dot{\varphi},\nonumber\\ &D_{310}=-(\dot{\upsilon}-\dot{\varphi  })e^{ \zeta -\upsilon},\label{dddij}
\end{align} 
are non null and enter into the connections terms
\begin{equation}
    \Gamma_{0\,ij} =-\frac{1}{2}\left ( D_{ij0}+D_{ji0} \right )\label{coned},
\end{equation}
which gives the last components of the connection
\begin{align}
&\Gamma _{011}=\dot{\zeta},\;\;\Gamma _{022}=\dot{\gamma },\;\;\Gamma _{033}=\dot{\varphi },\nonumber\\
&\Gamma _{013}=\frac{1}{2}\left ( \dot{\upsilon }-\dot{\varphi  }\right )e^{\zeta-\upsilon}. \label{gammas}
\end{align}

Now the covariant derivative of the time-like vector $u_a=(-1/H,0,0,0)$ is obtained considering the connection given by \eqref{H00_apendice} and \eqref{gammas}, as was done in Section \ref{sec2} 
\begin{equation*}
    \nabla _au_b=e^c_a\partial_c u_b-\Gamma^c_{ba}u_c,
\end{equation*}
\begin{equation*}
    \nabla _au_b=\delta _{a0}\delta _{b0}\frac{\dot{H}}{H^2}-\Gamma ^c_{ab}u_c=\delta _{a0}\delta _{b0}\frac{\dot{H}}{H^2}+\frac{\Gamma ^0_{ab}}{H}.
\end{equation*}
From the scalar product 
\[
u^a\nabla_a u_b=u^b\left( \delta _{a0}\delta _{b0}\frac{\dot{H}}{H^2}+\frac{\Gamma ^0_{ab}}{H} \right)=\delta _{a0}\frac{\dot{H}}{H}+\Gamma ^0_{a0}=0,
\] 
using \eqref{H00_apendice}, is possible to see that the vector $u^a$ is geodesic, $u^a\nabla_au_b=0 $. While for the vorticity tensor $\omega_{ab}=\nabla_au_b-\nabla_bu_a$ the only directly non null possibility would be $\omega_{ij}=(\Gamma^0_{ji}-\Gamma^0_{ij})/H$ which is zero because of \eqref{coned}. So that the vector $u^a$ is geodesic and vorticity free.

For the spatial part, $\nabla _iu_j=\Gamma ^0_{ij}/H=-H\Gamma _{0ij}$ if $i \neq 0$ and $j \neq 0$, using \eqref{cinva} and \eqref{gammas} it turns into
\begin{align}
    &\nabla _iu_j=-H\begin{pmatrix}
\dot{\zeta} & 0 &(\dot{\upsilon}-\dot{\varphi  })e^{ \zeta -\upsilon}/2 \\ 
0 & \dot{\gamma } & 0\\ 
 (\dot{\upsilon}-\dot{\varphi  })e^{ \zeta -\upsilon}/2& 0 & \dot{\varphi}\label{tri1}
\end{pmatrix}.
\end{align}
We impose that the expansion is related to the lapse function by $\nabla_au^a=\Theta=3H$ so that $H$ coincides with the Hubble parameter, 
\begin{equation}
    {\dot{\zeta }+\dot{\gamma }+\dot{\varphi }=-3},
\end{equation}
$\zeta$, $\gamma$ and $\varphi$ are connected by this relation, while $H$ remains an arbitrary function of time. Reminding that $\nabla _iu_j=\sigma_{ij}+\frac{1}{3}\Theta\delta_{ij}$ the shear is obtained as 
\begin{align}
    &\sigma _{ij}=-H\begin{pmatrix}
\dot{\zeta} +1 & 0 &(\dot{\upsilon}-\dot{\varphi  })e^{ \zeta -\upsilon}/2 \\ 
0 & \dot{\gamma } +1 & 0\\ 
 (\dot{\upsilon}-\dot{\varphi  })e^{ \zeta -\upsilon}/2& 0 & \dot{\varphi}+1\label{shear_ap}
\end{pmatrix}.
\end{align}
The shear given by \eqref{shear_ap} and \eqref{sig1} must coincide, such that
\begin{align}
    & \dot{\zeta}=-1+2\sigma_+/H,\nonumber\\
    &\dot{\gamma }=-1-\frac{\left (\sigma _++\sqrt{3}\sigma_ -  \right )}{H},\nonumber\\
   & \dot{\varphi  }=-1-\frac{\left (\sigma _+-\sqrt{3}\sigma_ -  \right )}{H},\nonumber\\
   & (\dot{\upsilon} -\dot{\varphi })e^{\zeta -\upsilon }=-2\phi _3/H.\label{78}
\end{align}
In the following we will show that the isotropic model is contained in \eqref{shear_ap}. It is obtained either through these above relations \eqref{78} setting $\sigma_+=0$, $\sigma_-=0$ and $\phi_3=0$ or by setting the tensor \eqref{shear_ap} to zero which give \newpage
\begin{align}
    &\dot{\zeta}+1=0, \,\,\dot{\gamma}+1=0, \,\,\dot{\varphi}+1=0 \implies \zeta=\gamma=\varphi=-t,\nonumber\\
    &\mbox{and } \upsilon\rightarrow \infty \label{cond_istropica}
\end{align}
each differing just by a constant. Reminding that dynamical time $dt=d\tau H$, where $H=a^\prime/a$ with $a^\prime=da/d\tau$ so that 
\[
e^\zeta\propto e^\gamma\propto e^\varphi=\exp\left(-\int \frac{a^\prime}{a} d\tau\right)=\frac{1}{a}, 
\]
and substituting $e^{-2\zeta}\propto a^2$, $e^{-2\varphi}\propto a^2$ and $e^{-2\gamma}\propto a^2$ and also $\upsilon\rightarrow \infty$ as in \eqref{cond_istropica} into \eqref{hiii} and into \eqref{cee}, results in the isotropic metric of Section \ref{subsec21} with scale factor $a$ and structure constants with the function $b(t)=1/a$ as given by \eqref{cee}.

\section{Appendix}\label{app}
The dynamical system for quadratic gravity is defined as follows and by the ENV first-order differential equations \eqref{deromegak}, \eqref{senv}.

\begingroup
\let\clearpage\relax
\onecolumn
\endgroup
\begin{align*}
& y_1 = B(t), y_2 = \Omega_K(t), y_3 = \Omega_m(t), y_4 = \Phi_{3}(t), y_5 = \Phi_{3, 0}(t), y_6 = \Phi_{3, 1}(t), y_7 = Q_1(t), y_8 = Q_2(t), y_9 = \Sigma_{+}(t),\\&  y_{10} = \Sigma_{+1}(t), y_{11} = \Sigma_{+2}(t), y_{12} = \Sigma_{-}(t), y_{13} = \Sigma_{- 1}(t), y_{14}= \Sigma_{-2}(t), y_{15} = \eta(t), y_{16}=r(t).
\end{align*}
With the first-order differential equations:
\begin{align*}
 \dot{\Omega}_m=& \left\{ y_{{3}} \left[ 2\,\cos \left( y_{{15}} \right) \cosh \left( y_{{16}}\right) \sinh \left( y_{{16}} \right) \sqrt {3} \left( 1+w \right) \sqrt {y_{{2}}}+ \left(  \left( 3\,y_{{9}} \left( 1+w \right)  \left( \cos \left( y_{{15}} \right)  \right) ^{2}-2\,\sin \left( y_{{15}} \right) y_{{4}} \left( 1+w \right) \right.\right.\right.\right.\\& \left.\left.\left.\left. \cos \left( y_{{15}} \right) + \left( -2\,y_{{7}}-y_{{9}}+2 \right) w+2\,y_{{7}}-y_{{9}}+2 \right)  \left( \cosh \left( y_{{16}} \right)  \right) ^{2}-3\,y_{{9}} \left( 1+w \right)  \left( \cos \left( y_{{15}} \right)  \right) ^{2}+2\,\sin \left( y_{{15}} \right) y_{{4}} \left( 1+w \right) \right.\right.\right.\\& \left.\left.\left.\cos \left( y_{{15}} \right) + \left( 2\,y_{{7}}+y_{{9}}+1 \right) w+1+y_{{9}} \right) \sqrt {3}-3\,y_{{12}} \left( \cosh \left( y_{{16}} \right) -1 \right)  \left( \cosh \left( y_{{16}} \right) +1 \right)  \left( \cos \left( y_{{15}} \right) -1 \right)  \left( \cos \left( y_{{15}} \right) +1 \right) \right.\right.\\& \left.\left. \left( 1+w \right)  \right] \right\}/\left\{ \left( 3\,w-3 \right)  \left( \cosh \left( y_{{16}} \right)  \right)^{2}-3\,w\right\},\\
\dot{\Phi}_{3, 1}=&\left\{  \left[ 4\,y_{{12}} \left( \chi+8 \right) {y_{{4}}}^{3}+ \left(  \left( 75\,{y_{{12}}}^{3}+ \left( 3\,{y_{{9}}}^{2}-111\,y_{{2}}+120\,y_{{7}}+9\,y_{{10}}+249 \right) y_{{12}}+9\,y_{{9}}y_{{13}}+9\,y_{{13}}y_{{7}}+36\,y_{{13}}\right. \right.\right.\right.\\& \left.\left.\left.\left.+9\,y_{{14}} \right)\chi+33\,{y_{{12}}}^{3}+ \left( 105\,{y_{{9}}}^{2}+27\,y_{{1}}+3\,y_{{2}}-12\,y_{{7}}-9\,y_{{10}}-33 \right) y_{{12}}-9\,y_{{9}}y_{{13}}-9\,y_{{13}}y_{{7}}-36\,y_{{13}}-9\,y_{{14}} \right) y_{{4}} \right.\right.\\& \left.\left. - \left( \chi-1 \right)\left(  \left(  \left( 18\,y_{{9}}-3\,y_{{7}}-18 \right) y_{{5}}-3\,y_{{6}} \right) y_{{12}}-9\,y_{{13}}y_{{5}} \right)  \right] \sqrt {3}+81\,\sin \left( y_{{15}} \right) y_{{1}}y_{{3}} \left( \cosh \left( y_{{16}} \right) -1 \right)  \left( \cosh \left( y_{{16}} \right) +1 \right)  \right.\\& \left. \left( 1+w \right)\cos \left( y_{{15}} \right) -12\, \left( \chi+8 \right)  \left( y_{{9}}+1 \right) {y_{{4}}}^{3}-12\,y_{{5}} \left( \chi+8 \right) {y_{{4}}}^{2}+ \left[  \left(  \left( -171\,y_{{9}}-252 \right) {y_{{12}}}^{2}-177\,y_{{12}}y_{{13}}-171\,{y_{{9}}}^{3}\right.\right.\right.\\& \left.\left.\left.-252\,{y_{{9}}}^{2}+ \left( -195\,y_{{10}}+777\,y_{{2}}-360\,y_{{7}}-747 \right) y_{{9}}-111\,{y_{{7}}}^{2}+ \left( -27\,y_{{10}}-777 \right) y_{{7}}-111\,y_{{8}}-108\,y_{{10}}-27\,y_{{11}}\right.  \right.\right.\\& \left.\left.\left.+111\,y_{{2}}-666 \right)\chi+\left( -153\,y_{{9}}-72 \right) {y_{{12}}}^{2}-39\,y_{{12}}y_{{13}}-153\,{y_{{9}}}^{3}-72\,{y_{{9}}}^{2}+ \left( -21\,y_{{10}}-81\,y_{{1}}-21\,y_{{2}}+36\,y_{{7}}\right. \right.\right.\\& \left.\left.\left. +99 \right)y_{{9}}+3\,{y_{{7}}}^{2}+ \left( 27\,y_{{10}}+21 \right) y_{{7}}+3\,y_{{8}}+108\,y_{{10}}+27\,y_{{11}}-81\,y_{{1}}-3\,y_{{2}}+18 \right] y_{{4}}+ \left( -75\,{y_{{12}}}^{2}y_{{5}}+ \left( -57\,{y_{{9}}}^{2} \right.\right.\right.\\& \left.\left.\left.\left.+( -9\,y_{{7}}-54 \right) y_{{9}}-3\,{y_{{7}}}^{2}-3\,y_{{8}}-27\,y_{{10}}+111\,y_{{2}}-138\,y_{{7}}-249 \right) y_{{5}}-3\,y_{{6}} \left( 3\,y_{{9}}+4\,y_{{7}}+6 \right)  \right) \chi-33\,{y_{{12}}}^{2}y_{{5}} \right.\\& \left.+ \left( -51\,{y_{{9}}}^{2}+ \left( 9\,y_{{7}}+54 \right)y_{{9}}+3\,{y_{{7}}}^{2}+3\,y_{{8}}+27\,y_{{10}}-27\,y_{{1}}-3\,y_{{2}}+30\,y_{{7}}+33 \right) y_{{5}}+3\,y_{{6}} \left( 3\,y_{{9}}+4\,y_{{7}}+6 \right) \right\}\\&/\left\{ 3\,\left(\chi-1\right) \right\}, \\
\dot{Q}_2=&\left\{-6\, \left( \chi-1 \right) y_{{4}} \left[  \left( 3\,y_{{9}}y_{{12}}-y_{{13}} \right) y_{{4}}+y_{{12}}y_{{5}} \right] \sqrt {3}-27\,y_{{3}}y_{{1}} \left( 1+w \right)  \left( \cosh \left( y_{{16}} \right)  \right) ^{2}+ \left[ -2\,{y_{{4}}}^{4}+ \left( -57\,{y_{{9}}}^{2}-75\,{y_{{12}}}^{2}\right.\right.\right.\\& \left.\left.\left.+55\,y_{{2}}-74\,y_{{7}}-18\,y_{{10}}-111 \right) {y_{{4}}}^{2}+ \left( -74\,y_{{5}}y_{{7}}+18\,y_{{9}}y_{{5}}-148\,y_{{5}}-74\,y_{{6}} \right) y_{{4}}-126\,{y_{{9}}}^{4}+ \left( -252\,{y_{{12}}}^{2}+1860\,\right.\right.\right.\\& \left.\left.\left.y_{{2}}-222\,y_{{7}}-333 \right) {y_{{9}}}^{2}+ \left( -222\,y_{{10}}y_{{7}}-876\,y_{{2}}-444\,y_{{10}}-222\,y_{{11}} \right) y_{{9}}-126\,{y_{{12}}}^{4}+ \left( 120\,y_{{2}}-222\,y_{{7}}-333 \right) {y_{{12}}}^{2} \right.\right.\\& \left.\left.+\left( -222\,y_{{13}}y_{{7}}-444\,y_{{13}}-222\,y_{{14}} \right) y_{{12}}-108\,{y_{{7}}}^{3}-1134\,{y_{{7}}}^{2}+ \left( -540\,y_{{8}}-216\,y_{{2}}-972 \right) y_{{7}}+54\,{y_{{2}}}^{2}+ \left( 420\,\right. \right.\right.\\& \left.\left.\left.y_{{10}} -108 \right)y_{{2}}-213\,{y_{{10}}}^{2}-213\,{y_{{13}}}^{2}-71\,{y_{{5}}}^{2}-648\,y_{{8}} \right] \chi-16\,{y_{{4}}}^{4}+ \left( -51\,{y_{{9}}}^{2}-33\,{y_{{12}}}^{2}-27\,y_{{1}}-19\,y_{{2}}+2\,y_{{7}}\right.\right.\\& \left.\left.+18\,y_{{10}}+3 \right) {y_{{4}}}^{2}+ \left( 2\,y_{{5}}y_{{7}}-18\,y_{{9}}y_{{5}}+4\,y_{{5}}+2\,y_{{6}} \right) y_{{4}}-36\,{y_{{9}}}^{4}+ \left( -72\,{y_{{12}}}^{2}-81\,y_{{1}}-24\,y_{{2}}+6\,y_{{7}}+9 \right) {y_{{9}}}^{2}\right.\\& \left.+ \left( 6\,y_{{10}}y_{{7}}+12\,y_{{2}}+12\,y_{{10}}+6\,y_{{11}} \right)y_{{9}}-36\,{y_{{12}}}^{4}+ \left( -81\,y_{{1}}-12\,y_{{2}}+6\,y_{{7}}+9 \right) {y_{{12}}}^{2}+ \left( 6\,y_{{13}}y_{{7}}+12\,y_{{13}}+6\,y_{{14}} \right) \right.\\& \left.y_{{12}}-54\,y_{{7}}y_{{1}}+ \left( 12\,y_{{10}}+27\,y_{{1}} \right) y_{{2}}-3\,{y_{{10}}}^{2}+ \left( -81+ \left( -54\,w+27 \right) y_{{3}} \right) y_{{1}}-3\,{y_{{13}}}^{2}-{y_{{5}}}^{2} \right\}/\left\{ 108\, \chi \right\},\end{align*}\begin{align*}
\dot{\Sigma}_{+2}=&\left\{ 12\,y_{{4}} \left(  \left( 3\,y_{{9}}y_{{12}}-3\,y_{{12}}-2\,y_{{13}} \right) y_{{4}}-y_{{12}}y_{{5}} \right)  \left( \chi-1 \right) \sqrt {3}-27\, \left( 3\, \left( \cos \left( y_{{15}} \right)  \right) ^{2}-1 \right)  \left( 1+w \right) y_{{1}}y_{{3}} \left( \cosh \left( y_{{16}} \right)  \right) ^{2}\right.\\& \left.+81\,y_{{3}}y_{{1}} \left( 1+w \right)  \left( \cos \left( y_{{15}} \right)  \right) ^{2}+ \left( 8\,{y_{{4}}}^{4}+ \left( 114\,{y_{{9}}}^{2}+150\,{y_{{12}}}^{2}-222\,y_{{2}}+240\,y_{{7}}-60\,y_{{9}}+16\,y_{{10}}+498 \right) {y_{{4}}}^{2}\right.\right.\\& \left.\left.+ \left( 24\,y_{{5}}y_{{7}}-76\,y_{{9}}y_{{5}}+108\,y_{{5}}+24\,y_{{6}} \right) y_{{4}}-504\,{y_{{9}}}^{3}+ \left( -504\,y_{{10}}+864\,y_{{2}} \right) {y_{{9}}}^{2}+ \left( -222\,{y_{{7}}}^{2}-504\,{y_{{12}}}^{2}-336\,\right.\right.\right.\\& \left.\left.\left.y_{{12}}y_{{13}}+216\,y_{{2}}-1554\,y_{{7}}-222\,y_{{8}}-1332 \right) y_{{9}}+ \left( -6\,{y_{{7}}}^{2}-168\,{y_{{12}}}^{2}+216\,y_{{2}}-276\,y_{{7}}-6\,y_{{8}}-498 \right) y_{{10}}-24\,\right.\right.\\& \left.\left.y_{{7}}y_{{11}}+18\,{y_{{5}}}^{2}-36\,y_{{11}} \right) \chi+64\,{y_{{4}}}^{4}+ \left( 102\,{y_{{9}}}^{2}+66\,{y_{{12}}}^{2}+54\,y_{{1}}+ 6\,y_{{2}}-24\,y_{{7}}-156\,y_{{9}}-88\,y_{{10}}-66 \right) {y_{{4}}}^{2}\right.\\& \left.+ \left( -24\,y_{{5}}y_{{7}}-68\,y_{{9}}y_{{5}}-108\,y_{{5}}-24\,y_{{6}} \right) y_{{4}}-144\,{y_{{9}}}^{3}-144\,{y_{{9}}}^{2}y_{{10}}+ \left( 6\,{y_{{7}}}^{2}-144\,{y_{{12}}}^{2}-96\,y_{{12}}y_{{13}}-162\,y_{{1}}\right.\right.\\& \left.\left.+42\,y_{{7}}+6\,y_{{8}}+36 \right) y_{{9}}+ \left( 6\,{y_{{7}}}^{2}-48\,{y_{{12}}}^{2}-54\,y_{{1}}+60\,y_{{7}}+6\,y_{{8}}+66 \right) y_{{10}}+24\,y_{{7}}y_{{11}}-27\,y_{{3}}y_{{1}} \left( 1+w \right) -18\,{y_{{5}}}^{2}\right.\\& \left.+36\,y_{{11}} \right\}/\left\{ 6\,(\chi-1) \right\},\\
\dot{\Sigma}_{-2}=&\left\{  \sqrt {3} \left[  \left(  \left(  \left(  \left( -36\,y_{{9}}-132 \right) y_{{12}}-32\,y_{{13}} \right) {y_{{4}}}^{2}-100\,y_{{4}}y_{{12}}y_{{5}}-504\,{y_{{12}}}^{3}-504\,{y_{{12}}}^{2}y_{{13}}+ \left( -504\,{y_{{9}}}^{2}+ \left( 816\,y_{{2}}-336\,\right.\right.\right.\right.\right.\right.\\& \left.\left.\left.\left.\left.\left.y_{{10}} \right) y_{{9}}-222\,{y_{{7}}}^{2}-1554\,y_{{7}}+240\,y_{{2}}-222\,y_{{8}}-444 \right) y_{{12}}+ \left( -6\,{y_{{7}}}^{2}-168\,{y_{{9}}}^{2}+240\,y_{{2}}-276\,y_{{7}}-6\,y_{{8}}-498 \right)\right.\right.\right.\right.\\& \left.\left.\left.\left. y_{{13}}-12\,y_{{14}} \left( 2\,y_{{7}}+3 \right)  \right) \chi+ \left(  \left( 36\,y_{{9}}-84 \right) y_{{12}}-40\,y_{{13}} \right) {y_{{4}}}^{2}-44\,y_{{4}}y_{{12}}y_{{5}}-144\,{y_{{12}}}^{3}-144\,{y_{{12}}}^{2}y_{{13}}+ \left( -144\,{y_{{9}}}^{2}\right.\right.\right.\right.\\& \left.\left.\left.\left.+ \left( 48\,y_{{2}}-96\,y_{{10}} \right) y_{{9}}+6\,y_{{8}}+6\,{y_{{7}}}^{2}+36-162\,y_{{1}}-24\,y_{{2}}+42\,y_{{7}} \right) y_{{12}}+ \left( 6\,{y_{{7}}}^{2}-48\,{y_{{9}}}^{2}-54\,y_{{1}}-24\,y_{{2}}+60\,y_{{7}}\right.\right.\right.\right.\\& \left.\left.\left.\left.+6\,y_{{8}}+66 \right) y_{{13}}+12\,y_{{14}} \left( 2\,y_{{7}}+3 \right)  \right) \sqrt {3}+81\,y_{{3}}y_{{1}} \left( \cos \left( y_{{15}} \right) -1 \right)  \left( \cos \left( y_{{15}} \right) +1 \right)  \left( w+1 \right)  \left( \cosh \left( y_{{16}} \right)  \right) ^{2}-81\,y_{{3}}y_{{1}}\right.\right.\\& \left.\left. \left( w+1 \right)  \left( \cos \left( y_{{15}} \right)  \right) ^{2}+ \left( -8\,{y_{{4}}}^{4}+ \left( -114\,{y_{{9}}}^{2}-150\,{y_{{12}}}^{2}+258\,y_{{2}}-240\,y_{{7}}-108\,y_{{9}}-72\,y_{{10}}-498 \right) {y_{{4}}}^{2}+ \left( -81\,\right.\right.\right.\right.\\& \left.\left.\left.\left.y_{{7}}y_{{5}}-36\,y_{{9}}y_{{5}}-108\,y_{{5}}-24\,y_{{6}} \right) y_{{4}}-18\,{y_{{5}}}^{2} \right) \chi-64\,{y_{{4}}}^{4}+ \left( -102\,{y_{{9}}}^{2}-66\,{y_{{12}}}^{2}-54\,y_{{1}}-42\,y_{{2}}+24\,y_{{7}}+108\right.\right.\right.\\& \left.\left.\left.\,y_{{9}}+72\,y_{{10}}+66 \right) {y_{{4}}}^{2}+ \left( 24\,y_{{7}}y_{{5}}+36\,y_{{9}}y_{{5}}+108\,y_{{5}}+24\,y_{{6}} \right) y_{{4}}+81\,y_{{3}}y_{{1}} \left( w+1 \right) +18\,{y_{{5}}}^{2} \right]  \right\}/\left\{ 18\,(\chi-1)\right\},\\
\dot{\eta}=&\left\{ \left[  \left( - \left( \cosh \left( y_{{16}} \right)  \right) ^{2}+1 \right) \sqrt {3}\sqrt {y_{{2}}}+\cosh \left( y_{{16}} \right)  \left(  \left( 2\,y_{{4}}\sin \left( y_{{15}} \right) -3\,\cos \left( y_{{15}} \right) y_{{9}} \right) \sqrt {3}+3\,y_{{12}}\cos \left( y_{{15}} \right)  \right) \sinh \left( y_{{16}} \right)  \right] \right.\\& \left.\sqrt {3}\sin \left( y_{{15}} \right)  \right\}/\left\{ 3\,\cosh \left( y_{{16}} \right) \sinh \left( y_{{16}} \right)  \right\},\\
\dot{r}=&-\left\{ \left[ 2\,\cos \left( y_{{15}} \right) \sqrt {3}w \left(\left( \cosh \left( y_{{16}} \right) \right)^2-1 \right)   \sqrt {y_{{2}}}+\sinh \left( y_{{16}} \right)  \left(  \left( 3\, \left( \cos \left( y_{{15}} \right)  \right) ^{2}y_{{9}}+3\,w-y_{{9}}-2\,\cos \left( y_{{15}} \right)\sin \left( y_{{15}} \right) \right.\right.\right.\right.\\& \left.\left.\left.\left. y_{{4}}-1 \right) \sqrt {3}-2\, \left( \cos \left( y_{{15}} \right)  \right) ^{2}y_{{12}}+2\,y_{{12}} \right) \cosh \left( y_{{16}} \right)  \right] \sqrt {3} \right\}/\left\{ 3\, \left[ \left( w-1 \right)  \left( \cosh \left( y_{{16}} \right)  \right) ^{2}-w\right] \right\} .
  \end{align*}
Subject to the constraints:
      \begin{align*}
E_{00}=&\left\{ -6\, \left( \chi-1 \right)  \left(  \left( 3\,y_{{9}}y_{{12}}-y_{{13}} \right) y_{{4}}+y_{{12}}y_{{5}} \right) y_{{4}}\sqrt {3}-27\,y_{{1}}y_{{3}} \left( 1+w \right)  \left( \cosh \left( y_{{16}} \right)  \right) ^{2}+ \left( -2\,{y_{{4}}}^{4}+ \left( -57\,{y_{{9}}}^{2}-75\,{y_{{12}}}^{2}\right.\right.\right.\\& \left.\left.\left.+55\,y_{{2}}-74\,y_{{7}}-18\,y_{{10}}-111 \right) {y_{{4}}}^{2}+ \left( -2\,y_{{5}}y_{{7}}+18\,y_{{9}}y_{{5}}+68\,y_{{5}}-2\,y_{{6}} \right) y_{{4}}-126\,{y_{{9}}}^{4}+ \left( -252\,{y_{{12}}}^{2}+132\,y_{{2}}\right.\right.\right.\\& \left.\left.\left.-222\,y_{{7}}-333 \right) {y_{{9}}}^{2}+ \left( -6\,y_{{10}}y_{{7}}-444\,y_{{2}}+204\,y_{{10}}-6\,y_{{11}} \right) y_{{9}}-126\,{y_{{12}}}^{4}+ \left( 120\,y_{{2}}-222\,y_{{7}}-333 \right) {y_{{12}}}^{2}\right.\right.\\& \left.\left.+ \left( -6\,y_{{7}}y_{{13}}+204\,y_{{13}}-6\,y_{{14}} \right) y_{{12}}+54\,{y_{{7}}}^{2}+324\,y_{{7}}+54\,{y_{{2}}}^{2}+ \left( -12\,y_{{10}}+108 \right) y_{{2}}+3\,{y_{{10}}}^{2}+3\,{y_{{13}}}^{2}+{y_{{5}}}^{2}\right.\right.\\& \left.\left.+108\,y_{{8}} \right) \chi-16\,{y_{{4}}}^{4}+ \left( -51\,{y_{{9}}}^{2}-33\,{y_{{12}}}^{2}-9\,y_{{1}}-19\,y_{{2}}+2\,y_{{7}}+18\,y_{{10}}+3 \right) {y_{{4}}}^{2}+ \left( 2\,y_{{5}}y_{{7}}-18\,y_{{9}}y_{{5}}+4\,y_{{5}}\right. \right.\\&\left. \left.+2\,y_{{6}} \right)y_{{4}}-36\,{y_{{9}}}^{4}+ \left( -72\,{y_{{12}}}^{2}-27\,y_{{1}}-24\,y_{{2}}+6\,y_{{7}}+9 \right) {y_{{9}}}^{2}+ \left( 6\,y_{{10}}y_{{7}}+12\,y_{{2}}+12\,y_{{10}}+6\,y_{{11}} \right) y_{{9}}-36\,{y_{{12}}}^{4}\right.\\& \left.+ \left( -27\,y_{{1}}-12\,y_{{2}}+6\,y_{{7}}+9 \right) {y_{{12}}}^{2}+ \left( 6\,y_{{7}}y_{{13}}+12\,y_{{13}}+6\,y_{{14}} \right) y_{{12}}+ \left( 12\,y_{{10}}-27\,y_{{1}} \right) y_{{2}}+27\,y_{{3}}wy_{{1}}-3\,{y_{{10}}}^{2}\right.\\& \left.-3\,{y_{{13}}}^{2}-{y_{{5}}}^{2}+27\,y_{{1}} \right\}/\left\{ 9\, y_{1} \right\},\\
E_{01}=&\left\{ \left[216\,{y_{{2}}}^{3/2}\chi\,\sqrt {3}y_{{9}}+ \left(  \left( \left( -168\,\chi-48 \right) {y_{{9}}}^{3}+ \left(  \left( -38\,{y_{{4}}}^{2}-168\,{y_{{12}}}^{2}-222\,y_{{7}}-444 \right) \chi+12-34\,{y_{{4}}}^{2}-54\,y_{{1}}\right.\right.\right.\right.\right.\\& \left.\left.\left.\left.\left.+6\,y_{{7}}-48\,{y_{{12}}}^{2} \right) y_{{9}}-6 \, \left( \chi-1 \right)  \left( -3\,{y_{{4}}}^{2}-3\,y_{{4}}y_{{5}}+y_{{10}} \left( y_{{7}}+3 \right) +y_{{11}} \right)  \right) \sqrt {3}-18\,y_{{12}}{y_{{4}}}^{2} \left( \chi-1 \right)  \right) \sqrt {y_{{2}}}+27\,y_{{3}}\right.\right.\\& \left.\left. \cosh \left( y_{{16}} \right)\sqrt {3}\sinh \left( y_{{16}} \right) \cos \left( y_{{15}} \right) y_{{1}} \left( 1+w \right)  \right] \sqrt {\chi}\right\}/\left\{  27\,{y_{{1}}}^{3/2}\sqrt {\beta} \right\},\\
E_{03}=&\left\{  \left[ -3\,y_{{4}} \left( 37\,\chi-1 \right) \sqrt {3}{y_{{2}}}^{3/2}+ \left(  \left(  \left( 32+4\,\chi \right) {y_{{4}}}^{3}+ \left(  \left( 57\,{y_{{9}}}^{2}+75\,{y_{{12}}}^{2}+111\,y_{{7}}+27\,y_{{9}}+27\,y_{{10}}+222 \right) \chi-27\,y_{{9}}\right.\right.\right.\right.\right.\\& \left.\left.\left.\left. \left.+33\,{y_{{12}}}^{2}-6+51\,{y_{{9}}}^{2}-27\,y_{{10}}+27\,y_{{1}}-3\,y_{{7}} \right) y_{{4}}+3\, \left( \chi-1 \right)  \left(  \left( y_{{7}}+3 \right) y_{{5}}+y_{{6}} \right)  \right) \sqrt {3}+27\, \left(  \left( -1+2\,y_{{9}} \right) y_{{12}}\right. \right.\right.\right.\\& \left.\left.\left.\left.-y_{{13}} \right) \left( \chi-1 \right) y_{{4}} \right) \sqrt {y_{{2}}}+27\,y_{{3}}\cosh \left( y_{{16}} \right) \sqrt {3}\sin \left( y_{{15}} \right) \sinh \left( y_{{16}} \right) y_{{1}} \left( 1+w \right)  \right] \sqrt {\chi}\beta \right\}/\left\{ 27\, \left( \beta\,y_{{1}} \right) ^{3/2} \right\}.
  \end{align*}

\twocolumn
\balance
 \bibliographystyle{utcaps}


\end{document}